\begin{document}
	
	\begin{center} {\bf\Large  Revealing Cluster Structures Based on\\ Mixed Sampling Frequencies\footnote{Opinions expressed herein are those of the authors alone and do not necessarily reflect the views of the Federal Reserve System. 
	We thank Gianni Amisano, Eric Ghysels, Michael Owyang and the participants of  Midwest Econometrics Group 2018 Meeting, the joint meeting of 11th International Conference of the ERCIM WG on Computational and Methodological Statistics and 12th International Conference on Computational and Financial Econometrics, and the Federal Reserve System Econometrics conference in 2020. Rho and Liu are partially supported by NSF-CPS grant \#1739422.
	A primitive  version of this paper had been circulated under the title ``Panel Nonparametric MIDAS Model: A Clustering Approach" by the first two authors.}	
}\end{center}
	\centerline{\textsc{Yeonwoo Rho$^1$, Yun Liu$^{2}$, and Hie Joo Ahn$^3$
			}\footnote{
	 Emails: Y. Rho (yrho@mtu.edu), Y. Liu (AnnaLiu@quickenloans.com), and H. Ahn (HieJoo.Ahn@frb.gov)
				}}
		
		\bigskip
	\centerline {$^1$Michigan Technological University}
		\centerline {$^2$Quicken Loans}
	\centerline {$^3$Federal Reserve Board}
	\bigskip
	\centerline{\today}
	\bigskip

	\begin{abstract}
This paper proposes a new linearized mixed data sampling (MIDAS) model and develops a framework to infer clusters in a panel regression with mixed frequency data. The linearized MIDAS estimation method is more flexible and substantially simpler to implement than competing approaches. We show that the proposed clustering algorithm successfully recovers true membership in the cross-section, both in theory and in simulations, without requiring prior knowledge of the number of clusters. This methodology is applied to a mixed-frequency Okun’s law model for state-level data in the U.S. and uncovers four meaningful clusters based on the dynamic features of state-level labor markets.	
\\
		\textit{key words}: Clustering; forecasting; mixed data sampling regression model; panel data; penalized regression.
	\end{abstract}
	
	%% To check the section titles  

%	\tableofcontents

\bigskip
%\pagebreak		
\section{Introduction}

%Electric power demand forecasting using interval time series: a comparison between VAR and iMLP

Following technological advances, the diffusion of social media, and the efforts of statistical agencies and private companies, new data sources have recently become available for empirical research in economics. In many cases, these data are characterized by large time series and cross sectional dimensions, with detailed information on economic agents often at a level of disaggregation more granular than that of traditional data sources. Due to the increasing availability of richer cross-section data, it has become particularly important to efficiently summarize and identify the most important features of subjects in the cross-section.
With the increased capacity in handling higher sampling frequency data, the mixed data sampling (MIDAS) models (e.g., \cite{lags}) are now widely used in practical applications in forecasting variables that inherently have a low sampling frequency. One of the purposes of MIDAS models is to understand how high-frequency variables are related to the low-frequency variables of interest, which is essentially captured by the shape of MIDAS weight function. With the aim of properly accounting for unit heterogeneity, this paper proposes a new empirical method to identify distinct groups in a panel data based on their MIDAS weights.

Previous studies on MIDAS models with panel data (e.g. \cite{andreou}) arbitrarily divided observations into different groups. This approach crucially depends on prior knowledge and homogeneity within each group is not necessarily guaranteed. To avoid such issues, we propose to construct an entirely data-driven clustering algorithm by adapting \cite{Ma&Huang:2017}'s clustering idea to a panel setting. This clustering is based on how close coefficient estimations are between subjects in the panel. Closer estimates are penalized, resulting in unanimous coefficient estimates within the same cluster. An important feature of this clustering idea is that the clustering procedure is entirely data-driven, including the number of clusters. 

The proposed clustering algorithm can be combined with, in principle, any MIDAS models.\footnote{This is because the proposed clustering algorithm can be applied to a general panel data setting. In this paper, we focus on its applications to mixed frequency settings for brevity.} In practice, however, the accompanying MIDAS estimation  procedure needs to be computationally efficient. This is because the penalized regression idea of \cite{Ma&Huang:2017} requires a complicated optimization process as well as grid searches of a few tuning parameters. Unfortunately, existing MIDAS models are either too complicated or too time-consuming to be used in combination with the clustering methods with penalized regression.  In parametric MIDAS models, arbitrary parametric functions (e.g., exponential Almon lag function, beta function) are used to model the coefficients on high-frequency variables. As these parametric functions are highly nonlinear in general, complicated numerical optimization is required for estimation.  As this estimation is numerically costly and challenging, practitioners often give MIDAS models a wide berth. The recently proposed nonparametric MIDAS model by  \cite{nonparametric} does not require any arbitrary choice of parametric functional forms in specifying the distributed lags structure of the coefficients on high frequency variables. However, the tuning parameters of \cite{nonparametric}’s methodology require demanding numerical search, which adds heavy computational burden to the proposed clustering algorithm. 
 
This paper fills this gap in the literature  by devising a novel linearized MIDAS model based on the Fourier flexible form and polynomials. The Fourier flexible form and polynomials allow the trajectory of coefficients on high-frequency variables to be flexibly determined by the data. Our model requires just ordinary least squares (OLS), which eschews estimation difficulty. %\footnote{A recent study by \cite{ghysels2019} develops a general framework with dictionaries for machine-learning time-series regression models, and recommends to use Legendre polynomials for the coefficients on high-frequency variables in a MIDAS model. Our use of trigometric functions for the MIDAS coefficients is a special case of \cite{ghysels2019}'s framework, and inherits advantages similar to those of Legendre polynomials in the estimation.} 
Unlike  \cite{nonparametric}’s methodology, the tuning parameters involved with our method do not require heavy computations. In addition, an arbitrary choice of these tuning parameters work reasonably well, a powerful feature when combining MIDAS models with the proposed clustering method.
 
%\cite{nonparametric} have recently proposed a nonparametric MIDAS model that, like ours, does not require to specify any functional form for the coefficients on the high-frequency.  Unlike our approach, however, the nature of the tuning parameter involved in \cite{nonparametric}’s methodology requires a more careful setting, which often is much more computationally demanding.

Simulations conducted in this paper show many desirable features of our method.  First, the proposed linearized MIDAS model tends to provide better one-step-ahead forecasts than \cite{nonparametric}.\footnote{Since \cite{nonparametric} reported better estimation and forecasting results of their method compared to parametric MIDAS models, we can infer that our method would also have a similar advantage to the parametric MIDAS.} Second, for our MIDAS method, an arbitrary choice of  tuning parameters works reasonably well, thought it can be further improved by a data-driven choice. Third, our MIDAS clustering method is faster in computation and yields more precise parameter estimation and better forecasting properties than other clustering approaches in the data environment of mixed-sampling frequencies.

As a relevant empirical application, we use our method to explore heterogeneity in labor market dynamics across states in the U.S. using a mixed-frequency panel Okun’s law model. Okun’s law is an empirical relationship that relates changes in unemployment rate to GDP growth.  Usually an Okun’s law model is specified at quarterly frequency, as GDP growth is available only quarterly. In our application, we include weekly initial claims of unemployment insurance (UI) benefits as the high-frequency indicator, which is known as the most timely indicator of job losses. By doing so, the model can better characterize the sudden rise in unemployment rate at the onset of a recession, picking up sudden bursts in layoffs. An additionally desirable feature of the mixed-frequency Okun’s law model is that it can be used to nowcast the unemployment rate at the state level on a weekly basis.

The algorithm identifies four clusters among states based on the responsiveness of unemployment rate to GDP growth and on the pattern of coefficients on weekly initial claims within the quarter. The coefficients on GDP growth and initial claims most likely reflect the structural aspects of state-level labor markets (e.g. the industry composition) and the local labor-market practices of hirings and layoffs.  Hence, the clusters identified by the model most likely capture relevant heterogeneity in the functioning of labor markets in different states.

We relate the identified clusters to observable state–level attributes such as the small-firms employment share, industry composition, the relevance of oil production, and the share of long-term unemployment out of total unemployment. Each cluster exhibits multi-dimensional attributes, suggesting that the differences in labor-market dynamics across states cannot be determined or accurately summarized by one or two observable factors. Another way to say this is that the clustering algorithm is able to capture a state’s unobserved attributes which are not fully reflected in the data but are nevertheless crucial for unemployment dynamics. In this regard, our proposed methodology can reveal similarities and differences across states in the functioning of their labor markets purely based on the data, and can provide a new understanding of regional heterogeneity in labor market dynamics.

The rest of this paper is organized as follows. 
Section \ref{sec:nonparaMIDAS} introduces the proposed linearized MIDAS approach using the Fourier flexible form and polynomials. 
Subsection \ref{parameter} introduces  the linearized MIDAS estimation in a non-panel setting. Subsection \ref{SParameter} demonstrates our linearized MIDAS method’s estimation and forecasting accuracy in finite samples.
Section \ref{sec:panelMIDAS} presents the clustering algorithm. The proposed clustering approach accompanied with our MIDAS method delivers accurate estimates, as proven in theory and shown in the finite sample simulations. 
Section \ref{Application} provides an empirical application of the method.
Details on algorithms used in simulations and technical proofs are relegated to Sections \ref{appen:algorithms} and  \ref{appen:proofs}, respectively, in the supplementary material.

The following notation will be used throughout the paper. The $p$-norm of a vector $x=(x_1,\ldots,x_m)'$ is $||x||_p=(\sum_{i=1}^m |x_i|^p)^{1/p}$.
For an $m\times n$ matrix $A$ with its $(i,j)$th element being $a_{ij}$, $||A||_p$ indicates the $p$-norm induced by the corresponding vector norm. That is, $||A||_p=\sup_{x\neq 0}||Ax||_p/||x||_p$. In particular, $||A||_1=\max_{j=1,\ldots,n}\sum_{i=1}^m|a_{ij}|$ and $||A||_\infty=\max_{i=1,\ldots,m}\sum_{j=1}^n|a_{ij}|$.
For a symmetric and positive definite matrix $A$, let $\lambda_{\min}(A)$ and $\lambda_{\max}(A)$ indicate the smallest and largest eigenvalues of $A$, respectively. 
It is worth noting that
$||A||_2=\lambda_{\max}(A)$.
$I_p$ is a $p\times p$  identity matrix and $\otimes$ denotes the Kronecker product.
For any real number $x$, $\lfloor x\rfloor$ denotes the largest integer that is smaller than or equal to $x$.
%The symbol $C$ indicates a constant that does not depend on any parameters. The values of $C$ may be different from line to line.
The symbol ${\bf 1}\{\cdot\}$ denotes the indicator function.

\section{Linearized MIDAS}\label{sec:nonparaMIDAS}

In this section, we  introduce our linearized MIDAS approach using the Fourier flexible form  \citep{GALLANT1981211}. We first introduce the framework, then  confirm that the proposed linearized MIDAS model  is a good approximation of popular parametric MIDAS models in finite samples.

\subsection{Linearized MIDAS with the Fourier flexible form and polynomials}\label{parameter}
Consider the following MIDAS model with the forecast lead $h\geq0$:
\begin{equation}\label{midas}
y_{t+h}=\sum_{i=1}^q\alpha_iz_{t,i}
+\sum_{j=0}^{m-1}\beta_{j/m}^*x_{t,j}+\varepsilon_{t+h}=\mathbf{z}_t'\boldsymbol{\alpha}+{\mathbf{x}_t}'\boldsymbol\beta^*+\varepsilon_{t+h},
\end{equation}
for $t=1,\dots,T$.
Here, $\mathbf{z}_{t}$ is the $q$-vector of low-frequency covariates at time $t$, and $\boldsymbol{\alpha}=(\alpha_1,\ldots,\alpha_q)'$ is the corresponding coefficient vector. The vector $\mathbf{x}_t=(x_{t,0},\dots,x_{t,m-1})'$ is the high-frequency variable at $t$  and $\boldsymbol\beta^*=(\beta_{0/m}^*,\dots,\beta_{m-1/m}^*)'$ are the coefficients that aggregate $\mathbf{x}_t$ to the low-frequency.
In a parametric MIDAS model, the coefficients $\beta_{j/m}^*$ can be written as a multiple of $\omega_j(\boldsymbol{\theta})$, where weights  $\omega_j(\boldsymbol{\theta})$ are assumed to be generated by, for example, an exponential Almon lag function
$$\beta_{j/m}^*=\alpha^*\omega_j(\boldsymbol{\theta})=\frac{\alpha^*\exp(\theta_1 j+\theta_2j^2+\cdots+\theta_Qj^Q)}{\sum_{i=0}^{m-1}\exp(\theta_1 i+\theta_2i^2+\cdots+\theta_Qi^Q)},$$
and $\alpha^*$ and $\boldsymbol{\theta}=(\theta_1,\theta_2,\ldots,\theta_Q)$ are parameters that need to be  estimated from data.
However, the  form of $\omega_j(\cdot)$ is somewhat limited, and it requires nonlinear estimation.
In this paper, we propose to model $\beta_{j/m}^*$ using the Fourier flexible form and polynomials.
The MIDAS coefficients $\beta^*_{j/m}$ are assumed to be generated by 
\begin{equation}\label{eq:fourier}
\beta_{j/m}^*=\sum_{l=0}^L\beta_l(j/m)^l+\sum_{k=1}^K\left\{\beta_{1,k}\sin(2\pi k\cdot j/m) +\beta_{2,k}\cos(2\pi k\cdot j/m)\right\},
\end{equation}
 for some positive integers $L$ and $K$.
The Fourier flexible form has been frequently used in macroeconomics and finance since \cite{GALLANT1981211}. It has been demonstrated that the Fourier flexible form is capable of approximating most forms of nonlinear time trends to any degree of accuracy if a sufficient number of parameters is used, and that a small $K$ is often enough to reasonably approximate smooth functions with finite numbers of breaks \citep{Becker2004,Becker2006,Enders2012,Taylor2012,Burak2017,perron2017testing}. 
In addition to  the Fourier flexible form, we also consider a few polynomial trends to cover wider range of  nonlinear functions, following suggestions in \cite{perron2017testing}.\footnote{Our approach assumes that the underlying nonlinear MIDAS weight function can be approximated by (\ref{eq:fourier}) with fixed $K$ and $L$. If $K,L\to\infty$ as $m\to\infty$, it is well-known that any bounded function $\beta_{\cdot}^*$ can be precisely approximated by (\ref{eq:fourier}); in this case, the proposed MIDAS model can be considered nonparametric. However, as seen in our simulations, (\ref{eq:fourier}) works reasonably well with an arbitrary choice of relatively small $K$ and $L$. In this paper, we choose to make our argument with fixed $K$ and $L$, and call our method ``linearlized" to emphasize its computational advantage and its ability to concisely represent a wide range of nonlinear MIDAS weight functions.}

The MIDAS model (\ref{midas}) with the Fourier flexible form (\ref{eq:fourier}) can be expressed as
\begin{equation*}\label{combine}
\mathbf{y}=\mathbf{Z}\boldsymbol\alpha+\mathbf{X}\boldsymbol\beta^*+\boldsymbol{\varepsilon}=\mathbf{Z}\boldsymbol\alpha+\widetilde{\mathbf{X}}\boldsymbol\beta+\boldsymbol{\varepsilon}=\mathbf{W}\boldsymbol\gamma+\boldsymbol{\varepsilon},
\end{equation*}
where $\mathbf{y}=(y_{1+h},\ldots,y_{T+h})'$, $\boldsymbol{\varepsilon}=(\varepsilon_{1+h},\ldots,\varepsilon_{T+h})'$,  $\mathbf{Z}=\left[\mathbf{z}_1,\cdots,\mathbf{z}_T\right]'$,  $\mathbf{X}=\left[\mathbf{x}_1,\cdots,\mathbf{x}_T\right]'$, $\mathbf{W} = (\mathbf{Z}, \widetilde{\mathbf{X}})$, $\widetilde{\mathbf{X}}=\mathbf{XM'}=\left[\widetilde{\mathbf{x}}_1,\cdots,\widetilde{\mathbf{x}}_T\right]'$, 
and
\begin{equation}\label{M}
{M}=\left[\begin{matrix}
 (0/m)^0 & (1/m)^0 & \cdots & ((m-1)/m)^0\\
 \vdots & \vdots &  & \vdots\\
 (0/m)^L & (1/m)^L & \cdots & ((m-1)/m)^L\\
 \sin(2\pi\cdot1\cdot0/m) & \sin(2\pi\cdot1\cdot1/m) & \cdots & \sin(2\pi\cdot1\cdot(m-1)/m)\\
 \cos(2\pi\cdot1\cdot0/m) & \cos(2\pi\cdot1\cdot1/m) &  \cdots & \cos(2\pi\cdot1\cdot(m-1)/m)\\
 \vdots & \vdots &  & \vdots\\
 \sin(2\pi\cdot K\cdot0/m) & \sin(2\pi\cdot K\cdot1/m) & \cdots & \sin(2\pi\cdot K\cdot(m-1)/m)\\
\cos(2\pi\cdot K\cdot0/m) & \cos(2\pi\cdot K\cdot1/m) & \cdots & \cos(2\pi\cdot K\cdot(m-1)/m)\\
\end{matrix}\right].
\end{equation}
Here, the matrix $M$ can be understood as a Fourier transform operator. This Fourier transformation summarizes the information in an $m$-dimensional vector $\mathbf{x}_t$ into a $(2K+L+1)$-dimensional vector $\widetilde{\mathbf{x}}_t=\mathbf{Mx}_t=(\widetilde{x}_{t,0},\widetilde{x}_{t,1},\cdots,\widetilde{x}_{t,L},\widetilde{x}^{(s)}_{t,1},\widetilde{x}^{(c)}_{t,1},\cdots,\widetilde{x}^{(s)}_{t,K},\widetilde{x}^{(c)}_{t,K})',$ where $\widetilde{x}_{t,l}$, $\widetilde{x}^{(s)}_{t,k}$ and $\widetilde{x}^{(c)}_{t,k}$ are transformed high-frequency data for $l=0,\dots,L$ and $k=1,\dots,K$, and are defined as
$\widetilde{x}_{t,l}=\sum_{j=0}^{m-1}(j/m)^lx_{t,j}$, $\widetilde{x}^{(s)}_{t,k}=\sum_{j=0}^{m-1}\sin(2\pi kj/m)x_{t,j}$, and $\widetilde{x}^{(c)}_{t,k}=\sum_{j=0}^{m-1}\cos(2\pi kj/m)x_{t,j}.$\footnote{Note that $\widetilde{\mathbf{x}}_t$ can effectively summarize the information in $\mathbf{x}_t$, because relatively small $K$ and $L$ are enough to capture main characteristics of a nonlinear trend function. For instance, \cite{Enders2012} reported that even a single frequency $K=1$ allows for multiple smooth breaks.}.

Unlike parametric MIDAS models, this model is linear. Noting that  $\boldsymbol{\beta}^*=\mathbf{M'}\boldsymbol{\beta}$, the ordinary least squares (OLS) estimator of $\boldsymbol{\beta}^*$ can be written as  \begin{equation}\label{beta} \widehat{\boldsymbol\beta^*}=\mathbf{M'D}\widehat{\boldsymbol\gamma}=\mathbf{M'D}(\mathbf{W'W})^{-1}\mathbf{W'y}=\boldsymbol\beta^*+\mathbf{M'D}\left(\dfrac{1}{T}\mathbf{W'W}\right)^{-1}\left(\dfrac{1}{T}\mathbf{W'}\boldsymbol{\varepsilon}\right),\end{equation}
where $\mathbf{D}=\left[\mathbf{0}_{(L+1+2K)\times q}, {I}_{L+1+2K}\right]$, and $\mathbf{I}_{L+1+2K}$ is an identity matrix. 
Under some regularity conditions,  $\boldsymbol\beta$ can be estimated consistently by the OLS estimator $\widehat{\boldsymbol{\beta}^*}$.

\subsection{Simulation: linearized MIDAS}\label{SParameter}
%\subsubsection{Settings and Comparable Method}
This subsection consists of two parts. The first part compares our proposed linearized MIDAS estimation with an existing  nonparametric MIDAS \citep{nonparametric}. \cite{nonparametric} imposes a smoothness condition  on $\beta_{j/m}^*$, which involves a tuning parameter. In our simulations, their tuning parameter is chosen by a modified Akaike information criterion (AIC) as proposed in \cite{nonparametric}. See Section  \ref{appen:BB} in the supplementary material  for more details. Our method is based on a generic choice of $L$ and $K$ with $L=2$ and $K=3$.
The second part of this subsection investigates if data-driven choices of $L$ and $K$ can improve the quality of our linearized MIDAS estimation.

The simulation setting considered in the first part is similar to that of  \cite{nonparametric}. For $j=0,\dots,m-1,\ t=1,\dots,T$,
\begin{equation}\label{DGP}
y_{t+h}=\alpha_0+\sum_{j=0}^{m-1}\beta^*_{j}x_{t,j}+\varepsilon_{t+h}, ~~x_{t,j}=c+d x_{t,j-1}+u_{t,j},
\end{equation}
where $\varepsilon_{t+h}\stackrel{iid}{\sim} N(0,0.125)$, $u_{t,j}\stackrel{iid}{\sim} N(0,1)$, $\alpha_0=0.5$, $\beta^*_j=\alpha_1\omega_j(\boldsymbol\theta)$, $\alpha_1\in\{0.2, 0.3, 0.4\}$, $T\in\{100, 200, 400\}$, and the frequency ratio $m\in\{20, 40, 60, 150, 365\}$. For the AR(1) high-frequency regressor, $c=0.5$ and $d=0.9$ are considered.
Five MIDAS weight functions $\omega_j(\boldsymbol{\theta})$ are considered: 
\begin{itemize}
	\item Exponential Decline: $\omega_{j}(\theta_1,\theta_2)=\dfrac{\exp\{\theta_1j+\theta_2j^2\}}{\sum_{i=0}^{m-1}\exp\{\theta_1i+\theta_2i^2\}},~\theta_1=7\times10^{-4},~\theta_2=-6\times10^{-3}$;
	\item Hump-Shaped: $\omega_{j}(\theta_1,\theta_2)=\dfrac{\exp\{\theta_1j-\theta_2j^2\}}{\sum_{i=0}^{m-1}\exp\{\theta_1i-\theta_2i^2\}},~\theta_1=0.08,~\theta_2=2\theta_1/m$;
	\item Linear Decline: $\omega_{j}(\theta_1,\theta_2)=\dfrac{\theta_1+\theta_2(j-1)}{\theta_1 m+\theta_2 m(m+1)/2},~\theta_1=1,~\theta_2=0.05$;
	\item Cyclical: $\omega_{j}(\theta_1,\theta_2)=\dfrac{\theta_1}{m}\left\{\sin\left(\theta_2+2\pi\dfrac{j}{m-1}\right)\right\},~\theta_1=100/m,~\theta_2=0.01$;
	\item Discrete: $\omega_{j}=(0,0,\cdots,0,5/m,\cdots,5/m)$ where the  value $5/m$ is assigned to the last one fifth elements and 0 to the rest.
\end{itemize}

%The first weight function is also known as the exponential Almon lag polynomial function proposed by \cite{lags}, which is capable of mimicking various shapes with a small number of parameters. Here,  the simple two-parameter models are considered for the exponential Almon lag polynomial. The last one among all weight functions illustrates the flexibility of the methods more according to \cite{nonparametric}. All weights are positive and normalized to sum up to one. 
For the evaluation of the estimation accuracy, the root mean square errors (RMSE) of estimators of $\boldsymbol{\beta}^*=(\beta_0^*,\dots,\beta_{m-1}^*)'$ are considered.
Our estimator $\widehat{\boldsymbol{\beta}}$ is brought back to the original scale by taking $\mathbf{M'}\widehat{\boldsymbol{\beta}}$. The RMSE of our method is calculated as $RMSE = \|\mathbf{M'}\widehat{\boldsymbol{\beta}}-{\boldsymbol{\beta}^*}\|_2.$
 The number of Monte-Carlo (MC) replications is  1000.
 For the comparison of forecasting accuracy, the root mean square forecast error (RMSFE) of the one-step-ahead forecast is considered. The number of MC replications is 250. 
The RMSFE is calculated as  following:
\begin{enumerate}
\item Obtain the estimated parameter  $\widehat{\boldsymbol{\beta}^*}_{T/2}$ in the regression model $y_{t+h}={\mathbf{x}_{t}}'\boldsymbol{\beta}^*+\varepsilon_{t+h}$ for $t=1,\cdots,T/2$.% Denote the estimated parameter as $\widehat{\boldsymbol{\beta}^*}$.
\item Calculate the one-step-ahead forecast using $\widehat{\boldsymbol{\beta}^*}_{T/2}$, that is,  $\widehat{y}_{T/2+h+1}={\mathbf{x}_{T/2+1}}'\widehat{\boldsymbol{\beta}^*}_{T/2}$.
\item Repeat steps 1-2 and obtain $\widehat{y}_{T/2+h+k}=\mathbf{x}_{T/2+k}\widehat{\boldsymbol{\beta}^*}_{T/2+k-1}$ for $k=2,\dots,T/2$. Here, $\widehat{\boldsymbol{\beta}^*}_{T/2+k-1}$ is calculated using $(y_{t+h},\mathbf{x}_t')$ for all $t=k,\ldots,T/2+k-1$.
\item Once the estimated responses $\widehat{y}_{t+h}$ for $t=T/2+1,\dots,T$ are calculated, calculate the RMSFE of the predicted response: $RMSFE = \sqrt{(2/T)\sum_{k=1}^{T/2}(\widehat{y}_{T/2+h+k}-{y}_{T/2+h+k})^2}.$
\end{enumerate}

Table \ref{MSE} presents the medians of  RMSEs of $\boldsymbol{\beta}$ estimation using  \cite{nonparametric} (B\&R) and our method (Fourier). For both methods, the estimation accuracy generally increases as the frequency ratio or the sample size become larger. For all five shapes of MIDAS weights,  our approach substantially improves estimation accuracy compared with B\&R's method. This improvement is more substantial when the sample size $T$ or the frequency ratio $m$ is relatively large. This finding implies that our approach tends to  capture the flexibility of various shapes of MIDAS weights more precisely than B\&R's approach. 
Another notable feature is that $\alpha_1$ does not have much effect on the accuracy of the estimation for both methods. It seems that  the MIDAS shape matters,  but not the magnitude of the signal.
Table \ref{OOS} presents the median one-step ahead RMSFEs. For both methods, the forecasts become more accurate as the sample size $T$, or the frequency ratio $m$ increases for all five MIDAS shapes. In general, the Fourier flexible form tends to provide slightly more precise forecasts compared with the B\&R's method. These results show that the proposed the Fourier flexible form approach tends to deliver more accurate estimation and forecasting compared with a competing  method.

It is remarkable that our linearized approach using the Fourier flexible form and polynomials generally outperforms \cite{nonparametric}'s method in terms of estimation and forecasting accuracy despite its disadvantage in the tuning parameter selections. The tuning parameter in \cite{nonparametric}, $\lambda_{BR}$ in  Section  \ref{appen:BB} in the supplementary material, requires a careful choice of its range, for which trial-and-error is often the only option. Then the objective function $Q_{BR}$ in Section \ref{appen:BB} should be optimized over a fine grid within that range. For instance, if one searches over $\lambda_{BR}\in(0,100)$ and considers 100 equally-spaced grid points, \cite{nonparametric}'s MIDAS needs to solve at least 100 quadratic programmings with constraints. This process can be computationally demanding. On the contrary, our linearized MIDAS with an arbitrary choice of $(L,K)=(2,3)$ requires only one OLS estimation.

The second part of this subsection shows that  our method  can be further improved by using a data-driven choice of the tuning parameters $(L,K)$. Notice that even if we consider data-driven $(L,K)$, the search for the optimal tuning parameters is not as demanding as that of \cite{nonparametric}. This is because $L$ and $K$ should be nonnegative integers, rather than real numbers. In addition, relatively small $L$ and $K$ would be enough to approximate most forms of MIDAS weights as mentioned in, for instance, \cite{Enders2012}. The simulation setting in the second part is the same as previous one. Results for $m=20$, $T=100$, $\alpha_1=0.2$ are  presented, as other cases delivered similar results in unreported simulations. The number of MC replications is 1000. The tuning parameters $L,K=0,1,2,3,4$ are considered, which requires only 25 OLS estimations per each MC replication.

Three popular information criteria\textemdash AIC,  modified AIC, and Bayesian information criterion (BIC)\textemdash are considered:
\begin{equation}\label{nonpara:AIC}
    AIC_{L,K}=\log\left\{(\mathbf{y}-W\widehat{\boldsymbol{\gamma}})'(\mathbf{y}-W\widehat{\boldsymbol{\gamma}})\right\}+2\left(K+L+3\right)/T.
\end{equation}
\begin{equation}\label{nonpara:AICc}
    AICc_{L,K}=\log\left\{(\mathbf{y}-W\widehat{\boldsymbol{\gamma}})'(\mathbf{y}-W\widehat{\boldsymbol{\gamma}})\right\}+2\left(K+L+3\right)/(T-K-L-4).
\end{equation}
\begin{equation}\label{nonpara:BIC}
    BIC_{L,K}=\log\left\{(\mathbf{y}-W\widehat{\boldsymbol{\gamma}})'(\mathbf{y}-W\widehat{\boldsymbol{\gamma}})\right\}+log(T)\left(K+L+3\right)/T.
\end{equation}
$AICc_{L,K}$ is similar to the modified AIC considered in \cite{nonparametric}.  

The left half of Table \ref{table:LK} presents the median RMSEs of the estimation accuracy of the proposed linearized MIDAS method with $L,K=0,1,2,3,4$. The last rows report  the median RMSEs with ($L,K$) optimized by the three information criteria along with the average of optimal choices of ($L,K$) in  parentheses. The right half of Table \ref{table:LK} reports the median RMSFEs of the forecasting accuracy. The last rows report the median RMSFEs with optimal ($L,K$)s: each $\widehat{\beta}^*_{T/2+k-1}$ for $k=1,2,\ldots,T/2$ in step 1 of RMSFE calculation is estimated using the optimal $(L,K)$  with the first $T/2$ data.  The average of mean optimal ($L,K$) is reported in parentheses.

For most MIDAS curves, the information criteria help reduce estimation errors, compared to our generic choice of $(L,K)=(2,3)$ in the previous simulation. The case that benefits the most is the linear decline MIDAS weight function, where $(L,K)=(1,0)$ has an obvious advantage over other choices.
Exponential decline and hump shaped weight functions also improve with data-driven choices of $(L,K)$. This could be due to the fact that these shapes can roughly be approximated by a second order polynomial. In this case, the trigometric functions would prevent the model from capturing the reducing-to-zero behavior at one end of the MIDAS curve. The information criteria likely help make the right decision to drop the trigometric part. On the contrary, for the discrete weight function, there is no obvious winner in $(L,K)$. Information criteria still does a reasonable job and so as our arbitrary choice $(L,K)=(2,3)$. The forecasting part of the table unveils similar patterns, although  RMSFEs do not vary as drastically as RMSEs according to the choices of ($L,K$). The three information criteria deliver similar results, although the BIC tends to achieve the smallest RMSEs and RMSFEs.

\begin{landscape}
	% Table generated by Excel2LaTeX from sheet 'BetaMSE'
	\begin{table}[ht]
		\setlength\tabcolsep{4pt}
		\centering
		\footnotesize
		\caption[Caption for LOF]{\footnotesize  Parameter Estimation Accuracy of B\&R's nonparametric MIDAS and our linearized MIDAS}
		\begin{tabular}{ccc|ccc|ccc|ccc|ccc|ccc}
			\toprule
			&       &       & \multicolumn{3}{c|}{$m=20$} & \multicolumn{3}{c|}{$40$} & \multicolumn{3}{c|}{$60$} & \multicolumn{3}{c|}{$150$} & \multicolumn{3}{c}{$365$} \\
			& T     & \multicolumn{1}{c|}{Method} & \multicolumn{1}{c}{$\alpha_1=0.2$} & \multicolumn{1}{c}{$0.3$} & \multicolumn{1}{c|}{$0.4$} & \multicolumn{1}{c}{$0.2$} & \multicolumn{1}{c}{$0.3$} & \multicolumn{1}{c|}{$0.4$} & \multicolumn{1}{c}{$0.2$} & \multicolumn{1}{c}{$0.3$} & \multicolumn{1}{c|}{$0.4$} & \multicolumn{1}{c}{$0.2$} & \multicolumn{1}{c}{$0.3$} & \multicolumn{1}{c|}{$0.4$} & \multicolumn{1}{c}{$0.2$} & \multicolumn{1}{c}{$0.3$} & \multicolumn{1}{c}{$0.4$} \\
			\midrule
			& \multirow{2}[2]{*}{100} & \multicolumn{1}{c|}{B\&R} & 0.9066 & 0.9195 & 0.9168 & 0.6568 & 0.6561 & 0.6534 & 0.5704 & 0.5618 & 0.5700 & 0.4858 & 0.4811 & 0.4820 & 0.4435 & 0.4455 & 0.4432 \\
			&       & Fourier & 0.5695 & 0.5829 & 0.5851 & 0.2340 & 0.2340 & 0.2299 & 0.1341 & 0.1306 & 0.1322 & 0.0717 & 0.0931 & 0.1188 & 0.0618 & 0.0898 & 0.1179 \\
			\cmidrule{2-18}    \multicolumn{1}{l}{Exp} & \multirow{2}[2]{*}{200} & \multicolumn{1}{c|}{B\&R} & 0.8630 & 0.8790 & 0.8801 & 0.5811 & 0.5806 & 0.5911 & 0.4940 & 0.4914 & 0.4941 & 0.3962 & 0.3962 & 0.3989 & 0.3814 & 0.3795 & 0.3786 \\
			\multicolumn{1}{l}{Decline} &       & Fourier & 0.4162 & 0.4157 & 0.4068 & 0.1560 & 0.1560 & 0.1596 & 0.0918 & 0.0920 & 0.0917 & 0.0627 & 0.0873 & 0.1127 & 0.0594 & 0.0872 & 0.1156 \\
			\cmidrule{2-18}          & \multirow{2}[2]{*}{400} & \multicolumn{1}{c|}{B\&R} & 0.8383 & 0.8441 & 0.8489 & 0.5435 & 0.5435 & 0.5378 & 0.4441 & 0.4421 & 0.4443 & 0.3410 & 0.3407 & 0.3418 & 0.3130 & 0.3143 & 0.3126 \\
			&       & Fourier & 0.2850 & 0.2818 & 0.2851 & 0.1086 & 0.1086 & 0.1093 & 0.0649 & 0.0641 & 0.0649 & 0.0583 & 0.0840 & 0.1104 & 0.0583 & 0.0862 & 0.1146 \\
			\midrule
			& \multirow{2}[2]{*}{100} & \multicolumn{1}{c|}{B\&R} & 0.9052 & 0.9172 & 0.9172 & 0.6563 & 0.6554 & 0.6537 & 0.5692 & 0.5696 & 0.5620 & 0.4868 & 0.4828 & 0.4820 & 0.4465 & 0.4411 & 0.4437 \\
			&       & Fourier & 0.5695 & 0.5829 & 0.5851 & 0.2339 & 0.2339 & 0.2298 & 0.1341 & 0.1341 & 0.1307 & 0.0465 & 0.0468 & 0.0459 & 0.0227 & 0.0232 & 0.0240 \\
			\cmidrule{2-18}    \multicolumn{1}{l}{Hump} & \multirow{2}[2]{*}{200} & \multicolumn{1}{c|}{B\&R} & 0.8639 & 0.8796 & 0.8776 & 0.5817 & 0.5804 & 0.5913 & 0.4935 & 0.4935 & 0.4915 & 0.4022 & 0.3993 & 0.4014 & 0.3779 & 0.3795 & 0.3777 \\
			\multicolumn{1}{l}{Shaped} &       & Fourier & 0.4162 & 0.4157 & 0.4069 & 0.1560 & 0.1560 & 0.1597 & 0.0920 & 0.0920 & 0.0921 & 0.0322 & 0.0321 & 0.0325 & 0.0163 & 0.0164 & 0.0176 \\
			\cmidrule{2-18}          & \multirow{2}[2]{*}{400} & \multicolumn{1}{c|}{B\&R} & 0.8390 & 0.8439 & 0.8490 & 0.5435 & 0.5442 & 0.5378 & 0.4440 & 0.4440 & 0.4420 & 0.3407 & 0.3412 & 0.3392 & 0.3133 & 0.3152 & 0.3133 \\
			&       & Fourier & 0.2850 & 0.2818 & 0.2851 & 0.1085 & 0.1122 & 0.1092 & 0.0651 & 0.0650 & 0.0640 & 0.0222 & 0.0226 & 0.0221 & 0.0116 & 0.0124 & 0.0136 \\
			\midrule
			& \multirow{2}[2]{*}{100} & \multicolumn{1}{c|}{B\&R} & 0.9064 & 0.9191 & 0.9147 & 0.6406 & 0.6511 & 0.6448 & 0.5686 & 0.5613 & 0.5661 & 0.4868 & 0.4828 & 0.4822 & 0.4465 & 0.4411 & 0.4437 \\
			&       & Fourier & 0.5694 & 0.5829 & 0.5851 & 0.2234 & 0.2201 & 0.2164 & 0.1341 & 0.1307 & 0.1350 & 0.0465 & 0.0468 & 0.0459 & 0.0222 & 0.0225 & 0.0223 \\
			\cmidrule{2-18}    \multicolumn{1}{l}{Linear} & \multirow{2}[2]{*}{200} & \multicolumn{1}{c|}{B\&R} & 0.8635 & 0.8786 & 0.8787 & 0.5836 & 0.5829 & 0.5854 & 0.4935 & 0.4917 & 0.4953 & 0.4023 & 0.3993 & 0.4014 & 0.3779 & 0.3795 & 0.3777 \\
			\multicolumn{1}{l}{Decline} &       & Fourier & 0.4162 & 0.4157 & 0.4068 & 0.1551 & 0.1537 & 0.1498 & 0.0920 & 0.0920 & 0.0922 & 0.0321 & 0.0321 & 0.0325 & 0.0158 & 0.0153 & 0.0155 \\
			\cmidrule{2-18}          & \multirow{2}[2]{*}{400} & \multicolumn{1}{c|}{B\&R} & 0.8379 & 0.8441 & 0.8483 & 0.5294 & 0.5314 & 0.5369 & 0.4433 & 0.4416 & 0.4416 & 0.3406 & 0.3412 & 0.3393 & 0.3133 & 0.3152 & 0.3133 \\
			&       & Fourier & 0.2850 & 0.2818 & 0.2851 & 0.1046 & 0.1052 & 0.1060 & 0.0651 & 0.0640 & 0.0649 & 0.0222 & 0.0226 & 0.0221 & 0.1087 & 0.0109 & 0.0110 \\
			\midrule
			& \multirow{2}[2]{*}{100} & \multicolumn{1}{c|}{B\&R} & 0.9144 & 0.9256 & 0.9257 & 0.6578 & 0.6538 & 0.6569 & 0.5698 & 0.5611 & 0.5662 & 0.4870 & 0.4828 & 0.4820 & 0.4465 & 0.4411 & 0.4437 \\
			&       & Fourier & 0.5689 & 0.5825 & 0.5694 & 0.2340 & 0.2297 & 0.2304 & 0.1341 & 0.1307 & 0.1350 & 0.0465 & 0.0468 & 0.0459 & 0.0222 & 0.0225 & 0.0223 \\
			\cmidrule{2-18}    \multicolumn{1}{l}{Cyclical} & \multirow{2}[2]{*}{200} & \multicolumn{1}{c|}{B\&R} & 0.8677 & 0.8774 & 0.8807 & 0.5796 & 0.5897 & 0.5893 & 0.4935 & 0.4915 & 0.4964 & 0.4022 & 0.3992 & 0.4014 & 0.3779 & 0.3795 & 0.3777 \\
			&       & Fourier & 0.4163 & 0.4159 & 0.4061 & 0.1560 & 0.1597 & 0.1599 & 0.9196 & 0.0920 & 0.0922 & 0.0321 & 0.0321 & 0.0325 & 0.0158 & 0.0153 & 0.0155 \\
			\cmidrule{2-18}          & \multirow{2}[2]{*}{400} & \multicolumn{1}{c|}{B\&R} & 0.8426 & 0.8456 & 0.8472 & 0.5480 & 0.5390 & 0.5340 & 0.4435 & 0.4421 & 0.4405 & 0.3407 & 0.3411 & 0.3391 & 0.3133 & 0.3152 & 0.3133 \\
			&       & Fourier & 0.2848 & 0.2818 & 0.2850 & 0.1085 & 0.1092 & 0.1101 & 0.6507 & 0.0640 & 0.0649 & 0.0222 & 0.0226 & 0.0221 & 0.0109 & 0.0109 & 0.0110 \\
			\midrule
			& \multirow{2}[2]{*}{100} & \multicolumn{1}{c|}{B\&R} & 1.0838 & 1.2615 & 1.4540 & 0.6833 & 0.7113 & 0.7555 & 0.5797 & 0.5868 & 0.6062 & 0.4876 & 0.4839 & 0.4849 & 0.4465 & 0.4411 & 0.4440 \\
			&       & Fourier & 0.7854 & 0.9965 & 1.2264 & 0.3704 & 0.4870 & 0.6196 & 0.2356 & 0.3177 & 0.4095 & 0.0908 & 0.1262 & 0.1634 & 0.0393 & 0.0533 & 0.0683 \\
			\cmidrule{2-18}    \multicolumn{1}{l}{Discrete} & \multirow{2}[2]{*}{200} & \multicolumn{1}{c|}{B\&R} & 1.0064 & 1.1517 & 1.3293 & 0.6044 & 0.6375 & 0.6785 & 0.5019 & 0.5105 & 0.5275 & 0.4033 & 0.4008 & 0.4048 & 0.3781 & 0.3796 & 0.3781 \\
			&       & Fourier & 0.6779 & 0.9027 & 1.1426 & 0.3260 & 0.4562 & 0.5928 & 0.2132 & 0.3028 & 0.3956 & 0.0841 & 0.1209 & 0.1589 & 0.0356 & 0.0503 & 0.0657 \\
			\cmidrule{2-18}          & \multirow{2}[2]{*}{400} & \multicolumn{1}{c|}{B\&R} & 0.9444 & 1.0658 & 1.2004 & 0.5636 & 0.5837 & 0.6115 & 0.4505 & 0.4584 & 0.4713 & 0.3408 & 0.3429 & 0.3414 & 0.3135 & 0.3153 & 0.3134 \\
			&       & Fourier & 0.6053 & 0.8492 & 1.1063 & 0.3055 & 0.4420 & 0.5810 & 0.2030 & 0.2958 & 0.3903 & 0.0807 & 0.1185 & 0.1569 & 0.0336 & 0.0490 & 0.0645 \\
			\bottomrule
		\end{tabular}%
		\label{MSE}%
		\begin{tablenotes}
			\item \scriptsize Each cell reports the median of RMSEs of 1000 MC samples, which is further multiplied by 100.
		\end{tablenotes}
	\end{table}%

	\begin{table}[ht]
		\setlength\tabcolsep{4pt}
		\centering
		\footnotesize
		\caption[Caption for LOF]{\footnotesize
		One-Step-Ahead Forecasting Accuracy of B\&R's nonparametric MIDAS and our linearized MIDAS
		}
		\begin{tabular}{ccc|ccc|ccc|ccc|ccc|ccc}
			\toprule
			&       &       & \multicolumn{3}{c|}{$m=20$} & \multicolumn{3}{c|}{$40$} & \multicolumn{3}{c|}{$60$} & \multicolumn{3}{c|}{$150$} & \multicolumn{3}{c}{$365$} \\
			& T     & \multicolumn{1}{c|}{Method} & \multicolumn{1}{c}{$\alpha_1=0.2$} & \multicolumn{1}{c}{$0.3$} & \multicolumn{1}{c|}{$0.4$} & \multicolumn{1}{c}{$0.2$} & \multicolumn{1}{c}{$0.3$} & \multicolumn{1}{c|}{$0.4$} & \multicolumn{1}{c}{$0.2$} & \multicolumn{1}{c}{$0.3$} & \multicolumn{1}{c|}{$0.4$} & \multicolumn{1}{c}{$0.2$} & \multicolumn{1}{c}{$0.3$} & \multicolumn{1}{c|}{$0.4$} & \multicolumn{1}{c}{$0.2$} & \multicolumn{1}{c}{$0.3$} & \multicolumn{1}{c}{$0.4$} \\
			\midrule
			& \multirow{2}[2]{*}{100} & \multicolumn{1}{c|}{B\&R} &  0.2106 & 0.2076 & 0.2107 & 0.1990 & 0.2001 & 0.2000 & 0.1967 & 0.1984 & 0.2021 & 0.2424 & 0.2320 & 0.2381 & 0.3187 & 0.3186 & 0.3220\\
			&       & Fourier & 0.1358 & 0.1366 & 0.1367 & 0.1375 & 0.1380 & 0.1378 & 0.1380 & 0.1366 & 0.1409 & 0.1515 & 0.1630 & 0.1880 & 0.2598 & 0.3504 & 0.4526 \\
			\cmidrule{2-18}    Exp   & \multirow{2}[2]{*}{200} & \multicolumn{1}{c|}{B\&R} & 0.2044 & 0.2056 & 0.2062 & 0.1888 & 0.1887 & 0.1881 & 0.1848 & 0.1842 & 0.1830 & 0.1897 & 0.1890 & 0.1895 & 0.2602 & 0.2631 & 0.2603 \\
			Decline &       & Fourier & 0.1310 & 0.1319 & 0.1303 & 0.1317 & 0.1295 & 0.1296 & 0.1316 & 0.1309 & 0.1310 & 0.1343 & 0.1370 & 0.1421 & 0.2457 & 0.3346 & 0.4289 \\
			\cmidrule{2-18}          & \multirow{2}[2]{*}{400} & \multicolumn{1}{c|}{B\&R} & 0.2015 & 0.2028 & 0.2032 & 0.1837 & 0.1857 & 0.1857 & 0.1760 & 0.1757 & 0.1743 & 0.1680 & 0.1684 & 0.1681 & 0.1926 & 0.1923 & 0.1924 \\
			&       & Fourier & 0.1280 & 0.1283 & 0.1277 & 0.1280 & 0.1289 & 0.1282 & 0.1276 & 0.1280 & 0.1283 & 0.1316 & 0.1349 & 0.1400 & 0.2376 & 0.3257 & 0.4220 \\
			\midrule
			& \multirow{2}[2]{*}{100} & \multicolumn{1}{c|}{B\&R} & 0.2105 & 0.2077 & 0.2107 & 0.1990 & 0.2001 & 0.2001 & 0.1968 & 0.1984 & 0.2021 & 0.2423 & 0.2320 & 0.2381 & 0.3181 & 0.3189 & 0.3228 \\
			&       & Fourier & 0.1358 & 0.1365 & 0.1367 & 0.1374 & 0.1379 & 0.1374 & 0.1376 & 0.1363 & 0.1397 & 0.1380 & 0.1360 & 0.1371 & 0.1379 & 0.1385 & 0.1416 \\
			\cmidrule{2-18}    Hump  & \multirow{2}[2]{*}{200} & \multicolumn{1}{c|}{B\&R} & 0.2045 & 0.2057 & 0.2063 & 0.1887 & 0.1884 & 0.1880 & 0.1850 & 0.1841 & 0.1830 & 0.1897 & 0.1890 & 0.1895 & 0.2601 & 0.2631 & 0.2601 \\
			Shaped &       & Fourier & 0.1310 & 0.1320 & 0.1302 & 0.1317 & 0.1294 & 0.1296 & 0.1314 & 0.1309 & 0.1305 & 0.1311 & 0.1312 & 0.1298 & 0.1296 & 0.1324 & 0.1326 \\
			\cmidrule{2-18}          & \multirow{2}[2]{*}{400} & \multicolumn{1}{c|}{B\&R} & 0.2017 & 0.2028 & 0.2031 & 0.1837 & 0.1859 & 0.1856 & 0.1757 & 0.1756 & 0.1743 & 0.1680 & 0.1684 & 0.1681 & 0.1926 & 0.1923 & 0.1925 \\
			&       & Fourier & 0.1279 & 0.1283 & 0.1277 & 0.1279 & 0.1287 & 0.1278 & 0.1273 & 0.1274 & 0.1273 & 0.1281 & 0.1279 & 0.1278 & 0.1275 & 0.1285 & 0.1294 \\
			\midrule
			& \multirow{2}[2]{*}{100} & \multicolumn{1}{c|}{B\&R} & 0.2106 & 0.2076 & 0.2107 & 0.1991 & 0.2001 & 0.2002 & 0.1974 & 0.1986 & 0.2021 & 0.2423 & 0.2320 & 0.2381 & 0.3181 & 0.3189 & 0.3228 \\
			&       & Fourier & 0.1359 & 0.1366 & 0.1368 & 0.1375 & 0.1378 & 0.1377 & 0.1377 & 0.1360 & 0.1402 & 0.1380 & 0.1362 & 0.1376 & 0.1378 & 0.1377 & 0.1405 \\
			\cmidrule{2-18}    Linear & \multirow{2}[2]{*}{200} & \multicolumn{1}{c|}{B\&R} & 0.2045 & 0.2056 & 0.2062 & 0.1887 & 0.1883 & 0.1881 & 0.1851 & 0.1844 & 0.1830 & 0.1897 & 0.1890 & 0.1895 & 0.2601 & 0.2631 & 0.2183 \\
			Decline &       & Fourier & 0.1311 & 0.1319 & 0.1304 & 0.1317 & 0.1296 & 0.1296 & 0.1313 & 0.1309 & 0.1306 & 0.1314 & 0.1313 & 0.1300 & 0.1292 & 0.1307 & 0.1314 \\
			\cmidrule{2-18}          & \multirow{2}[2]{*}{400} & \multicolumn{1}{c|}{B\&R} & 0.2015 & 0.2028 & 0.2073 & 0.1838 & 0.1857 & 0.1856 & 0.1757 & 0.1757 & 0.1743 & 0.1680 & 0.1684 & 0.1681 & 0.1926 & 0.1923 & 0.1925 \\
			&       & Fourier & 0.1281 & 0.1284 & 0.1277 & 0.1280 & 0.1288 & 0.1280 & 0.1274 & 0.1276 & 0.1277 & 0.1281 & 0.1280 & 0.1281 & 0.1271 & 0.1275 & 0.1283 \\
			\midrule
			& \multirow{2}[2]{*}{100} & \multicolumn{1}{c|}{B\&R} & 0.2102 & 0.2078 & 0.2119 & 0.1990 & 0.1999 & 0.2000 & 0.1971 & 0.1983 & 0.2021 & 0.2423 & 0.2320 & 0.2381 & 0.3181 & 0.3189 & 0.3228 \\
			&       & Fourier & 0.1353 & 0.1367 & 0.1368 & 0.1376 & 0.1381 & 0.1371 & 0.1375 & 0.1367 & 0.1398 & 0.1380 & 0.1362 & 0.1373 & 0.1380 & 0.1374 & 0.1395 \\
			\cmidrule{2-18}    Cyclical & \multirow{2}[2]{*}{200} & \multicolumn{1}{c|}{B\&R} & 0.2035 & 0.2061 & 0.2063 & 0.1890 & 0.1890 & 0.1880 & 0.1849 & 0.1842 & 0.1832 & 0.1897 & 0.1890 & 0.1890 & 0.2601 & 0.2631 & 0.2601 \\
			&       & Fourier & 0.1310 & 0.1320 & 0.1301 & 0.1315 & 0.1295 & 0.1297 & 0.1314 & 0.1310 & 0.1305 & 0.1310 & 0.1309 & 0.1309 & 0.1293 & 0.1307 & 0.1308 \\
			\cmidrule{2-18}          & \multirow{2}[2]{*}{400} & \multicolumn{1}{c|}{B\&R} & 0.2017 & 0.2026 & 0.2029 & 0.1839 & 0.1859 & 0.1856 & 0.1758 & 0.1756 & 0.1745 & 0.1680 & 0.1684 & 0.1684 & 0.1926 & 0.1923 & 0.1925 \\
			&       & Fourier & 0.1280 & 0.1282 & 0.1278 & 0.1278 & 0.1286 & 0.1278 & 0.1273 & 0.1273 & 0.1274 & 0.1281 & 0.1278 & 0.1278 & 0.1270 & 0.1276 & 0.1276 \\
			\midrule
			& \multirow{2}[2]{*}{100} & \multicolumn{1}{c|}{B\&R} & 0.2113 & 0.2116 & 0.2140 & 0.1998 & 0.2011 & 0.2005 & 0.1967 & 0.1989 & 0.2029 & 0.2420 & 0.2325 & 0.2382 & 0.3177 & 0.3188 & 0.3236 \\
			&       & Fourier & 0.1365 & 0.1406 & 0.1427 & 0.1400 & 0.1432 & 0.1447 & 0.1409 & 0.1437 & 0.1507 & 0.1442 & 0.1479 & 0.1566 & 0.1433 & 0.1512 & 0.1626 \\
			\cmidrule{2-18}    \multicolumn{1}{l}{Discrete} & \multirow{2}[2]{*}{200} & \multicolumn{1}{c|}{B\&R} & 0.2039 & 0.2083 & 0.2090 & 0.1891 & 0.1898 & 0.1890 & 0.1853 & 0.1845 & 0.1836 & 0.1899 & 0.1888 & 0.1894 & 0.2603 & 0.2633 & 0.2604 \\
			&       & Fourier & 0.1320 & 0.1349 & 0.1354 & 0.1335 & 0.1338 & 0.1368 & 0.1336 & 0.1367 & 0.1411 & 0.1361 & 0.1409 & 0.1498 & 0.1360 & 0.1452 & 0.1532 \\
			\cmidrule{2-18}          & \multirow{2}[2]{*}{400} & \multicolumn{1}{c|}{B\&R} & 0.2023 & 0.2039 & 0.2049 & 0.1844 & 0.1861 & 0.1857 & 0.1758 & 0.1757 & 0.1750 & 0.1680 & 0.1682 & 0.1682 & 0.1927 & 0.1923 & 0.1923 \\
			&       & Fourier & 0.1290 & 0.1312 & 0.1317 & 0.1296 & 0.1331 & 0.1355 & 0.1300 & 0.1328 & 0.1371 & 0.1329 & 0.1387 & 0.1464 & 0.1320 & 0.1411 & 0.1511 \\
			\bottomrule
		\end{tabular}%
		\label{OOS}%
		\begin{tablenotes}
			\item \scriptsize Each cell reports the median of RMSFEs of 250 MC samples.
		\end{tablenotes}
	\end{table}%
	
\end{landscape}

\begin{table}[ht]\label{table:LK}
\centering
\footnotesize
\caption[Caption for LOF]{\footnotesize
		The effect of different choices of tuning parameters $L$ and $K$ for our linearized MIDAS
		}
\begin{tabular}{cc|ccccc|ccccc}
  \toprule
   &&\multicolumn{5}{c|}{Estimation Accuracy (RMSE)} &\multicolumn{5}{c}{Forecasting Accuracy (RMSFE)} 
  \\
 $L$& $K$& \multicolumn{1}{c}{Exp} & \multicolumn{1}{c}{Hump}& \multicolumn{1}{c}{Lin} & \multicolumn{1}{c}{Cyc}& \multicolumn{1}{c|}{Disc} & \multicolumn{1}{c}{Exp}& \multicolumn{1}{c}{Hump} & \multicolumn{1}{c}{Lin}& \multicolumn{1}{c}{Cyc} & \multicolumn{1}{c}{Disc}\\ 
  \midrule
0 & 0 & 0.2608 & 0.3692 & 0.1908 & 3.4477 & 2.0037 & 0.1362 & 0.1460 & 0.1328 & 0.6442 & 0.3357 \\ 
  0 & 1 & 0.1572 & 0.1977 & 0.1646 & 0.3943 & 1.5726 & 0.1302 & 0.1313 & 0.1294 & 0.1340 & 0.1912 \\ 
  0 & 2 & 0.2444 & 0.2636 & 0.2595 & 0.3494 & 1.1081 & 0.1333 & 0.1332 & 0.1316 & 0.1330 & 0.1461 \\ 
  0 & 3 & 0.3935 & 0.4024 & 0.4024 & 0.4511 & 0.8412 & 0.1356 & 0.1361 & 0.1348 & 0.1356 & 0.1384 \\ 
  0 & 4 & 0.5680 & 0.5896 & 0.5906 & 0.5975 & 0.8697 & 0.1393 & 0.1396 & 0.1386 & 0.1388 & 0.1398 \\ 
  1 & 0 & 0.1583 & 0.1794 &{\bf  0.0603} & 2.4440 & 1.4500 & 0.1300 & 0.1305 & {\bf 0.1277} & 0.2637 & 0.2046 \\ 
  1 & 1 & 0.1643 & 0.1628 & 0.1647 & {\bf 0.1652} & 0.8425 & 0.1314 & 0.1316 & 0.1302 & {\bf  0.1298} & 0.1371 \\ 
  1 & 2 & 0.2958 & 0.2989 & 0.3131 & 0.3046 & 0.7507 & 0.1346 & 0.1345 & 0.1334 & 0.1332 & {\bf 0.1364} \\ 
  1 & 3 & 0.4634 & 0.4773 & 0.4774 & 0.4757 & {\bf 0.7360} & 0.1374 & 0.1378 & 0.1363 & 0.1368 & 0.1380 \\ 
  1 & 4 & 0.6722 & 0.6739 & 0.6857 & 0.6680 & 0.8299 & 0.1414 & 0.1413 & 0.1402 & 0.1400 & 0.1415 \\ 
  2 & 0 & {\bf 0.1088} & {\bf 0.1272} & 0.1044 & 2.4506 & 0.9642 &{\bf  0.1297} & {\bf 0.1305} & 0.1287 & 0.2667 & 0.1449 \\ 
  2 & 1 & 0.2301 & 0.2317 & 0.2431 & 0.2343 & 0.8490 & 0.1329 & 0.1334 & 0.1314 & 0.1320 & 0.1367 \\ 
  2 & 2 & 0.3882 & 0.3993 & 0.4093 & 0.4000 & 0.7503 & 0.1359 & 0.1361 & 0.1348 & 0.1352 & 0.1374 \\ 
  \rowcolor[gray]{.85}2 & 3 & 0.5747 & 0.5754 & 0.5956 & 0.5833 & 0.7969 & 0.1392 & 0.1396 & 0.1384 & 0.1384 & 0.1397 \\ 
  2 & 4 & 0.7836 & 0.7951 & 0.8010 & 0.7841 & 0.9171 & 0.1433 & 0.1430 & 0.1420 & 0.1416 & 0.1434 \\ 
  3 & 0 & 0.1624 & 0.1651 & 0.1679 & 0.4340 & 0.8755 & 0.1314 & 0.1316 & 0.1301 & 0.1320 & 0.1363 \\ 
  3 & 1 & 0.3112 & 0.3168 & 0.3285 & 0.3197 & 0.7931 & 0.1342 & 0.1346 & 0.1333 & 0.1334 & 0.1369 \\ 
  3 & 2 & 0.4869 & 0.4971 & 0.4971 & 0.5039 & 0.7697 & 0.1377 & 0.1377 & 0.1365 & 0.1368 & 0.1382 \\ 
  3 & 3 & 0.6895 & 0.7004 & 0.7231 & 0.6837 & 0.8700 & 0.1412 & 0.1416 & 0.1405 & 0.1397 & 0.1418 \\ 
  3 & 4 & 0.9070 & 0.9206 & 0.9306 & 0.9074 & 0.9963 & 0.1453 & 0.1452 & 0.1443 & 0.1440 & 0.1452 \\ 
  4 & 0 & 0.2358 & 0.2371 & 0.2476 & 0.4726 & 0.8863 & 0.1327 & 0.1334 & 0.1313 & 0.1338 & 0.1374 \\ 
  4 & 1 & 0.3995 & 0.4139 & 0.4266 & 0.4131 & 0.7408 & 0.1359 & 0.1360 & 0.1349 & 0.1352 & 0.1370 \\ 
  4 & 2 & 0.5924 & 0.6089 & 0.6109 & 0.6145 & 0.8497 & 0.1392 & 0.1397 & 0.1384 & 0.1384 & 0.1401 \\ 
  4 & 3 & 0.8116 & 0.8167 & 0.8309 & 0.8148 & 0.9270 & 0.1432 & 0.1432 & 0.1421 & 0.1419 & 0.1436 \\ 
  4 & 4 & 1.0166 & 1.0345 & 1.0370 & 1.0279 & 1.1010 & 0.1475 & 0.1469 & 0.1465 & 0.1460 & 0.1474 \\ 
  \midrule
  \multicolumn{2}{c|}{\multirow{2}{*}{AIC}}  & 0.1645 & 0.1867 & 0.1073 & 0.3660 & 0.8623 &  0.1347 &0.1353& 0.1344& 0.1363& 0.1414\\
  &&(1.2,0.7)&(1.2,0.6)&(1.2,0.5)&(1.4,1.3)&(1.9,2.1)&(1.2,0.8)  &  (1.2,0.8)& (1.1,0.7) &(1.4,1.4)  & (1.9,1.8)\\
  \multicolumn{2}{c|}{\multirow{2}{*}{AICc}}  & 0.1601 & 0.1831 & 0.0931 & 0.3275 & {\bf 0.8575} &  0.1324& 0.1326& 0.1308& 0.1337& 0.1393\\ 
  &&(1.1,0.5)&(1.2,0.5)&(1.0,0.4)&(1.3,1.2)&(1.8,1.7)& (1.0,0.4) & (1.0,0.4) & (0.9,0.3) & (1.2,1.0) & (1.8,1.2)\\
  \multicolumn{2}{c|}{\multirow{2}{*}{BIC}}  & {\bf 0.1590} & {\bf 0.1803} & {\bf 0.0694} & {\bf 0.2465} & { 0.8650} & {\bf 0.1322} & {\bf 0.1318} & {\bf 0.1302} & {\bf 0.1333} & {\bf 0.1393}\\ 
  &&(0.9,0.2)&(1.0,0.2)&(0.8,0.0)&(1.0,0.9)&(1.9,0.9)&  (0.7,0.1)& (0.9,0.2) &  (0.6,0.1)&(0.9,0.9)  & (1.8,0.8)\\
  \bottomrule
\end{tabular}
\begin{tablenotes}
			\item \scriptsize Each cell reports median RMSEs and RMSFEs of 1000 MC replications when $T=100$, $m=20$, and $\alpha_1=0.2$. RMSEs are divided by 100, as in Table \ref{MSE}. The smallest RMSEs (or RMSFEs) of each column among the fixed and IC-driven $(L,K)$s are boldfaced. Our arbitrary choice of $(L,K)=(2,3)$ in the previous simulation is lightly shaded. The last rows present median RMSEs and RMSFEs when  $(L,K)$s are chosen by information criteria. The averages of optimal $(L,K)$s are reported in parentheses.
		\end{tablenotes}
\end{table}

%In the next section, we shall introduce a specification test of the flat aggregation against MIDAS model. Local alternatives are presented in order to illustrate the test empirically.

\section{Panel Data and Clustering}\label{sec:panelMIDAS}
%\section{}
%\subsection{}
In this section, a clustering procedure of MIDAS coefficients for panel data is  proposed. The high-frequency regressors are aggregated using the linearized MIDAS coefficient functions introduced in Section \ref{sec:nonparaMIDAS} for each cross-section object. These coefficients are further clustered using a penalized regression approach.
The linearity of our MIDAS model confers a great advantage to the proposed clustering procedure, as the clustering alone would require quite heavy computations. We first review relevant literature on clustering.

\subsection{Literature Review on Clustering Based on Penalized Regression}\label{appen:litreview}

In this paper, we propose to adapt \cite{Ma&Huang:2017}'s clustering idea to a panel setting.\footnote{There is a large literature on clustering in a panel setting using other clustering methods such as $K$-means. For brevity, we focus only on those with penalized regression idea in this literature review.}
\cite{Ma&Huang:2017} introduced a penalized method for cross-sectional data. Their clustering is based on intercepts. The penalty functions used in \cite{Ma&Huang:2017} are minimax concave penalty (MCP) \citep{zhang2010} and smoothly clipped absolute deviations penalty (SCAD) \citep{Fan&Li:2001}, which not only share the sparsity properties like Lasso but are also  asymptotically unbiased. Later on, \cite{ma2016estimating} extended their work, increasing the number of parameters used in clustering. However, neither  \cite{Ma&Huang:2017} nor \cite{ma2016estimating} can be applied to a panel data setting. Indeed, their method is based on strong assumptions that make it nontrivial to extend to panel data.

\cite{Zhu&Qu:2018} is the only study, to the best of our knowledge, that extends Ma and Huang's clustering procedure to a data environment similar to panel data. \cite{Zhu&Qu:2018} applied \cite{Ma&Huang:2017}'s algorithm to repeated cross-section data with one dependent variable and one covariate. In their model, the dependent variable is assumed to vary smoothly in response to the covariate, and this smooth function is estimated using a nonparametric B-spline. Strictly speaking, \cite{Zhu&Qu:2018}'s method is not designed for panel data, but if their covariate is allowed to vary over time, \cite{Zhu&Qu:2018}'s setting can be viewed as a simple panel setup. \cite{lv_etal:sinica} further extended  \cite{Zhu&Qu:2018}'s approach allowing for one random effect as an additional covariate. 

The clustering procedure we propose is based on \cite{Ma&Huang:2017} and \cite{ma2016estimating}. It should be noted that this extension is nontrivial. In particular, the assumption (C3) in \cite{ma2016estimating} requires all variables on the right-hand side of the equation to be non-random and have length exactly 1. This assumption may be appropriate for a clinical trial setting, for which \cite{Ma&Huang:2017}, \cite{ma2016estimating}, \cite{Zhu&Qu:2018}, \cite{lv_etal:sinica}, and other related papers are developed. However, this assumption is too strong for a more general panel data setting where time-varying regressors are included. 
The theory we present circumvent this issue. 

\cite{Su:2016}, to our best knowledge, is the first study that developed a clustering algorithm using penalized regression based on similarity in the coefficients in a truly panel setting.  \cite{Su:2016} modified the traditional Lasso penalty in regression models into classifier-Lasso (C-Lasso) that penalizes the difference between the estimated parameters of each subject and the estimated average parameters of groups. C-Lasso requires a predetermined maximum for the number of groups and a choice of tuning parameter.

Our clustering algorithm based on \cite{Ma&Huang:2017}'s idea can actually be an appealing alternative to \cite{Su:2016}'s method in a general panel setting for the following two reasons. First, \cite{Su:2016}'s method requires to pre-specify a possible range for the number of clusters. If there is no prior knowledge of the number of clusters and if the size of cross-section is large, finding right clusters can become computationally challenging. In addition, the form of their penalty function makes computation much heavier if the pre-specified maximum number of clusters is large. Our clustering method does not require any information on the number of clusters. When possible number of clusters is large or unknown, our method has a computational advantage. Second, in simulation studies in Section \ref{sim:panelMIDAS}, our clustering approach generally produces more accurate estimations and forecasts than \cite{Su:2016}'s counterpart.

%requires a few user-chosen parameters, but the range of choices is well-defined and does not yield intractable possibilities.  %Second, \cite{Su:2016}'s method requires the number of clusters to be fixed regardless of the size of cross-section. The theory behind our clustering algorithm does flexibly allow the number of groups to adjust to a change in the size of cross-section. 

The next subsection introduces our MIDAS clustering algorithm.

\subsection{Clustering based on MIDAS weights}\label{panelMIDAS}
%In the following framework, We consider a slightly general model to allow for low-frequency covariates.
Suppose there are $n$ subjects in the cross-section of panel data.
For simplicity,   all subjects are assumed to have the same sample size $T$ and frequency ratio $m$. 
For the $i$-th subject, 
let $\mathbf{z}_{i,t}$ be the $q$-vector of covariates including the intercept  at time $t$ ($t=1,\ldots,T$), and let $\boldsymbol{\alpha}_i$ be the  corresponding coefficient. 
Consider the following MIDAS model with  lead $h\geq0$:
\begin{equation*}
%y_{i,t+h}={\alpha}_{i,0}+\mathbf{x}_{i,t}'\boldsymbol{\beta}_i^*+\varepsilon_{t+h},~~~~t=1,\ldots,T_i,~~i=1,\ldots,n,
y_{i,t+h}=\mathbf{z}_{i,t}'\boldsymbol{\alpha}_i+\mathbf{x}_{i,t}'\boldsymbol{\beta}_i^*+\varepsilon_{i,t+h},~~~~t=1,\ldots,T,~~i=1,\ldots,n,
\end{equation*}
or equivalently,
\begin{equation}\label{eq:panel1}
\mathbf{y}_{i}=Z_i\boldsymbol{\alpha}_i+{X}_i\boldsymbol{\beta}_i^*+\boldsymbol{\varepsilon}_i,~~~~~i=1,\ldots,n,
\end{equation}
where $\mathbf{y}_i=(y_{i,1+h},\ldots,y_{i,T+h})'$, $\boldsymbol{\varepsilon}_i=(\varepsilon_{i,1+h},\ldots,\varepsilon_{i,T+h})'$,  $\boldsymbol{\beta}_i^*=(\beta_{i,0}^*,\ldots,\beta_{i,m-1}^*)'$,  $X_i$ is a $T\times m$ matrix with  $t$-th row being $\mathbf{x}_{i,t}'=(x_{i,t,0},x_{i,t,1},\ldots,x_{i,t,m-1})$, and
 $Z_i$ is a $T\times q$ matrix with  $t$-th row being $\mathbf{z}_{i,t}'=(z_{i,t,1},\ldots,z_{i,t,q})$.

We assume that the  MIDAS coefficients $\boldsymbol{\beta}^*_i$ takes  the Fourier flexible form as in (\ref{eq:fourier}). For each subject $i=1,\ldots,n$,  $\widetilde{X}_i=X_i{M}'$, where $M$ is from (\ref{M}).
%For all $i$, $X_i\boldsymbol{\beta}_i^*\approx\widetilde{X}_i\boldsymbol{\beta}_i$, as long as $L$ and $K$ are large enough and the underlying MIDAS weight functions $\boldsymbol{\beta}_i^*(\cdot)$ satisfy the Dirichlet conditions.
Let $W_i=(Z_i,\widetilde{X}_i)$ and $\boldsymbol{\gamma}_i=(\boldsymbol\alpha_i',\boldsymbol\beta_i')'$. The equation  (\ref{eq:panel1}) can be rewritten as
\begin{equation}\label{eq:panel1-1}
\mathbf{y}_i= (Z_i,{X}_i)\left(\begin{matrix}\boldsymbol{\alpha}_i\\ \boldsymbol{\beta}_i^*\end{matrix}\right)+\boldsymbol{\varepsilon}_i= (Z_i,\widetilde{X}_i)\left(\begin{matrix}\boldsymbol{\alpha}_i\\ \boldsymbol{\beta}_i\end{matrix}\right)+\boldsymbol{\varepsilon}_i=W_i\boldsymbol{\gamma}_i+\boldsymbol{\varepsilon}_i
\end{equation}
Concatenating the $\mathbf{y}_i$ in (\ref{eq:panel1-1}) into $\mathbf{y}$, a vector of length $nT$, we have:
\begin{equation}\label{eq:panel2}
%\mathbf{y}=\alpha_{i,0}\mathbf{j}_{T}+{X}\boldsymbol{\beta}^*+\boldsymbol{\varepsilon},
\mathbf{y}= {W}\boldsymbol{\gamma}+\boldsymbol{\varepsilon},
\end{equation}
where  $\mathbf{y}=(\mathbf{y}_1',\ldots,\mathbf{y}_n')'$, $W={\rm diag}(W_1,\ldots,W_n)$, $\boldsymbol{\gamma}=(\boldsymbol{\gamma}_{1}',\ldots,\boldsymbol{\gamma}_{n}')'$, and $\boldsymbol{\varepsilon}=(\boldsymbol{\varepsilon}_1',\ldots,\boldsymbol{\varepsilon}_n')'$.
Let $p=q+2K+L+1$. In our formulation, $\boldsymbol{\gamma}_i$ is a vector of length $p$ and $\boldsymbol{\gamma}$ is of length $np$\footnote{The framework introduced in this section and in Section \ref{parameter} considers only one high-frequency variable. However, our framework can easily extend to accommodate more than one high-frequency variables. For instance, if two high-frequency variables are considered, the length of $\boldsymbol{\gamma}_i$ will be $q+2K+L+1+2K'+L'+1$, where $2K'$ and $L'+1$ are the numbers of trigonometric functions and polynomials considered for the second high-frequency variable. The subsequent clustering procedure is also straightforward. %The only major complication due to allowing more than one high-frequency variable is that in the clustering procedure,  choosing which coefficients to consider in the clustering, which is subject to a practitioner's decision anyway.
}.

%where  $\mathbf{y}=(\mathbf{y}_1',\ldots,\mathbf{y}_n')'$, $Z={\rm diag}(Z_1,\ldots,Z_n)$, $\boldsymbol{\alpha}=(\boldsymbol{\alpha}_{1}',\ldots,\boldsymbol{\alpha}_{n}')'$, $X={\rm diag}(X_1,\ldots,X_n)$, $\boldsymbol{\beta}^*=({\boldsymbol{\beta}^*_1}',\ldots,{\boldsymbol{\beta}^*_n}')'$, $\boldsymbol{\varepsilon}=(\boldsymbol{\varepsilon}_1',\ldots,\boldsymbol{\varepsilon}_n')'$, $\widetilde{X}={\rm diag}(\widetilde{X}_1,\ldots,\widetilde{X}_n)$, and $\boldsymbol{\beta}=({\boldsymbol{\beta}_1}',\ldots,{\boldsymbol{\beta}_n}')'$.

\begin{remark}\label{remark3}{\rm
The arguments in this section should still be valid with different sample sizes and different frequency ratios for different subjects/time periods, at the expense of more complicated notation and slight changes in the results.  The major complication arises from the need of using different $M_{i,t}$ for each $i$ and $t$. That is, $\widetilde{\mathbf{x}}_{i,t}=\mathbf{M}_{i,t}\mathbf{x}_{i,t}$, where
\begin{equation*}
{M}_{i,t}=\left[\begin{matrix}
 (0/m_{i,t})^0 & (1/m_{i,t})^0 & \cdots & \{(m_{i,t}-1)/m_{i,t}\}^0\\
 \vdots & \vdots &  & \vdots\\
 (0/m_{i,t})^L & (1/m_{i,t})^L & \cdots & \{(m_{i,t}-1)/m_{i,t}\}^L\\
 \sin(2\pi\cdot1\cdot0/m_{i,t}) & \sin(2\pi\cdot1\cdot1/m_{i,t}) & \cdots & \sin\{2\pi\cdot1\cdot(m_{i,t}-1)/m_{i,t}\}\\
 \cos(2\pi\cdot1\cdot0/m_{i,t}) & \cos(2\pi\cdot1\cdot1/m_{i,t}) &  \cdots & \cos\{2\pi\cdot1\cdot(m_{i,t}-1)/m_{i,t}\}\\
 \vdots & \vdots &  & \vdots\\
 \sin(2\pi\cdot K\cdot0/m_{i,t}) & \sin(2\pi\cdot K\cdot1/m_{i,t}) & \cdots & \sin\{2\pi\cdot K\cdot(m_{i,t}-1)/m_{i,t}\}\\
\cos(2\pi\cdot K\cdot0/m_{i,t}) & \cos(2\pi\cdot K\cdot1/m_{i,t}) & \cdots & \cos\{2\pi\cdot K\cdot(m_{i,t}-1)/m_{i,t}\}\\
\end{matrix}\right]
\end{equation*}
should be used, and $\mathbf{y}$ is a vector of length $\sum_{i=1}^nT_i$ rather than $nT$.
As this makes the notation for the subsequent proofs more complicated without adding fundamental differences,  this generalization is not pursued in this paper. 
In contrast,
it is necessary to use the same $L$ and $K$ for all subjects $i=1,\ldots,n$, to allow for direct comparison of coefficients $\boldsymbol{\gamma}_i$.
}\end{remark}

%\begin{remark}\label{remark4}{\rm}\end{remark}

%\begin{remark}{\rm If all subjects have the same frequency ratio and have the same sample size, the relationship between the original high-frequency variables $X$ and its Fourier transformation $\widetilde{X}$  can be simplified. In this case,  $m_i=m_1$ and $T_i=T_1$ for all $i=1,\ldots,n$, which mean that the Fourier transformation matrix $M_i=M_1$  is also the same for all subjects.  This would lead to  $\widetilde{X}=
%{\rm diag}(\widetilde{X}_1,\ldots,\widetilde{X}_n)={\rm diag}(X_1,\ldots,X_n)(I_n\otimes M_1')= X(I_n\otimes M_1')$, where $\otimes$ denotes the Kronecker product.}\end{remark}

Consider the estimation of parameters in (\ref{eq:panel2}) if the subjects can be separated into a small number of groups. Denote the number of groups by $G$.
The advantage of the proposed procedure is that it does not require any prior knowledge of group information or the number of groups. {The only information required is the features of cluster. For example, if we are willing to assume that a cluster has the same parameters of interest--that is, all elements in $\boldsymbol{\gamma}_i$ are the same within a group--the clusters are identified solely based on parameter estimates.\footnote{It is possible to relax this assumption by letting some of $\boldsymbol{\gamma}_i$ be individual-specific, rather than assuming all parameters are strongly tied with groups. If there are subject-specific coefficients, a similar argument would still work, although some rates and conditions would change. In particular, the number of coefficients that are subject-specific should be added following a similar argument to \cite{Ma&Huang:2017,ma2016estimating}.  However, for brevity, this direction will be not elaborated in this paper.}} 

An OLS solution of (\ref{eq:panel2}) would minimize
%\begin{equation}\label{objectfunc}
$\frac{1}{2}||\mathbf{y}-W\boldsymbol{\gamma}||_2^2,$
but this would not reflect  the relevant group information. To reveal clusters, we propose a penalized
regression method  to force all elements in $\boldsymbol{\gamma}_i$ to have similar values within a group.
%Groups can be found by pairwisely comparing the  group-specific parameters. 
Our method is based on the assumption that  if two subjects $i$ and $j$ belong to the same group, the difference of their group-specific parameter would be zero, i.e.,
 $\boldsymbol{\eta}_{ij}=\boldsymbol{\gamma}_i-\boldsymbol{\gamma}_j=\mathbf{0}$.
In this case, the OLS estimator of $\boldsymbol{\eta}_{ij}$ would also be somewhat close to a zero vector, though it would not be exactly zero. 
However, since $i$ and $j$ are in the same group, $\boldsymbol{\eta}_{ij}$ should be better estimated to be exactly zero, rather than ``somewhat close" to zero. This can be forced by imposing a penalty for small values of $\boldsymbol{\eta}_{ij}$.
In particular, if the number of groups $N$ is much smaller than the number of  subjects $n$,  only a small number of $\boldsymbol{\eta}_{ij}$ would be nonzero. 
The following penalized objective function is considered:
\begin{equation}\label{objectfunc2}
Q(\boldsymbol{\gamma};\theta,\lambda_1)=\frac{1}{2}||\mathbf{y}-W\boldsymbol{\gamma}||_2^2+\sum_{1\leq i<j\leq n}\rho_\theta\left(||\boldsymbol{\gamma}_i-\boldsymbol{\gamma}_j||_2,\lambda_1\right),
\end{equation}
where $\rho_\theta(\cdot,\lambda_1)$ is an appropriate penalty function, and $\theta$ and $\lambda_1$ are tuning parameters that discipline clustering. 
Clustering using a penalized regression as in (\ref{objectfunc2}) has been explored in a number of papers \citep{Ma&Huang:2017,Zhu&Qu:2018,lv_etal:sinica,ma2016estimating}. As illustrated in the previously-mentioned papers, this optimization problem can be solved using the  alternating direction method of multipliers (ADMM) algorithm, which can also be implemented in  our setting. Section  \ref{appen:our} in  the supplementary material introduces the ADMM algorithm in our setting, proving that the proposed algorithm is convergent.
The tuning parameters $\theta$ and $\lambda_1$ can be chosen by minimizing information criteria such as
\begin{equation*}\label{BIC}
    BIC_{\theta,\lambda_1}=\log\left(\dfrac{\|\mathbf{y}-W\widehat{\boldsymbol{\gamma}}\|_2^2}{n}\right)+\dfrac{\log(n)\cdot\left(\widehat{G}p\right)}{n},
\end{equation*}
where the estimated coefficients $\widehat{\boldsymbol{\gamma}}$ and the estimated number $\widehat{G}$  of groups  are obtained by minimizing (\ref{objectfunc2}) and depend on the choice of the two tuning parameters $\theta$ and $\lambda_1$.
%\begin{equation}\label{AIC}
%    AIC_{\lambda_1}=\log\left(\dfrac{\|\mathbf{y}-W\widehat{\boldsymbol{\gamma}}\|_2^2}{n}\right)+\dfrac{2\left(\widehat{G}+\frac{1}{n}\sum_{i=1}^n tr\{W_i(W_i'W_i)^{-1}W_i'\}\right)}{n}.
%\end{equation}
%In (\ref{BIC}) and (\ref{AIC}), $tr(\cdot)$ indicates the trace of matrix. Here,  $tr\{W_i(W_i'W_i)^{-1}W_i'\}=p$.
%\todo{These are essentially the same as}
%\begin{equation}\label{BIC}
%    BIC_{\lambda_1}=\log\left(\dfrac{\|\mathbf{y}-W\widehat{\boldsymbol{\gamma}}\|_2^2}{n}\right)+\dfrac{\log(n)\cdot\left(\widehat{G}+p\right)}{n}.
%\end{equation}
%\begin{equation}\label{AIC}
%    AIC_{\lambda_1}=\log\left(\dfrac{\|\mathbf{y}-W\widehat{\boldsymbol{\gamma}}\|_2^2}{n}\right)+\dfrac{2\left(\widehat{G}+p\right)}{n}.
%\end{equation}
%Note that in (\ref{BIC}) and (\ref{AIC}),  $\frac{1}{n}\sum_{i=1}^n tr\{W_i(W_i'W_i)^{-1}W_i'\}=p$, so (\ref{BIC}) and (\ref{AIC}) can be simplified by using  

The rest of this section presents theoretical properties of the estimators that solve  the optimization problem  in (\ref{objectfunc2}).
Let $G$ be the true number of groups and   $\mathcal{G}_g$ be the set of subject indices that corresponds to the $g$-th group, for $g=1,\ldots,G$.
Assume that each subject belongs to exactly one group; that is,
$\mathcal{G}_1,\ldots,\mathcal{G}_G$ are  mutually exclusive and $\mathcal{G}_1\cup\ldots\cup\mathcal{G}_G=\{1,\ldots,n\}$.
Denote $|\mathcal{G}_g|$ be the number of elements in $\mathcal{G}_g$, for $g=1,\ldots,G$.
Define $g_{\min}=\min_{g=1,\ldots,G}|\mathcal{G}_g|$ and $g_{\max}=\max_{g=1,\ldots,G}|\mathcal{G}_g|$.
Let $\boldsymbol\gamma_i^0$ be the true parameter of the $i$-th subject, and $\boldsymbol\varphi_g^0$  the true common vector for the group $\mathcal{G}_g$.  
The common value for the $\boldsymbol\gamma_i$s of the group $\mathcal{G}_g$ is denoted by $\boldsymbol\varphi_g$; that is, $\boldsymbol\gamma_i=\boldsymbol\varphi_g$ for all $i\in\mathcal{G}_g$ and for any $g=1,\cdots,G$. 
%In other words, $\boldsymbol\varphi_g$ indicates the $g$-th panel-specific parameter. 
Set $\boldsymbol{\gamma}^0=({\boldsymbol\gamma_1^0}',\cdots,{\boldsymbol\gamma_n^0}')'$, $\boldsymbol{\varphi}^0=({\boldsymbol\varphi_1^0}',\cdots,{\boldsymbol\varphi_G^0}')'$, and
$\boldsymbol\varphi=(\boldsymbol\varphi_1',\cdots,\boldsymbol\varphi_G')'$. 
 Denote the estimated group by $\widehat{\mathcal{G}}_g=\{i: \widehat{\boldsymbol{\gamma}}_i=\widehat{\boldsymbol{\varphi}}_g, 1\leq i\leq n\}$, for $g=1,\dots,\widehat{G}$, where $\widehat{G}$ is the estimated number of groups. 
For an estimate  $\widehat{\boldsymbol\gamma}$ of $\boldsymbol{\gamma}$, the corresponding estimated group parameter for the $g$-th group is defined as $\widehat{\boldsymbol{\varphi}}_g=|\widehat{\mathcal{G}}_g|^{-1}\sum_{i\in\widehat{\mathcal{G}}_g}\widehat{{\boldsymbol{\gamma}}}_i$.
Note that 
$\widehat{\boldsymbol\varphi}_1,\cdots,\widehat{\boldsymbol\varphi}_{\widehat{G}}$ are the distinct values, since the clustering algorithm would lead to $\widehat{\boldsymbol{\eta}}_{ij}=\mathbf{0}$ \citep{Ma&Huang:2017}.

Let $\Pi$ be an $n\times G$ matrix with  $(i,g)$-th element being 1 if $i$-th subject belongs to the $g$-th group, and 0 otherwise. 
Then $\boldsymbol{\gamma}=(\Pi\otimes I_p)\boldsymbol{\varphi}=\Gamma\boldsymbol{\varphi},$ where $\Gamma=(\Pi\otimes I_p)$. 
An oracle estimator of $\boldsymbol{\gamma}^0$ can be  defined as
$\widehat{\boldsymbol{\gamma}}^{or}=\Gamma\widehat{\boldsymbol{\varphi}}^{or}$, where $ \widehat{\boldsymbol{\varphi}}^{or}=\displaystyle{\rm argmin}_{\boldsymbol{\varphi}\in\mathbb{R}^{Gp}}\frac{1}{2}||\mathbf{y}-W\Gamma\boldsymbol{\varphi}||_2^2=(\Gamma'W'W\Gamma)^{-1}\Gamma'W'\mathbf{y}$.
The matrix $\Gamma'W'W\Gamma$ is invertible as long as $n\gg G$.
Here, the estimator $\widehat{\boldsymbol{\gamma}}^{or}$ is called an oracle estimator since it utilizes  the knowledge of the true group memberships in $\Pi$, which is not feasible in practice.
Asymptotic properties of this oracle estimator $\widehat{\boldsymbol{\gamma}}^{or}$
 will be presented in Theorem \ref{thm:oracle}. Then the asymptotic equivalence of our estimator $\widehat{\boldsymbol{\gamma}}$ and the oracle estimator will be introduced in Theorem \ref{thm:main}.

\begin{assump}\label{ass:sparsity1}
The number of clusters is much smaller than the number of subjects, i.e., $G\ll n$. In this paper,  the case with $G\geq 2$ is considered. The smallest group size $g_{min}$ is  smaller than $n/G$.
\end{assump}
\begin{assump}\label{ass:lambda}
Assume $\lambda_{\min}(\sum_{i\in\mathcal{G}_g}W_i'W_i)\geq c|\mathcal{G}_g|T$, 
$\lambda_{\max}(\sum_{i\in\mathcal{G}_g}W_i'W_i)\leq c'nT$,
and 
$\max_{1\leq i\leq n} \allowbreak\lambda_{\max}(W_i'W_i)\leq c''T$
for some constants $c$, $c'$ and $c''$ that do not depend on $g=1,\ldots,G$.
Further, assume that for any $\epsilon>0$, there exist $M_1,\ldots,M_4>0$ such that 
\begin{equation*}
    \begin{aligned}
     &P\left(\sup_{i=1,\ldots,n} ||Z_i'Z_i||_\infty>\sqrt{qT}M_1\right)<\epsilon,~~P\left(\sup_{i=1,\ldots,n} ||X_i'X_i||_\infty>\sqrt{mT}M_2\right)<\epsilon,\\
        &P\left(\sup_{i=1,\ldots,n} ||Z_i'X_i||_\infty>\sqrt{mT}M_3\right)<\epsilon,~~P\left(\sup_{i=1,\ldots,n} ||X_i'Z_i||_\infty>\sqrt{qT}M_4\right)<\epsilon.
       % &\sup_{i=1,\ldots,n} ||Z_i'Z_i||_\infty=O_p((qT)^{1/2}),~~\sup_{i=1,\ldots,n} ||X_i'X_i||_\infty=O_p((mT)^{1/2}),\\
       % &\sup_{i=1,\ldots,n} ||Z_i'X_i||_\infty=O_p((mT)^{1/2}),~~\sup_{i=1,\ldots,n} ||X_i'Z_i||_\infty=O_p((qT)^{1/2}).
    \end{aligned}
\end{equation*}
\end{assump}
\begin{assump}\label{ass:penalty}
The penalty function $\rho(t)=\lambda^{-1}\rho_\theta(t,\lambda)$ is symmetric, nondecreasing, and concave in $t$, on $t\in[0,\infty)$.
There exists a positive constant $c_\rho$ such that $\rho(t)$ is constant for all $t\geq c_\rho\lambda$. Assume that $\rho(t)$ is differentiable,  $\rho'(t)$ is continuous except for a finite number of $t$, $\rho(0)=0$, and $\rho'(0+)=1$.
\end{assump}
\begin{assump}\label{ass:subgauss}
There exists a constant $\tilde{c}>0$ such that
$$E\left\{\exp\left(\sum_{i=1}^n\sum_{t=1}^T\nu_{i,t}\varepsilon_{i,t}\right)\right\}\leq \exp\left(\tilde{c}\sum_{i=1}^n\sum_{t=1}^T\nu_{i,t}^2\right)$$
for any real numbers $\nu_{i,t}$, for $i=1,\ldots,n$ and $t=1,\ldots,T$.
\end{assump}

%%%%
%Assumption \ref{ass:sparsity}  essentially states the Dirichlet condition, which allows our Fourier transformation to be a valid nonparametric estimation tool for the MIDAS weights. 
Assumption \ref{ass:sparsity1} assures sparsity, which is often necessary for the validity of the penalized regression such as (\ref{objectfunc2}). We also limit our interest to the case with more than one cluster, but similar arguments also works for the homogeneous case\footnote{The extension to a homogeneous case can be done similarly to that of \cite{Ma&Huang:2017}.}.
Assumption \ref{ass:lambda} is reasonable considering  the usual assumption that the smallest eigenvalue of $W_i'W_i$ is bounded by $cT$ where $T$ is the sample size and $c$ is some constant. 
This condition can be relaxed allowing different $c_g$ for different groups. In such a case, our results would not hold if the number of clusters $G$ grows to infinity. It would still work as long as $G$ is finite, by choosing $c=\min_{g=1,\ldots,G}c_g$ in the statement of Theorem \ref{thm:oracle}.
Assumption \ref{ass:penalty} is  adapted from \cite{Ma&Huang:2017} and is conventional in the literature. Popular penalty functions such as MCP and SCAD penalty satisfy this  assumption.
Assumption \ref{ass:subgauss} holds for any independent subgaussian vector $\boldsymbol\varepsilon$, which is commonly assumed in high-dimensional settings.

\begin{remark}\label{remark:c3}{\rm
Assumption \ref{ass:lambda} is more appropriate for time series data than those in \cite{Ma&Huang:2017,ma2016estimating}. For instance, assumption (C3) in \cite{ma2016estimating} requires, for a given $t$, $\sum_{i=1}^nz_{i,t,l}^2=n$, for $l=1,\ldots,q$ and $\sum_{i=1}^n\tilde{x}_{i,t,j}^2{\bf 1}\{i\in\mathcal{G}_g\}=|\mathcal{G}_g|$ for $j=1,\ldots,2K+L+1$, if the clustering is solely based on $\tilde{x}_{i,t,j}$, but not on $z_{i,t,l}$. 
Here, $\widetilde{x}_{i,t,j}$ are the elements of Fourier transformed high-frequency variable $\widetilde{\mathbf{x}}_{i,t}=\mathbf{M}\mathbf{x}_{i,t}$\footnote{Note that this setting is slightly different from our setting, where both $z_{i,t,l}$ and $\tilde{x}_{i,t,l}$ are considered in the clustering procedure. Treatments for variables that are not included in the clustering is described in Section \ref{appendix:Cmat} in the supplementary material.}.
If we were to extend these assumptions to a panel setting, one might modify them to
$\sum_{t=1}^T\sum_{i=1}^nz_{i,t,l}^2=nT$, for $l=1,\ldots,q$ and $\sum_{t=1}^T\sum_{i=1}^n\tilde{x}_{i,t,j}^2{\bf 1}\{i\in\mathcal{G}_g\}=|\mathcal{G}_g|T$ for $j=1,\ldots,2K+L+1$.
These assumptions are unnecessarily strong for panel data.
The former assumption, $\sum_{t=1}^T\sum_{i=1}^nz_{i,t,l}^2=nT$, cannot be satisfied for a time series $z_{i,t,l}$ unless it is properly standardized. Standardizing relevant variables before clustering is often necessary, but only for the variables that are involved in clustering. In this case, clustering is not based on $z_{i,t,l}$, standardizing this variable would add a redundant step that would not even affect the clustering results. The latter assumption, $\sum_{t=1}^T\sum_{i=1}^n\tilde{x}_{i,t,j}^2{\bf 1}\{i\in\mathcal{G}_g\}=|\mathcal{G}_g|T$,
 is also too strong, as it requires standardizing  
 $\tilde{x}_{i,t,j}$ within its true cluster, even before any clustering can be done. To remedy the issues in  \cite{Ma&Huang:2017,ma2016estimating}, we lifted these strong assumptions and replaced them with Assumption \ref{ass:lambda} above, which is more appropriate for time series. Lemmas in  Section  \ref{appen:lemmasection} of the supplementary material address the issues in proofs due to the absence of these strong  assumptions.
%This assumption is too strong for time series data. For instance, $\sum_{l=1}^nz_{i,t,l}^2=n$ can be achieved by proper standardization. Let $\tilde{z}_{i,t,l}$ be the raw values for $z_{i,t,l}$. Then $z_{i,t,l}=\sqrt{n}\tilde{z}_{i,t,l}/\sqrt{\sum_{k=1}^n\tilde{z}_{k,t,l}^2}$ should satisfy the condition. However, since $\sum_{k=1}^n\tilde{z}_{k,t,l}^2\neq\sum_{k=1}^n\tilde{z}_{k,t+1,l}^2$ in general, the dependence structure of between $z_{i,t,l}$ and $z_{i,t+1,l}$ will be different from that between $\tilde{z}_{i,t,l}$ and $\tilde{z}_{i,t+1,l}$. The same issue will arise for $\tilde{x}_{i,t,j}$.
}
\end{remark}
%%%%%

The following theorem provides conditions for the convergence of the oracle estimator $\widehat{\boldsymbol{\gamma}}^{or}$.

%\begin{thm}\label{thm:oracle}
%If Assumptions \ref{ass:sparsity}--\ref{ass:subgauss} hold,
%then
%$$P(||\widehat{\boldsymbol{\gamma}}^{or}-\boldsymbol{\gamma}^0||_\infty\leq \phi_{n,T,G,\zeta_{n,T,G}})\geq 1-e^{-\zeta_{n,T,G}},$$
%where $\phi_{n,T,G}=\dfrac{\tilde{c}}{c}C_{q,m}\dfrac{(mng_{\max})^{1/2}(Gp)^{3/4}}{g_{\min}T^{3/4}}(Gp+2\sqrt{Gp}\sqrt{\zeta}+2\zeta)^{1/2}$, and $C_{q,m}=[q^{1/2}+m^{1/2}(L+1+2K)]^{1/2}$.
%\end{thm}

\begin{thm}\label{thm:oracle}
If Assumptions \ref{ass:sparsity1}--\ref{ass:subgauss} hold,
then
$$P(||\widehat{\boldsymbol{\gamma}}^{or}-\boldsymbol{\gamma}^0||_\infty\leq \phi_{n,T,G,\zeta})\geq 1-e^{-\iota},$$
where $\phi_{n,T,G,\zeta}=\dfrac{\sqrt{2\tilde{c}}}{c}B_{q,m}^{1/2}\dfrac{(m\tilde{M}g_{\max})^{1/2}(Gp)^{3/4}}{g_{\min}T^{3/4}}(Gp+2\sqrt{Gp}\sqrt{\zeta}+2\zeta)^{1/2}$, $B_{q,m}=[q^{1/2}+m^{1/2}(L+1+2K)]^{1/2}$, $\tilde{M}=\max\{M_1,M_2,M_3,M_4\}$, $\iota=\min\{\zeta,-\log(\epsilon)\}-\log(2)$, for $\epsilon$ chosen in Assumption \ref{ass:lambda}.
Furthermore, if $g^3_{\rm min}/g_{\rm max}\gg n^{5/3}T^{1/3}$, for any vector $c_n\in\mathbb{R}^{Gp}$ such that $\|c_n\|_2=1$, the asymptotic distribution of $\hat{\boldsymbol{\gamma}}^{or}$ is
$$c_n'(\hat{\boldsymbol{\gamma}}^{or}-\boldsymbol{\gamma}^0)\to N(0,\sigma^2_{\gamma}),$$
where $\sigma^2_{\gamma}=Var(\hat{\boldsymbol{\gamma}}^{or}-\boldsymbol{\gamma}^0)$.
\end{thm}

The proof of Theorem \ref{thm:oracle} can be found in Section \ref{appen:proofs} in the supplementary material.  Theorem \ref{thm:oracle} implies that with an appropriate choice of $\zeta_{n,T,G}$,  the oracle estimator converges to the true parameter in probability. 

\begin{corollary}\label{cor1}
Under the assumptions of Theorem \ref{thm:oracle}, the oracle estimator $\widehat{\boldsymbol{\gamma}}^{or}$ converges to the true parameter $\boldsymbol{\gamma}^{0}$ in probability if one of the following conditions holds:
\begin{enumerate}
    \item The number $n$ is fixed, and $T\to\infty$.
    \item The number $n\to\infty$, and $G$ is fixed. The number $T$ is either fixed or $T\rightarrow\infty$.  Further, the size of the smallest group is large enough such that $g_{min}=O(n^{1/2+\tilde\alpha_4})$ for a positive constant $\tilde{\alpha}_4<1/2$.
x
    \item The number $n\to\infty$, and  $G\rightarrow\infty$. The number $T$ is either fixed or $T\rightarrow\infty$.
    Further, the size of the smallest group is large enough such that
     $g_{min}=O(n^{5/7+\tilde{\alpha}_5})$ for a positive constant $\tilde{\alpha}_5<2/7$.
\end{enumerate}
\end{corollary}
Corollary \ref{cor1} states that the oracle estimator is consistent if $n$ is fixed, or if the size of the smallest group grows somewhat comparably to the increase of $n$. More specifically, if $n$ is fixed, increasing information across time is necessary for consistent estimation. On the contrary, when increasing information across panel can be obtained, $T$ can be held fixed, as long as all the groups have reasonable sizes.
% Some of the listed conditions specify the minimum number of subjects in groups. 

Theorem \ref{thm:main} demonstrates that the proposed estimator of the parameter $\boldsymbol{\gamma}$ is equivalent to the oracle estimator with probability approaching to 1, which  implies that our estimator converges to the true parameter without  prior knowledge of the true group memberships. For our clustering algorithm to work properly,  groups should be distinctive enough.
Assumption \ref{ass:min} states that the pairwise differences of the true parameters  should be large enough for different groups.

\begin{assump}\label{ass:min}
The minimal difference of the common values between two panels is
    $$b_{n,T,G}=\min_{i\in\mathcal{G}_g,j\in\mathcal{G}_{g'},g\neq g'}\|\boldsymbol\gamma_i^0-\boldsymbol\gamma_j^0\|_2=\min_{g\neq g'}\|\boldsymbol\varphi_g^0-\boldsymbol\varphi_{g'}^0\|_2>a\lambda_1+2p\phi_{n,T,G},$$
for some constant $a>0$.
\end{assump}

\begin{thm}\label{thm:main}%cor
Assume the conditions of Theorem \ref{thm:oracle} and Assumption \ref{ass:min} hold. %Define $\zeta^*$  as in the proof of  Theorem \ref{thm:main} in  the supplementary material. 
For $\lambda_1\gg p\phi_{n,T,G}$, where $\phi_{n,T,G}$ is given in Theorem \ref{thm:oracle},
the local minimizer  $\widehat{\boldsymbol{\gamma}}$ of (\ref{objectfunc2}) is almost surely the same as the oracle estimator $\widehat{\boldsymbol{\gamma}}^{or}$, if one of the following conditions hold:
\begin{enumerate}
    \item Suppose $n\rightarrow\infty$, and $T$  is fixed. The size of the smallest group is large enough such that  $(p+2\sqrt{p}+2)^{1/2}n^{1/2}\ll g_{min}=O(n^{7/9+\tilde{\alpha}_0})$ for a positive constant $\tilde{\alpha}_0<2/9$;
    \item Suppose $n,T\rightarrow\infty$, and $G$ is fixed. The size of the smallest group  is large enough such that  $g_{min}=O(n^{1/2+\tilde{\alpha}_4})$ for a positive constant $\tilde{\alpha}_4<1/2$;
        \item Suppose $n,T,G\to\infty$. The size of the smallest group is large enough such that one of the following conditions is met:
        \begin{enumerate}
            \item For a positive constant  $\tilde{\alpha}_3<2/9$, $\max\left\{\frac{n^{7/13}}{T^{1/13}},(p+2\sqrt{p}+2)^{1/2}n^{1/2}\right\}\ll g_{min}=O(n^{7/9+\tilde{\alpha}_3})$; or,
            \item for a positive constant $\tilde{\alpha}_5<2/7$, $g_{min}=O(n^{5/7+\tilde{\alpha}_5})$.
        \end{enumerate}
    \end{enumerate}
That is, if one of the above conditions holds, as $nT\to \infty$,
$$P( \widehat{\boldsymbol{\gamma}}= \widehat{\boldsymbol{\gamma}}^{or})\to1.$$
\end{thm}
Theorem \ref{thm:main} demonstrates that our estimator with prior knowledge of the group information is, asymptotically, as good as the oracle estimator with probability 1, under the presented set of assumptions.\footnote{The  theoretical results presented in this section handle the case with $G\geq2$. In the homogeneous panel case, similar arguments can be made, following \cite{ma2016estimating}. The details are not presented in this paper, but are available upon request.} There are a couple of differences between the two estimators, $\widehat{\boldsymbol{\gamma}}$ and $\widehat{\boldsymbol{\gamma}}^{or}$. The first note-worthy difference is that the oracle estimator converges to the true parameter with probability 1, even when $n$ is fixed, whereas the non-oracle estimator does need $n\to\infty$. This is  expected since the oracle estimator already knows the true group membership, so that increasing the information in time domain only can make the estimator precise enough. On the contrary, the non-oracle estimator needs increasing information in cross-section to estimate the group memberships correctly.
The other difference is that the non-oracle needs stronger assumption on the minimum group size. Again, this is natural, since the non-oracle estimator lacks the group information.

%Throughout two steps of convergence, with properly chosen values of parameters, our estimator is shown to be consistent under different circumstances. Note that as $T\to\infty$ with other parameters fixed, the convergence of our estimator to the oracle estimator cannot be guaranteed.  
%This proves that our estimator, theoretically, converges to the true parameters. The next section demonstrates the performance of our method in finite samples.

\subsection{Simulation: Clustering}\label{sim:panelMIDAS}

The simulation settings in this section are designed to achieve two goals. The first goal is demonstrating the clustering accuracy of the proposed method in finite samples, and providing a guidance on the choice of the tuning parameters involved in our method, in particular, $\theta$ and $\lambda_1$.
The other goal is comparing the performance of the proposed method with other clustering algorithms using the penalized regression idea. To our best knowledge, the only other such clustering algorithm is \cite{Su:2016} (SSP, hereafter). This method is employed in our MIDAS context by taking the Fourier transformation and applying SSP's penalty function as in equation (\ref{su:penalty}) in Section \ref{appen:su} in the supplementary material, rather than (\ref{objectfunc2}). This method is labeled as ``Fourier-SSP." In addition, our penalty function (\ref{objectfunc2}) does not limit how the MIDAS part should be handled. Therefore,  B\&R's approach can be adapted in place of the  Fourier flexible form for our method. This method is labeled as ``B\&R-clust." Our method is labeled as ``F-clust".  Sections  \ref{appen:brclust} and \ref{appen:su} in the supplementary material provide algorithms and relevant details of theses two additional clustering methods.

Section \ref{performance} provides the finite sample performance of F-clust and B\&R-clust for different values of $\theta$. Section \ref{sec:tuningpara} provides guidance on the choice of $\theta$ and $\lambda_1$  for our method. Using the optimal $\theta$ suggested in Section \ref{sec:tuningpara}, Section \ref{sec:compare} compares the three methods, F-clust, B\&R-clust, and Fouier-SSP, in term of parameter estimation accuracy and forecasting accuracy.

In all simulation settings, two clusters with the exponential decline and the cyclical function shapes shown in Section \ref{SParameter} are considered. In each cluster, 15 independent time series are generated. That is, there are 30 coefficient vectors, and two groups are expected after clustering. Each data process follows (\ref{DGP}) shown in Section \ref{SParameter}; $\theta\in\{2, 2.5\}$, $\lambda_1\in\{1,1.5,\cdots,4.5\}$, $\beta_0=0$, $T\in\{100,200,400\}$, $m=\{20,40\}$, and $\alpha_1\in\{0.2, 0.3,0.4\}$ are considered. Notice that $\boldsymbol{\alpha}_i$ are all null-vectors, that is, $\boldsymbol{\gamma}_i=\boldsymbol{\beta}_i$. 
The ADMM algorithm used for the optimization problem (\ref{objectfunc2})  requires  one additional parameter, $\lambda_2$. 
See Algorithm \ref{algorithm} in Section \ref{appen:our} in the supplementary material for more details. Following the choice of \cite{Zhu&Qu:2018}, $\lambda_2=1$ is used.
The clustering algorithm was forced to stop at the 3,000-th iteration if the stopping conditions cannot be satisfied during the process. For the Fourier flexible form and polynomials, an arbitrary choice of $(L,K)=(2,3)$ is considered. The simulation results are already reasonable, so data-driven $(L,K)$s is not considered in this section.

\begin{remark}{\rm The B\&R-clust method involves an additional tuning parameter ($\theta_{\gamma^*}$ in Section \ref{appen:brclust} in the supplementary material) to mange the amount of smoothing. According to the simulation results in \cite{nonparametric}, the choice of $\theta_{\gamma^*}$ is sensitive to the sample size. 
In all our simulations, comparisons are made with the optimal choice of B\&R-clust for each pair of $\theta$ and $\lambda_1$. The optimal $\theta_{\gamma^*}$ is chosen    in $[0,100]$ that minimizes AIC.
It should be noted that by involving an additional parameter, the B\&R-clust method is much more time-consuming than our proposed method, F-clust. This  additional parameter  also tends to make it difficult to find the optimal values for other tuning parameters by increasing the dimension of the parameter space by one.
}\end{remark}

\begin{remark}{\rm
Besides the difficulties involved with the tuning parameter choices, there is yet another reason that our Fourier-based methods are much faster than other methods based on B\&R's MIDAS in general. This is because our the Fourier flexible form and polynomials reduces the number of parameters from $m$ to $q+2K+L+1$, where small values of $K$ and $L$ are generally acceptable. Since our linearized MIDAS model handles much smaller design matrices, it is natural that the estimation is much faster.
}\end{remark}

\subsubsection{Clustering Performance}\label{performance}

This section explores the clustering accuracy of the proposed method and B\&R-clust over a range of $\theta$ and $\lambda_1$. 
As measures of clustering accuracy,  the Rand index \citep{Ran71}, the adjusted Rand index (ARI) \citep{Hubert&Arabie:1985}, Jaccard Index \citep{jaccard1912distribution}, the estimated number of groups $\widehat{G}$, and the median of RMSE of $\widehat{\boldsymbol{\gamma}}$ are presented.
In particular, the first three measures (Rand, ARI, and Jaccard) assess the similarity of the estimated clusters and the true clusters, and defined as
 $Rand=\dfrac{TP+TN}{TP+TN+FP+FN},$
$ARI = \dfrac{Rand-E(Rand)}{\max(Rand)-E(Rand)},$ and
$Jaccard = \dfrac{TP}{TP+FP+FN}.$
Here, TP, TN, FP, and FN indicate true positives, true negatives, false positives, and false negatives, respectively.
The estimated number of clusters and median RMSE of estimated $\widehat{\boldsymbol{\gamma}}$ are also presented. The 
RMSE of  F-clust is calculated as  $RMSE = \sqrt{n^{-1}\sum_{i=1}^{n}\|\mathbf{M'}\widehat{\boldsymbol{\gamma}}_i-{\boldsymbol{\gamma}_i^*}\|^2_2}$;
200 MC samples are generated to evaluate the performance.

\begin{table}[htp]
\centering
\caption{The Influence of Tuning Parameters ($\theta$ and $\lambda_1$)  on the Clustering Performance}
\begin{tabular}{ccccccccc}
\hline
$\theta$              & $\lambda_1$          & Method  & Rand  & ARI   & Jaccard & Clusters & RMSE &$\lambda_{1,BIC}$\\ \hline
%\multirow{26}{*}{2}   & \multirow{2}{*}{0.5} & Our     & 0.524 & 0.015 & 0.015   & 27.52    & 6.27           \\ %\cline{3-8}
%                      &                      & B\&R    & 0.517 & 0.000 & 0.000   & 30.00    & 22.40          \\
\multirow{18}{*}{2}   & \multirow{2}{*}{1}   & F-clust     & 0.531 & 0.030 &  0.026  & 26.45    &  0.5246          \\
                      &                      & B\&R-clust   & 0.530 & 0.756 & 0.027   & 27.05     & 0.4270        \\
                      & \multirow{2}{*}{1.5} & F-clust     & 0.545 & 0.057 & 0.059   &  23.62    &  0.5741       \\
                      &                      & B\&R-clust    & 0.950 & 0.899 & 0.899   &  3.57    &  0.5984       \\
                      & \multirow{2}{*}{2}   & F-clust     & 0.526 & 0.020 & 0.021   & 26.32     & 0.6197           \\
                      &                      & B\&R-clust    & 0.950 & 0.899 & 0.899   &  3.63    &  0.7630         \\
                      & \multirow{2}{*}{2.5} & F-clust     & 0.483 & 0.000 & 0.483   &  1.00    &   0.6620         \\
                      &                      & B\&R-clust    & 0.995 & 0.989 & 0.989   &  2.17    &   0.8954        \\
                      & \multirow{2}{*}{3}   & F-clust     & 0.517 & 0.007 & 0.517   &  1.07    &   0.6937         \\
                      &                      & B\&R-clust    & 0.998 & 0.996 & 0.996   &  2.05    &   1.0139        \\
                      & \multirow{2}{*}{3.5} & F-clust     & 0.483 & 0.000 & 0.483   &  1.00    &   0.7408         \\
                      &                      & B\&R-clust    & 0.999 & 0.998 & 0.998   &  2.01    &   1.1296        \\
                      & \multirow{2}{*}{4}   & F-clust     & 0.483 & 0.000 & 0.480   &  1.20    &   0.7676         \\
                      &                      & B\&R-clust    & 0.995 & 0.989 & 0.989   &  2.12    &   1.2880        \\
                      & \multirow{2}{*}{4.5} & F-clust     & 0.483 & 0.000 & 0.483   &  1.00    &   0.8055         \\
                      &                      & B\&R-clust    & 0.984 & 0.967 & 0.966   &  2.47    &   1.3130        \\
                      & \multirow{2}{*}{BIC} & F-clust & 0.498 & 0.029 & 0.497 & 1.06 & 0.7094&3.157 \\
                      &                                    & B\&R-clust& 0.951 & 0.905 & 0.931 & 2.51 & 0.8385&2.226 \\ \hline
\multirow{18}{*}{2.5} & \multirow{2}{*}{1}   & F-clust     & 0.671 & 0.326 & 0.319   & 13.52   &   0.5308         \\
                      &                      & B\&R-clust    & 0.962 & 0.924 & 0.922   & 3.19    &   0.4368         \\
                      & \multirow{2}{*}{1.5} &F-clust     & 0.906 & 0.810 & 0.805   & 5.43    &   0.5789         \\
                      &                      & B\&R-clust    & 0.985 & 0.983 & 0.966   & 2.13    &   0.6120         \\
                      & \multirow{2}{*}{2}   & F-clust     & 0.968 & 0.935 & 0.933   & 3.00    &   0.6321         \\
                      &                      & B\&R-clust    & 0.998 & 0.996 & 0.995   & 2.06    &   0.7533         \\
                      & \multirow{2}{*}{2.5} & F-clust     & 0.999 & 0.998 & 0.998   & 2.01    &   0.6618         \\
                      &                      & B\&R-clust    & 1.000 & 1.000 & 1.000   & 2.00    &   0.8597         \\
                      & \multirow{2}{*}{3}   & F-clust     & 1.000 & 1.000 & 1.000   & 2.00    &   0.6897         \\
                      &                      & B\&R-clust    & 1.000 & 1.000 & 1.000   & 2.00    &   0.9593        \\
                      & \multirow{2}{*}{3.5} &F-clust     & 1.000 & 1.000 & 1.000   & 2.00    &   0.7325         \\
                      &                      & B\&R-clust    & 1.000 & 1.000 & 1.000   & 2.00    &   1.0465        \\
                      & \multirow{2}{*}{4}   &F-clust     & 1.000 & 1.000 & 1.000   & 2.00    &   0.7736         \\
                      &                      & B\&R-clust    & 1.000 & 1.000 & 1.000   & 2.00    &   1.1210        \\
                      & \multirow{2}{*}{4.5} & F-clust     & 1.000 & 1.000 & 1.000   & 2.00    &   0.8146         \\
                      &                      & B\&R-clust    & 1.000 & 1.000 & 1.000   & 2.00    &   1.1837        \\
                      & \multirow{2}{*}{BIC} & F-clust & 0.998 & 0.996 & 0.996 & 2.05 & 0.6534&2.190 \\
                      &                                    & B\&R-clust& 0.994 & 0.987 & 0.987 & 2.18 & 0.4388&1.107\\ 
\hline
\end{tabular}
\label{tab:clustering}
   \begin{tablenotes}
		\scriptsize 
		    \item 200 MC samples, $T=100$, $\alpha_1=0.4$, $m=20$.
\item Each cell in the ``RMSE" column reports the median of RMSEs of 200 MC samples, which is further multiplied by 100.
	\end{tablenotes}
\end{table}

Table \ref{tab:clustering} reports  clustering indexes,  the number of clusters, and medians of RMSE of estimated $\boldsymbol{\gamma}$,
for $T=100$, $m=20$, and $\alpha_1=0.4$. 
When $\theta=2$, B\&R-clust reveals much better clustering performance than our method in general. In particular, B\&R-clust presents almost perfect clustering, when $\lambda_1$ exceeds 2.5. On the contrary, our method exhibits poor clustering performance; the Rand and Jaccard indexes for our method are about half of that of B\&R-clust, and the ARI is almost zero.  Nonetheless, it is interestig that in terms of RMSE, our method still is comparable or sometimes better than B\&R-clust.
However, when $\theta=2.5$, with a proper choice of tuning parameter $\lambda_1$, the two clustering methods seem to have similar clustering performance. In particular, when  $\lambda_1$ exceeds 1.5, both methods result in almost perfect clustering. 
RMSEs seem to be comparable as well. 

Two conclusions can be drawn from this simulation exercise. One is that choosing the right range of tuning parameters affects the result greatly. The other is that upon the right choice of tuning parameters, both methods lead to reliable clusters, and they both estimate the coefficients quite precisely.

\subsubsection{Selection of Tuning Parameters}\label{sec:tuningpara}

The clustering performance shown above raises the need to carefully choose  the tuning parameters,  $\theta$ and $\lambda_1$, as they affect the clustering performance considerably. In particular, when $\theta=2$, our clustering method does not work well for the simulation settings we considered, no matter what $\lambda_1$ is. Therefore, it is important for users to make a wise choice of $\theta$. Due to the limited knowledge about the true groups in practice, the BIC alone cannot be the right guide to choose the parameters appropriately.
In this section, we shall propose a strategy to select the tuning parameters by calculating the globally convex interval.

Let $c^*_{\theta}(\lambda_1)$ be the minimal eigenvalue of the corresponding design matrix ${W(\Pi^*\otimes I_p)}/n$, where $\Pi^*$ contains the estimated group information with the given parameters. Note that $\Pi^*$ is similar to $\Pi$, except that it is built with estimated groups from data rather than the true groups.
Following the arguments in  \cite{Breheny:2011}, it can be shown that a subset of the globally convex regions of $\theta$ and $\lambda_1$ is given by $\lambda_1\geq \lambda_1^*$ and $\theta$ that satisfy:
\begin{equation}\label{thetaconvex}
\begin{aligned}
\lambda_1^*&=\inf\{\lambda_1: \theta>1/c^*_{\theta}(\lambda_1)\}~~~\text{if the MCP penalty is used,} \\
\lambda_1^*&=\inf\{\lambda_1: \theta>1+1/c^*_{\theta}(\lambda_1)\}~~~\text{if the  SCAD penalty is used.}\\
\end{aligned}
\end{equation}
Here is one strategy to find a convex region:
\begin{enumerate}[label=Step~\arabic*]
    \setlength\itemindent{20pt} \item\label{step1} For a given $\theta$, choose $\lambda_1$ that minimizes BIC. Denote it as $\lambda_{1,BIC}$.
    \setlength\itemindent{20pt} \item
    Find $c^*_{\theta}(\lambda_{1,BIC})$ and $\lambda_1^*$.
    \setlength\itemindent{20pt} \item Check if $\lambda_1>\lambda_1^*$ and $\theta$ satisfy (\ref{thetaconvex}).
    If not, increase the value of $\theta$ and go back to  \ref{step1}.
\end{enumerate}
Table \ref{table:tuning} presents examples of subsets of convex intervals for an MCP penalty, determined from the simulation settings in Section \ref{performance}.  Two random samples are considered.

\begin{table}
\centering
\caption{Selection of $\lambda_1$ given $\theta$}\label{table:tuning}
\begin{tabular}{c|c|ccc}
\hline
                                    & Sample & $\theta=2$      & $\theta>2$     &              \\  \hline
\multirow{2}{*}{$\lambda_{1,BIC}$} & 1      & 4.5               & 3.5              &               \\
                                    & 2      & 5.0               & 4.0              &               \\  \hline
\multirow{2}{*}{$c^*_{\theta}(\lambda_{1,BIC})$}   & 1      & 0.1452            & 0.0681           &            \\
                                    & 2      & 0.1420            & 0.0694           &            \\  \hline
\multirow{2}{*}{Globally Convex Interval of $\theta$}    & 1      & $(6.89,\infty)$   & $(14.69,\infty)$ &  \\
                                    & 2      & $(7.04,\infty)$   & $(14.41,\infty)$ &  \\  \hline
\end{tabular}
\label{selection}
\end{table}
Let us take sample 1 as an example. When $\theta=2$, the BIC-chosen $\lambda_1$ is 4.5, and the subset of the globally convex interval for $\theta$ is calculated as $(6.89,\infty)$. Since $\theta$ is not in this region, increase the value of $\theta$. Repeat the process with $\theta$=2.1. The convex interval for $\theta$ is $(14.68,\infty)$, which does not include $\theta$. We need to increase $\theta$ again. As a matter of fact, for our simulation setting, the clustering results was the same for all $\theta=2.1,2.2,\ldots,16$, which successfully identify the true clusters. The design matrices are also the same as a result, which leads to the (almost) same choice of $\lambda_{1,BIC}$ and $c^*_{\theta}(\lambda_{1,BIC})$.
Therefore, for this dataset, sample 1, as long as $\theta$ is greater 2, we would have an optimal clustering results with a BIC-chosen $\lambda_1$.
This observation is consistent with our simulation results. Our method performed well when $\theta>2$ but not when $\theta=2$.
%Note that this method is provided as one strategy of tuning parameter choice. The proposed interval may not cover all convex regions.

\subsubsection{Comparison of the three clustering methods}\label{sec:compare}

This section compares the three clustering methods (F-clust, B\&R-clust, and Fourier-SSP) and the subject-wise linearized MIDAS using the Fourier flexible form and polynomials (F-noclust). These methods are compared in terms of the accuracy for  parameter estimation in RMSE and for forecasting in RMSFE.
For F-clust and B\&R-clust,
 $\theta=2.5$ is considered, following the suggestion in Section \ref{sec:tuningpara}.
 The frequency ratios $m$ selected in Table \ref{tab:betaMSE} are 20 and 40 to save workload on B\&R's method.  250 samples are generated in MC simulation. 
Other than that, the sample size $T$ and the scale $\alpha_1$ of weights are the same as those considered in Sections \ref{SParameter} and \ref{performance}.
In Fourier-SSP method, the maximum number of groups is fixed as two for the grid search to save the calculation load, taking advantage of prior knowledge of the true number of clusters.
However, in practice, it could be a problem if this number is  improperly chosen.

% Table generated by Excel2LaTeX from sheet 'Sheet3'
\begin{table}[ht]
  \setlength\tabcolsep{5pt}
  \centering
  \caption{Parameter Estimation Accuracy in a Panel Setting}
    \begin{tabular}{cccccc@{\hskip 0.2in}ccc}
    \hline
         &    &$m$  & \multicolumn{3}{c}{$20$} & \multicolumn{3}{c}{$40$}\\
     \cline{3-9}
             $\alpha_1$    &   Method &$T$ & $100$ & $200$ & $400$  &$100$ & $200$ & $400$ \\
        \hline
   \multirow{4}[0]{*}{0.2} & F-clust      &  &  0.4031    & 0.3466     & 0.3059 & 0.1587    & 0.1442    & 0.1304 \\
                & B\&R-clust  &     & 0.3945   & 0.3279      & 0.2261  & 0.1487    & 0.1103     & 0.1005  \\
                & Fourier-SSP  & &  9.6804  &  8.4929     & 7.9253   &   8.8707    &  7.5578     &  6.2652 \\
               & F-noclust  & &  8.2571  &  5.5480     & 3.7683   &  13.7938   &  8.4324    &  5.5765  \\
    \hline
           \multirow{4}[0]{*}{0.3} & F-clust  &  & 0.5163    & 0.4691    & 0.4315 & 0.2152    & 0.2012    & 0.1828  \\
                & B\&R-clust &     & 0.4306    & 0.3531      & 0.2404 & 0.1670      & 0.1221   & 0.0922  \\
                 & Fourier-SSP  &     &   7.4175    &  6.8505     &  6.2241   &  7.2194     &  5.8779     &  4.3612\\
               & F-noclust  &  &  8.2573  &  5.5478    & 3.7685   &  13.7938    &  8.3948     &  5.5765\\
    \hline
           \multirow{4}[0]{*}{0.4} & F-clust  &     & 0.6392   & 0.5966    & 0.5558    & 0.2744    & 0.2603      & 0.2145\\
                 & B\&R-clust   &    & 0.4496    & 0.3663      & 0.2482  & 0.1789      & 0.1364     & 0.0959\\
               & Fourier-SSP    &   &    5.8207   &  5.4699     &  5.1157  &   5.8455    &   4.8590    &  4.6615 \\
                 & F-noclust   &     &    8.2573   &  5.5478     &  3.7685  &  13.7938    &   8.3948    &  5.5765\\
    \hline
     \end{tabular}%
  \label{tab:betaMSE}%
  \begin{tablenotes}
  	\centering
		\scriptsize 
		    \item Each cell reports the median of RMSEs of 250 MC samples, which is further multiplied by 100.
	\end{tablenotes}
\end{table}%

 Table \ref{tab:betaMSE} presents median RMSEs of $\widehat{\boldsymbol\gamma}$.
In particular, the RMSE of all Fourier-based methods are calculated as $RMSE = \sqrt{n^{-1}\sum_{i=1}^{n}\|\mathbf{M'}\widehat{\boldsymbol{\beta}}_i-{\boldsymbol{\beta}_i^*}\|^2_2}.$
Measures of clustering accuracy are not presented in this table, because all three clustering methods have perfectly identified the true clusters using BIC. In terms of estimation accuracy, F-clust and B\&R-clust tend to outperform Fourier-SSP and the subject-level linear regression. Fourier-SSP and the subject-level linear regression do become more accurate as the sample size increases, but not to the extent that they exceed the accuracy of the other two methods based on the penalized regression with (\ref{objectfunc2}).
The B\&R-clust seems  to have the best performance for all settings, while our approach is quite close to the B\&R-clust.
%It is reasonable that B\&R's method outperforms our method since applying Fourier approximation results in a two-layer estimation of parameters.This may have reduced the estimating accuracy.
All three cluster-based method tend to improve as the scale $\alpha_1$ increases, whereas F-noclust is not affected. This is consistent with the results in Table \ref{tab:betaMSE}, where Fourier method is not affected much by a different $\alpha_1$. In contrast, the three cluster-based methods tend to perform better if the signal is stronger.

Computation is the fastest in F-noclust since it does not involve the penalized optimization. F-clust is the next fastest method, followed by Fourier-SSP\footnote{In our simulations, these two methods have similar computation time. This is because we limit the maximum number of groups of Fourier-SSP to 2, utilizing the true group information, which saves the computation time considerably. In reality, our method is faster when the true group information cannot be used.}. B\&R-clust is the slowest, taking at least three times the computation time of our method. 
F-noclust does not utilize the group information, and parameter estimation tends to be less accurate than in other methods, especially when the sample size $T$ is smaller or the frequency ratio $m$ is larger.  
The quality does get better at a faster rate than Fourier-SSP as $T$ increases, but to achieve the same amount of accuracy as F-clust or B\&R cluster, one would need $T\gg 400$, which is often not possible in practice. 
When $T$ is relatively small for a given $m$, using the neighbor information in the same cluster can be one way to improve the quality of the parameter estimation. 
Therefore, our method (F-clust) successfully identifies true clusters and save computation time substantially, without loosing too much accuracy in parameter estimation.

% Table generated by Excel2LaTeX from sheet 'Sheet4'
\begin{table}[htp]
  \setlength\tabcolsep{3pt}
  \centering
  \caption{One-Step-Ahead Forecasting Accuracy in a Panel Setting}
    \begin{tabular}{cccccc@{\hskip 0.2in}ccc}
    \hline
         &    &$m$  & \multicolumn{3}{c}{$20$} & \multicolumn{3}{c}{$40$}\\
     \cline{3-9}
             $\alpha_1$    &   Method &$T$ & $100$ & $200$ & $400$  &$100$ & $200$ & $400$ \\
    \hline
    \multirow{4}[0]{*}{0.2} & F-clust  &  & 0.7700   & 0.7437   & 0.7164  & 0.7591   & 0.7173   & 0.7081\\
                 & B\&R-clust &  & 0.9942   & 0.7925 & 0.7214 & 0.7781   & 0.7336   & 0.7139  \\
                 & Fourier-SSP  & &  2.5779 & 2.6228  & 2.7425 & 2.6582  & 2.5276  & 2.4786  \\
               & F-noclust & & 0.1619   & 0.1401  &  0.1319 & 0.2916   & 0.1617  &  0.1398   \\
    \hline
          \multirow{4}[0]{*}{0.3} & F-clust &  & 0.7911   & 0.7591   & 0.7192  &  0.7836    & 0.7150   & 0.7051\\
                 & B\&R-clust   && 0.9937   & 0.8214  & 0.7197   & 0.8010    & 0.7243   & 0.7144\\
               & Fourier-SSP &  &  2.4774  & 2.4952  & 2.5131  & 2.5103  & 2.3803  & 2.3066 \\
                 & F-noclust &&  0.1619  & 0.1401  & 0.1319  &  0.2916  & 0.1621  &  0.1398 \\
    \hline
          \multirow{4}[0]{*}{0.4} & F-clust&    & 0.8072   & 0.7722   & 0.7290  & 0.8058     & 0.7252  & 0.7176  \\
                 & B\&R-clust &   & 1.0281   & 0.8336   & 0.7289  & 0.8166  & 0.7277    & 0.7257 \\
                & Fourier-SSP &  & 2.2377  & 2.2315  & 2.2493   & 2.2844  & 2.1698  & 2.0982  \\
                & F-noclust & & 0.1619   & 0.1401  &  0.1319 &  0.2916  &  0.1621 & 0.1398  \\
    \hline
    \end{tabular}%
  \label{tab:forecast}%
  	\begin{tablenotes}
  	\centering
		\scriptsize \item Each cell reports the median of RMSFEs of 250 MC samples.
	\end{tablenotes}
\end{table}%

Table \ref{tab:forecast} presents the median RMSFEs of the one-step-ahead forecast. The RMSFEs are computed in a similar way as presented in Section \ref{SParameter}, replacing $\widehat{\boldsymbol{\beta^*}}$ with the one obtained from the penalized regression (\ref{objectfunc2}), and
$RMSFE = \sqrt{(nT/2)^{-1}\sum_{k=1}^{T/2}\sum_{j=1}^{n}(\widehat{y}_{j,T/2+h+k}-{y}_{j,T/2+h+k})^2}.$
It is worth noting that F-noclust outperforms all the cluster-based approaches. 
%Clustering approaches tend to estimate the parameters as well as keeping the possible group information.
This is somewhat expected, as  $\widehat{\boldsymbol{\beta^*}}$ from the subject-level regression is supposed to be the most efficient estimator of $\boldsymbol{\beta^*}$ among all unbiased estimators under our set of assumptions. Nonetheless, F-clust and B\&R-clust provide reasonably accurate forecast compared to the Fourier-SSP method. This observation demonstrates that our penalty functions in  (\ref{objectfunc2}) may perform better than SSP's penalty functions, both in terms of estimation and forecast accuracy in a setting similar to ours.

Overall, if one is interested in identifying  clusters in a panel MIDAS data without prior knowledge on group structures, it seems that our method performs reasonably well without requiring too heavy computations.

\section{Heterogeneity in Labor Market Dynamics across States: Through the Lens of a Mixed-Frequency Okun’s Law Model }

With the new method, we explore the heterogeneity in labor market dynamics across states through the lens of a mixed-frequency Okun’s law model. 

\subsection{Panel Data of State-Level Labor Markets and Model Description }\label{Application}

Okun’s law refers to the empirical negative correlation between output growth  and unemployment rate. A popular specification often adopted in the literature (e.g., \citet{Knotek:2007}) is the following. Let $u_t$ be the first-differenced unemployment rate and $y_t$ be the growth rates of GDP. Okun’s law is a linear relationship between these two variables

\begin{equation*}
u_{t} = \delta + \alpha  y_{t} + \varepsilon_{t}, \label{okun}
\end{equation*}
where $\delta$ is a constant, $\varepsilon_{t}$ is an error term and the coefficient $\alpha$ has a negative sign.\footnote{This specification is often referred to as the differenced version of Okun's law.}

It has been observed that an Okun’s law model might encounter difficulties in dealing with a sudden and abrupt rise in the unemployment rate due to a burst of job losses at the inception of an economic downturn (e.g., \citet{LEE2000331,Moazzami:2011,emaj97}).  In other words, an Okun’s law model with GDP growth as the sole explanatory variable is likely to have difficulties in explaining the nonlinear feature of unemployment dynamics. Weekly initial claims have the highest frequency among the publicly available labor market indicators, and thus can capture the magnitude of job loss in a timely manner. In this regard, the Okun’s law model with weekly initial claims can better capture the non-linearity in unemployment dynamics and also can be used to nowcast the unemployment rate on a weekly basis in real time.

The variables that we use for the mixed-frequency Okun's law model are the quarterly growth rate of log GDP in state $i$ in quarter $t$ ($y_{i,t}$),
the first-differenced unemployment rate of state $i$ in quarter $t$ ($u_{i,t}$), and the log of initial claims in week $j$ of quarter $t$ in state $i$ ($x_{i,t,j}$).\footnote{The state-level GDP growth is from Bureau of Economic Analysis, the state-level unemployment rate is from Local Area Unemployment Statistics by Bureau of Labor Statistics, and the state-level initial claims are from Depart of Labor. 
We seasonally adjust initial claims using seasonal-trend decomposition using LOESS (STL), and use the seasonally adjusted claims for the estimation of Okun's law model.}

We consider 50 states and the District of Columbia, a total 51 cross-section units (or subjects).\footnote{In some states, there is a small number of weeks when the initial claims data are not released, due, for instance, to the shutdown of a local agency collecting the data.} The sample period is from 2005 to the second quarter of 2018, as the quarterly real GDP at the state level is available from 2005.

The mixed-frequency Okun's law model is specified as follows:
 \begin{equation*}
u_{i,t} = \delta_{i} + \alpha_{i}{y}_{i,t}+\mathbf{x}_{i,t}'\boldsymbol{\beta}_i^*+{\varepsilon}_{i,t},
\end{equation*}
where $\mathbf{x}_{i,t} = ({x}_{i,t,1},\cdots,{x}_{i,t,m_{t}})'$ is the collection of weekly initial claims of the corresponding quarter.  
One complication of the mixed-frequency Okun’s law model is that the distributed lag structure of weekly initial claims coefficient $\boldsymbol{\beta}^*_i$ is not well defined, as a quarter has a different number of weeks ranging from 12 to 14.
In this case, the construction of a MIDAS model usually requires a more complicated parameterization to cope with these irregular frequencies. Notably, our method does not require such a procedure. The proposed MIDAS model can flexibly handle the changing number of MIDAS parameters, as the algorithm allows the Fourier transformation matrix $M_{i,t}$ to vary over time as noted in Remark \ref{remark3}.   
The Fourier-transformed log initial claims can be written as $\widetilde{\mathbf{x}}_{i,t}=\mathbf{M}_{i,t}\mathbf{{x}}_{i,t}$, and now the model is re-specified as follows:
\begin{equation*}
%    y_{i,t} = \mathbf{z}_{i,t}'\boldsymbol{\alpha}_i+\mathbf{\tilde{x}}_{i,t}'\boldsymbol{\beta}_i+\varepsilon_{i,t},
    u_{i,t} = \delta_{i} + \alpha_{i}{y}_{i,t} +\mathbf{\tilde{x}}_{i,t}'\boldsymbol{\beta}_i+\varepsilon_{i,t},
\end{equation*}
where $\boldsymbol{\beta}_i=(\beta_{i,1},\cdots,\beta_{i,2K+L+1})'$ and $L$ and $K$ are the number of parameters in Fourier approximation. We cluster states based on $\alpha_{i}$ and $\beta_{i}$, as these parameters capture the dynamic features of unemployment in state $i$. States that share similar values for $({\alpha}_i,\boldsymbol{\beta}_i')$ are allocated to the same group.

In our clustering algorithm, $L=1$ and $K=2$ are chosen to effectively summarize the high-frequency information.\footnote{The clustering results were similar to unreported results with $L=2$ and $K=3$.} This selection ensures that the total number of parameters is smaller than in the conventional MIDAS model. 
For this dataset, we could not find a global convex area for  $\theta$ and $\lambda_1$.  Instead, we conducted a grid search on a range of $\lambda_1$ for each $\theta=2,3,5,8,10$. 
Among the values on the grid, we found that $\theta=2$ and $\lambda_1=2.6$ minimize both AIC and BIC criteria and therefore our reported results are based on these values.

\subsection{Clustering Analysis for the State-Level Labor Markets in the United States}

Table \ref{tab:summary} summarizes estimation results. The algorithm identifies four clusters. There are 24 states in cluster 1, 19 states in cluster 2, 7 states in cluster 3, and 1 state in cluster 4. Clusters 1, 2, and 3 account for 47.0\%, 36.4\%, and 15.3\% of national payroll employment, respectively. Cluster 4 consists of a single state—Louisiana —and constitutes 1.4\% of aggregate employment.
The clusters are determined jointly by the coefficients on GDP growth and the coefficients on log weekly initial claims. Based on the absolute size of coefficients on GDP growth and log initial claims (columns 5 and 6 of Table 7), the labor markets of cluster 3 are the most cyclically sensitive, while those in clusters 2 and 1 are moderately and weakly cyclical, respectively. 
Quite differently, the coefficient on GDP growth in cluster 4 is close to zero and statistically insignificant, but the sum of coefficients on log initial claims is positive and statistically significant: hence, in cluster 4 GDP growth rate does not affect the unemployment rate, while initial claims do. The clusters are further distinguished by the pattern of coefficients on log weekly initial claims. The estimated trajectories within the quarter are plotted in Figure 1, and summarized in Column 7 of Table 7. These trajectories are quite distinct across clusters as shown in Figure 1. The coefficients exhibit an uptrend in cluster 1, while those in cluster 2 and 3 show “W ” and “M ” shapes, respectively. The coefficients of Cluster 4 show an “N ” shape. This result suggests that both the trajectory and the size of these coefficients are important in distinguishing clusters.

\begin{table}[htbp]
\setlength\tabcolsep{3pt}
  \centering
  \caption{Summary of identified clusters}

    \begin{tabular}{c|c|c|c|c|c|c}
\hline\cline{1-5}\cline{6-7}
& {\small Number} & \multicolumn{1}{|c|}{\small Member} & 
\multicolumn{1}{|c|}{\small Emp.} & {\small GDP } & {\small Sum of IC } & 
\multicolumn{1}{|c}{\small IC coeff.'s} \\ 
& {\small of states } & {\small states} & {\small share} & {\small Coeff.} & 
{\small Coeff.} & {\small Shape} \\ \hline
{\small Cluster 1} & {\small 24} & \multicolumn{1}{|l|}{\footnotesize South
Carolina, North Carolina, Florida, } & {\small 47.0\%} & {\small -0.141} & 
{\small 0.862} & {\small Upward } \\ 
&  & \multicolumn{1}{|l|}{\footnotesize Wisconsin, Colorado, Rhode Island,}
&  & \small{(0.00925)} &  & {\small sloping } \\ 
&  & \multicolumn{1}{|l|}{\footnotesize Iowa, South Dakota, Kansas, North
Dakota,   } &  &  &  &  \\ 
&  & \multicolumn{1}{|l|}{\footnotesize Hawaii, Indiana,  Wyoming,
Oklahoma,   } &  &  &  &  \\ 
&  & \multicolumn{1}{|l|}{\footnotesize New Hampshire, New Jersey,  Maine,}
&  &  &  &  \\ 
&  & \multicolumn{1}{|l|}{\footnotesize Michigan, Vermont, Nebraska,} &  &  & 
&  \\ 
&  & \multicolumn{1}{|l|}{\footnotesize California, Delaware, New York,
Alaska } &  &  &  &  \\ \hline
{\small Cluster 2} & {\small 19} & \multicolumn{1}{|l|}{\footnotesize %
Georgia, Oregon, Ohio, Utah, Tennessee,} & {\small 36.4\%} & {\small -0.203}
& {\small 0.972} & {\small W-shape} \\ 
&  & \multicolumn{1}{|l|}{\footnotesize Texas, New Mexico, West Virginia, }
&  &\small{(0.0150)}&  &  \\ 
&  & \multicolumn{1}{|l|}{\footnotesize Missouri, Mississippi,
Arkansas, } &  &  &  &  \\ 
&  & \multicolumn{1}{|l|}{\footnotesize Massachusetts, Kentucky, } &  &  &  & 
\\ 
&  & \multicolumn{1}{|l|}{\footnotesize District of Columbia, Massachusetts, }
&  &  &  &  \\ 
&  & \multicolumn{1}{|l|}{\footnotesize  Idaho, Pennsylvania, Montana,
Connecticut  } &  &  &  &  \\ \hline
{\small Cluster 3} & {\small 7} & \multicolumn{1}{|l|}{\footnotesize %
Alabama, Arizona, Illinois, Washington, } & {\small 15.3\%} & {\small -0.313}
& {\small 1.468} & {\small M-shape} \\ 
&  & \multicolumn{1}{|l|}{\footnotesize Nevada, Minnesota, Virginia} &{\small } & {\small (0.0280)}  &  &  \\ 
\hline
{\small Cluster 4} & {\small 1} & \multicolumn{1}{|l|}{\footnotesize %
Louisiana} & {\small 1.4\%} & {\small 0.0250} & {\small 1.244} & {\small %
N-shape} \\ 
&  & \multicolumn{1}{|l|}{\footnotesize } &{\small } & {\small (0.0625)}  &  &  \\ 
\hline
\end{tabular}

  	\begin{tablenotes}
		\footnotesize \item The abbreviation ``Emp. share" refers to the share out of aggregate payroll employment; ``GDP Coeff." refers to the coefficient on GDP growth; ``IC coeff's shape refers to the shape of coefficients on the weekly initial claims through the corresponding quarter. Numbers in the parentheses are the standard errors.
	\end{tablenotes}
    \label{tab:summary}
\end{table}

\begin{figure}[ht]
\centering
\includegraphics[width=0.60\textwidth]{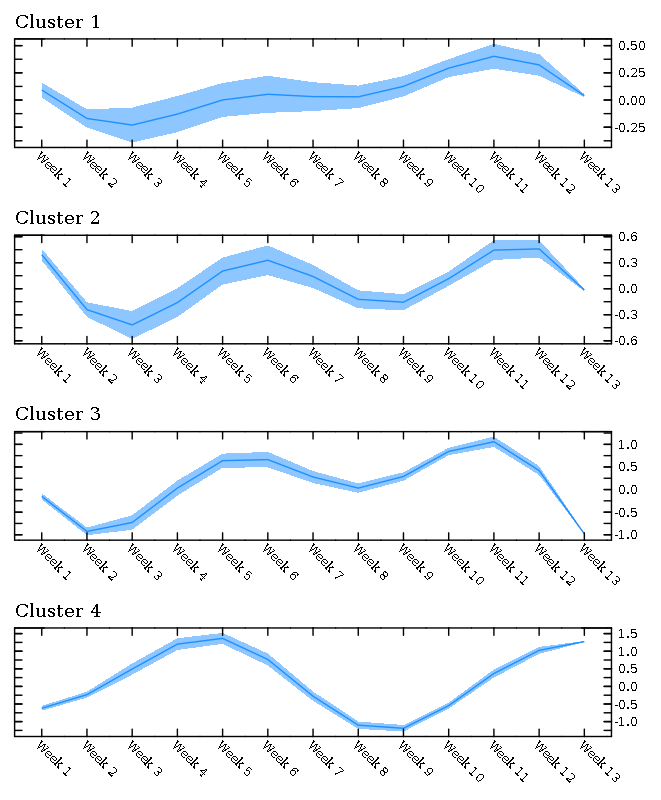}
\caption{Coefficients on log initial claims by cluster} {\footnotesize{The shaded area denotes the 95\% confidence intervals.}}\label{fig2:coef_IC} 
\end{figure}

The large coefficients on certain weeks’ initial claims suggest that those who file for UI benefits during these weeks are more likely to raise the state’s unemployment rate than others. This might be related hirings and layoffs practices and the timing of regular employment turnovers  in different states or regions. Layoffs related to temporary hirings might be concentrated in particular weeks in some states. Workers previously hired by the firms who periodically lay off and recall their workers typically file for UI claims during specific weeks of quarter and are pretty quickly re-employed. On the other hand, those who file for UI benefits outside these weeks might be more likely to be permanent job losers who tend to stay unemployed for a longer period. Therefore, the initial claims filed by these workers tend to be more strongly correlated with the unemployment rate than those filed by temporary job losers.  As an example, in cluster 2 more permanent job losers might file for UI claims during weeks 1, 5, 6, 11 and 12 than in other weeks.  Quite differently, in cluster 1 where the coefficients on initial claims exhibit an upward trend, temporary layoffs might be concentrated early in the quarter. In synthesis, each cluster’s coefficients pattern might reflect these institutional factors. Hence, the different shapes of coefficients can be interpreted as the outcomes of labor market conventions that  differ across clusters.\footnote{The large positive coefficients observed on week 5,6, 11, and 12 might also be related to the reference week of Current Population Survey that usually falls in the second week of each month. The number of those file for the UI claims in the first half of month might be highly correlated with a change in the unemployment rate captured by the survey, if the recent filing of UI claims make the survey respondent more likely to report unemployment as their labor force status. However, this feature is not clearly observed in other clusters. Therefore, the pattern of coefficients on initial claims is less likely to be the outcome of reference-week effect.}

\begin{figure}[ht]
\includegraphics[width=1\textwidth]{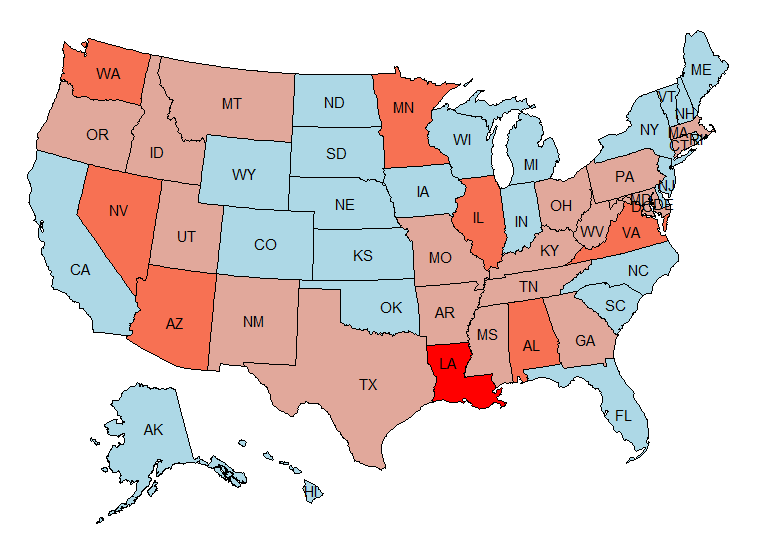}
\caption{States by cluster (cluster 1=light blue, cluster 2=pink, cluster 3 =  orange, cluster 4=red)} %{\footnotesize{ Note to Figure 2: Authors' calculations.}}
\label{fig1:usmap} 
\end{figure}

Figure \ref{fig1:usmap} displays the geographical locations of clusters. Cluster 1, denoted as light blue, is composed of (1) agricultural states in the Midwest region, (2) manufacturing states in the East-north central region, and (3) states in the Northeast. Far from the states in cluster 1, however, California, Alaska, Florida, and North and South Carolina also belong to Cluster 1.\footnote{We follow the Census Bureau's division of regions.} 
Cluster 2 denoted as pink is broadly composed of (1) agricultural states in the West  (mountain region), (2) states in the  central South region, and (3) manufacturing states in the middle Atlantic region of the Northeast. 
Overall, states that belong to the same cluster are adjacent with each other in cluster 1 and 2. Quite differently, states in cluster 3 denoted as orange are widely dispersed. This observation suggests that the geographical proximity is not a necessary condition for the identification of a cluster, but is only partially correlated with cluster membership.  This might be because adjacent states often share similar structural characteristics, such as available natural resources, oil production, and industrial structure.

We further relate the clusters to observable state characteristics in order to find an economic interpretation. To this end, we  consider five variables: (1) the  small firms  share, (2) the employment share of manufacturing,  (3) the employment share of finance industry, (4) the GDP share of oil production, and (5) the fraction of long-term unemployment on total unemployment. The first four characteristics are considered in \cite{Hamilton&Owyang:2012} as possible explanations for heterogeneous regional business cycles.\footnote{Following this study, we also analyze the clusters based on these attributes. The state-level data of four variables are from \cite{Hamilton&Owyang:2012}.} 
We also include the share of long-term unemployment, a component that is likely to reflect the structural unemployment.\footnote{The fraction is calculated based on the micro data from the Current Population Survey (CPS).} 
The unemployment rate of a state where the share of long-term unemployment is high might be less responsive to changes in labor demand.

\begin{figure}[ht]
\centering
\includegraphics[width=0.8\textwidth]{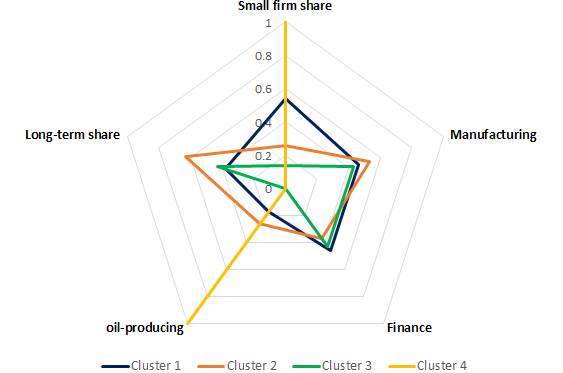}
\caption{Features of each cluster} %{\footnotesize{ Note to Figure 3: Authors' calculations.}}
\label{fig3:spider} 
\end{figure}

We find that the four clusters are moderately distinct in these five observable dimensions.  Table \ref{tab:feature} and Figure \ref{fig3:spider} summarize the observable features of each cluster.  The feature  of each cluster is computed from the fraction of states in the cluster whose particular observable characteristic is more prominent than the average of all states.  For example, according to the second column of Table 8, the fraction of states in cluster 1 whose small firms share is larger than the average of all states is 0.54, and that in cluster 2 is 0.26.

\begin{table}
    \centering
    \caption{Features of each cluster}
  
\begin{tabular}{c|c|c|c|c|c}
\hline\hline
& {\small Small-firm} & {\small Manufacturing} & {\small Finance} & {\small %
Oil-} & {\small Long-term} \\ 
& {\small share} & {\small intensive} & {\small intensive} & {\small %
producing} & {\small unemployment} \\ \hline
{\small Cluster 1} & {\small \textbf{\textcolor{red}{0.54}}} & {\small \textbf{\textcolor{blue}{0.46}}} & {\small \textbf{\textcolor{blue}{0.46}}} & {\small %
0.17} & {\small 0.38} \\ 
{\small Cluster 2} & {\small 0.26} & {\small \textbf{\textcolor{red}{0.53}}} & {\small 0.37} & {\small %
0.26} & {\small \textbf{\textcolor{red}{0.63}}} \\ 
{\small Cluster 3} & {\small 0.14} & {\small \textbf{\textcolor{blue}{0.43}}} & {\small \textbf{\textcolor{blue}{0.43}}} & {\small %
0} & {\small \textbf{\textcolor{blue}{0.43}}} \\ 
{\small Cluster 4} & {\small \textbf{\textcolor{red}{1}}} & {\small 0} & {\small 0} & {\small \textbf{\textcolor{red}{1}}} & 
{\small 0} \\ \hline\hline
\end{tabular}
    \label{tab:feature}
    
    \begin{tablenotes}
		\centering\footnotesize \item Numbers in red are larger than 0.5; those in blue are between 0.4 and 0.5. 
	\end{tablenotes}
	
\end{table}

Cluster 1 is summarily described as small-firm/manufacturing/finance intensive. More than half of the states in this cluster have the share of small firms higher than the average of all states. At the same time, a little less than a half of the states in this cluster have above average employment share of manufacturing.

Cluster 2 is characterized as long-term-unemployment prone and manufacturing intensive. About 60\% of states in this cluster have a higher than average share of long-term unemployment, and a little more than half of the states have manufacturing shares in employment above the average.

Cluster 3 is characterized as manufacturing-finance intensive and long-term unemployment prone. Three out of seven states have larger than average fraction of employment in manufacturing and finance. In addition, the three states have above-average long-term unemployment shares.

Louisiana (cluster 4) is an oil-producing state, whose share of small firms is larger than average.

Summing up, clusters are heterogeneous in multiple dimensions, characterized by differences in several observable attributes, as shown in Figure \ref{fig3:spider}. 
In synthesis, the empirical application demonstrates that our algorithm is be able to reveal meaningful heterogeneity in labor-market dynamics across states without requiring prior knowledge, which in many cases derives from data limitations or from theories lacking empirical support.

\section{Conclusion}

This paper proposed a new clustering method in a panel MIDAS setting, grouping subjects with similar MIDAS coefficients.   The clustering is based on penalized regression approach that is purely data-driven. The major advantage of our method is that it does not require prior knowledge of true group membership, not even of the number of groups. A penalized regression already requires at least two tuning parameters, which are often difficult to choose. A strategy for choosing these tuning parameter are proposed based on a convex region approach. 
%In our nonparametric MIDAS specification, based on the Fourier flexible form and polynomials, has an advantage over other completing methods, for its computational efficiency. %
We show that our proposed clustering method works well both asymptotically and in finite samples. The proposed clustering algorithm shines the most when it is combined with our linearized MIDAS model based on Fourier flexible form and polynomials. This novel linearized MIDAS model is simple, accurate, and computationally fast, making it suitable to use with the proposed clustering algorithm for mixed frequency panel data.  As an empirical example, we provide an application to labor market dynamics at state level in the United States. The application, based on a mixed frequency Okun’s law model, allows grouping the states in four meaningful clusters that correspond to relevant and measurable differences along different dimensions.

\bibliography{References}

\end{document}

\begin{center} {\bf\large Supplementary Material to ``Revealing Cluster Structures Based on Mixed Sampling Frequencies"
	}\end{center}
	\centerline{\textsc{Yeonwoo Rho$^1$*\footnote[0]{
				%Yun Liu is a PhD candidate and Yeonwoo Rho is Assistant Professor in Statistics at the Department of Mathematical Sciences.\\
				*Address of correspondence: Yeonwoo Rho, Department of Mathematical Sciences, Michigan Technological University, Houghton, MI 49931, USA. (yrho@mtu.edu)
				}, Yun Liu$^{2}$, and Hie Joo Ahn$^3$\footnote[0]{Emails: Y. Rho (yrho@mtu.edu), Y. Liu (AnnaLiu@quickenloans.com), and H. J. Ahn (HieJoo.Ahn@frb.gov)}
			}}
		
		\bigskip
	\centerline {$^1$Michigan Technological University}
		\centerline {$^2$Quicken Loans}
	\centerline {$^3$Federal Reserve Board}
	\bigskip
	\centerline{\today}
	\bigskip
	
	\bigskip
This supplementary material consists of two parts: Section \ref{appen:algorithms} consists of details of algorithms used in the main paper; and Section \ref{appen:proofs} presents all proofs.

\appendix\section{Algorithms}\label{appen:algorithms}
This section contains details of algorithms introduced in the main paper. Section \ref{appen:BB} introduce details of B\&R's nonparametric MIDAS in our setting in Section \ref{sec:nonparaMIDAS}. Section \ref{appen:our} present our clustering algorithm (F-clust). Details on how to solve the optimization problem in (\ref{objectfunc2}) in our setting is presented using the alternating direction method of multipliers (ADMM) algorithm. The proposed algorithm is also shown to be convergent. Sections \ref{appen:brclust} and \ref{appen:su} present the details of the two competing clustering methods.  In particular, Section \ref{appen:brclust} introduces how to combine the penalized regression approach with objective function (\ref{objectfunc2}) and the B\&R's method (B\&R-clust).
Section \ref{appen:su} present the algorithm combining Su's penalty function and the Fourier transformation for MIDAS (Fourier-Su). Section \ref{appendix:Cmat} presents the algorithm to exclude a part of parameters from clustering.

\subsection{\cite{nonparametric}'s Nonparametric MIDAS}\label{appen:BB}

The nonparametric MIDAS in \cite{nonparametric} is based on the discrete form of the cubic smooth spline. The least-squares objective function is penalized by the sum of the second difference of weights to balance the goodness of fit and the smoothness of weights. 
Assume that the MIDAS model is shown in (\ref{midas}). The penalized least-squares objective function is
$$Q_{BR} = \sum_{t=1}^{T}\left(y_{t+h}-\alpha_0-\sum_{i=0}^{m-1}{x}_{t,i}{\beta}^*_i\right)^2 + \lambda_{BR}\sum_{i=2}^m\left(\bigtriangledown^2\beta_i^*\right)^2,$$
where $\bigtriangledown^2\beta_i^*=(\beta_i^*-2\beta_{i-1}^*+\beta_{i-2}^*)$ indicates the second difference of weights. 
%For simplicity but without loss of generality, the constant term $\alpha_0$ was ignored and the 
The smoothed least-squares (SLS) estimator \citep{nonparametric} becomes
$$\widehat{\boldsymbol\beta}_{BR}^*=\arg \min_{\beta^*}\left\{(\mathbf{y}-\mathbf{X}\boldsymbol{\beta}^*)'(\mathbf{y}-\mathbf{X}\boldsymbol{\beta}^*)+\lambda_{BR}(\mathbf{D}\boldsymbol{\beta}^*)'\mathbf{D}\boldsymbol{\beta}^*\right\},$$
where $$D_{(m-2)\times (m+1)}=\left(\begin{matrix}
0 & 1 & -2 & 1 & 0 & \cdots & 0\\
0 & 0 & 1 & -2 & 1 & \cdots & 0\\
\vdots& \vdots & \vdots & \vdots & \vdots & \vdots & \vdots\\
0 & 0 & 0 & \cdots & 1 & -2 & 1
\end{matrix}\right).$$

The tuning parameter $\lambda_{BR}$ can be chosen using an information criteria. For example, \cite{nonparametric} proposed to use the modified Akaike information criterion (AIC), 
$$AIC_{\lambda_{BR}} = \log\left\{(\mathbf{y}-\widehat{\mathbf{y}}_{BR})'(\mathbf{y}-\widehat{\mathbf{y}}_{BR})\right\}+\dfrac{2(s_{\lambda_{BR}}+1)}{T-s_{\lambda_{BR}}+2},$$
where $\widehat{\mathbf{y}}_{BR} = \mathbf{X}(\mathbf{X}'\mathbf{X}+\lambda_{BR}D'D)^{-1}\mathbf{X}'\mathbf{y}$.

\subsection{Clustering algorithm for the Fourier Transformed data}\label{appen:our}
The optimization problem in (\ref{objectfunc2}) is not trivial. The alternating direction method of multipliers (ADMM) algorithm by \cite{Boyd_etal:2011} has been successfully employed solving this optimization problem \citep{Ma&Huang:2017,Zhu&Qu:2018}. This section introduces the ADMM algorithm in our setting and proves that it is convergent.

By introducing $\boldsymbol{\eta}_{ij}=\boldsymbol{\gamma}_i-\boldsymbol{\gamma}_j$, minimizing (\ref{objectfunc2}) is equivalent to minimizing
\begin{equation*}\label{objectfunc3}
Q(\boldsymbol{\gamma},\boldsymbol{\eta})=\frac{1}{2}||\mathbf{y}-W\boldsymbol{\gamma}||_2^2+\sum_{1\leq i<j\leq n}\rho(\boldsymbol{\eta}_{ij},\lambda_1)
~~\mbox{subject to}~~\boldsymbol{\eta}_{ij}=\boldsymbol{\gamma}_i-\boldsymbol{\gamma}_j,
\end{equation*}
where $\boldsymbol{\eta}=(\boldsymbol{\eta}_{12}',\ldots,\boldsymbol{\eta}_{n-1,n}')'$.
Following \cite{Boyd_etal:2011},
this constrained optimization problem can be solved using a variant of 
the augmented Lagrangian
\begin{equation}\label{objectfunc4}
\begin{aligned}
Q_{\lambda_2}(\boldsymbol{\gamma},\boldsymbol{\eta},\boldsymbol{\xi})&=\frac{1}{2}||\mathbf{y}-W\boldsymbol{\gamma}||_2^2+\sum_{i<j}\rho(\boldsymbol{\eta}_{ij},\lambda_1)+\frac{\lambda_2}{2}\sum_{i<j}
||\boldsymbol{\gamma}_i-\boldsymbol{\gamma}_j-\boldsymbol{\eta}_{ij}||_2^2+\sum_{i<j}\boldsymbol{\xi}_{ij}'(\boldsymbol{\gamma}_i-\boldsymbol{\gamma}_j-\boldsymbol{\eta}_{ij}),%\\
%&=\dfrac{1}{2}\|\mathbf{y}-W\boldsymbol{\gamma}\|_2^2+\dfrac{\lambda_2}{2}\|D\boldsymbol{\gamma}-(\boldsymbol{\eta}+\boldsymbol{\xi}/\lambda_2)\|_2^2+\sum_{i<j}\rho(\boldsymbol{\eta}_{ij},\lambda_1)-\dfrac{\boldsymbol{\xi}'\boldsymbol{\xi}}{2\lambda_2},
\end{aligned}
\end{equation}
where $\boldsymbol{\xi}=(\boldsymbol{\xi}_{12}',\boldsymbol{\xi}_{13}',\ldots,\boldsymbol{\xi}_{n-1,n}')'$  and $\boldsymbol{\xi}_{ij}$ are $p$-vectors of Lagrangian multipliers.
As proposed by \cite{Boyd_etal:2011}, the optimization problem in (\ref{objectfunc4}) can be solved using the alternating direction method of multipliers (ADMM) algorithm.
At the $(s+1)$-th step of the ADMM algorithm, estimated parameters $\boldsymbol{\gamma}^{s+1},$ $\boldsymbol{\eta}^{s+1}$ and $\boldsymbol{\xi}^{s+1}$ are updated as
\begin{equation}\label{ADMM}
\begin{aligned}
	\boldsymbol{\gamma}^{s+1}&=\arg\min_{\boldsymbol{\gamma}}Q_{\lambda_2}(\boldsymbol{\gamma},\boldsymbol{\eta}^s,\boldsymbol{\xi}^s),\\
	\boldsymbol{\eta}^{s+1}&=\arg\min_{\boldsymbol{\eta}}Q_{\lambda_2}(\boldsymbol{\gamma}^{s+1},\boldsymbol{\eta},\boldsymbol{\xi}^s),\\
	\boldsymbol{\xi}_{ij}^{s+1}&=\boldsymbol{\xi}_{ij}^{s}+\lambda_2(\boldsymbol{\eta}_{ij}^{s+1}-\boldsymbol{\gamma}_{i}^{s+1}+\boldsymbol{\gamma}_{j}^{s+1}),\\
\end{aligned}
\end{equation}
where $\boldsymbol{\eta}^s$ and $\boldsymbol{\xi}^s$ are the estimates in the $s$-th iteration.
By collecting terms related to $\boldsymbol{\gamma}$, the first function in (\ref{ADMM}) is equivalent to minimizing
\begin{equation*}\label{Objective}
    Q_{\lambda_2}^{\boldsymbol{\gamma}}(\boldsymbol{\gamma},\boldsymbol{\eta},\boldsymbol{\xi})=\dfrac{1}{2}\|\mathbf{y}-W\boldsymbol{\gamma}\|_2^2+\dfrac{\lambda_2}{2}\|D\boldsymbol{\gamma}-(\boldsymbol{\eta}+\boldsymbol{\xi}/\lambda_2)\|_2^2,
\end{equation*}
where $D_{ij}=(\boldsymbol{e}_i-\boldsymbol{e}_j)'\otimes I_{p}$, $D=(D_{12}', D_{13}',\cdots,D_{n-1,n}')'$, $\boldsymbol{e}_i$ is an $n$-dimension vector with the $i$-th element as one and the rest as zeros, and $I_{p}$ is an identity matrix with rank $p$. Therefore, $\boldsymbol{\gamma}^{s+1}=\left(W'W+\lambda_2D'D\right)^{-1}\left\{W'\mathbf{y}+\lambda_2D'(\boldsymbol{\eta}^s+\boldsymbol{\xi}^s/\lambda_2)\right\}$.

%Similarly, we define another function with terms only related to $\boldsymbol{\eta}$.
%$$Q_{\lambda_2}^{\boldsymbol{\eta}_{ij}}(\boldsymbol{\gamma},\boldsymbol{\eta},\boldsymbol{\xi})=\dfrac{1}{\lambda_2}\rho(\boldsymbol{\eta}_{ij},\lambda_1)+\dfrac{\lambda_2}{2}\|D\boldsymbol{\gamma}-(\boldsymbol{\eta}+\boldsymbol{\xi}/\lambda_2)\|_2^2-\dfrac{\boldsymbol{\xi}'\boldsymbol{\xi}}{2\lambda_2}.$$
The MCP is shown to be nearly unbiased and is applicable here to update $\boldsymbol{\eta}^{s+1}$ \citep{Zhu&Qu:2018}. The penalty function of MCP is $\rho(\boldsymbol{\gamma}_i-\boldsymbol{\gamma}_j,\lambda_1)=\rho_\theta(\|\boldsymbol{\gamma}_i-\boldsymbol{\gamma}_j\|_2,\lambda_1)$ where $\rho_\theta(x,t)=t\int_0^x(1-\frac{u}{\theta t})_+du$. As a consequence, when the MCP is selected, $\boldsymbol{\eta}_{ij}^{s+1}$ can be updated by
\begin{equation*}\label{MCP}
\boldsymbol{\eta}_{ij}^{s+1}=\left\{\begin{array}{ll}
\tilde{\boldsymbol{\eta}}_{ij}^{s+1} & \text{if }\|\tilde{\boldsymbol{\eta}}_{ij}^{s+1}\|_2\geq\theta\lambda_1,\\
\dfrac{\theta\lambda_2}{\theta\lambda_2-1}\left(1-\dfrac{\lambda_1/\lambda_2}{\|\tilde{\boldsymbol{\eta}}_{ij}^{s+1}\|_2}\right)_+\tilde{\boldsymbol{\eta}}_{ij}^{s+1}& \text{if }\|\tilde{\boldsymbol{\eta}}_{ij}^{s+1}\|_2<\theta\lambda_1,
\end{array}\right.
\end{equation*}
where $\tilde{\boldsymbol{\eta}}_{ij}^{s+1}=\boldsymbol{\gamma}_i^{s+1}-\boldsymbol\gamma_j^{s+1}-\boldsymbol\xi_{ij}^s/\lambda_2$ and $\theta>1/\lambda_2$ for the global convexity of the second minimization function in (\ref{ADMM}) \citep{Wang:2018}.

%Another penalty function that we consider to update $\boldsymbol{\eta}^{s+1}$, is SCAD.  We also present the formula for SCAD penalty \citep{ma2016estimating}. Note that $\theta$ is expected to be larger than $1+1/\lambda_1$ to guarantee the convexity of the second minimization function in (\ref{ADMM}).
%\begin{equation}\label{SCAD}
%\boldsymbol{\eta}_{ij}^{s+1}=\left\{\begin{array}{ll}
%\tilde{\boldsymbol{\eta}}_{ij}^{s+1} & \text{if }\|\tilde{\boldsymbol{\eta}}_{ij}^{s+1}\|_2\geq\theta\lambda_1,\\
%\dfrac{(\theta-1)\lambda_2}{(\theta-1)\lambda_2-1}\left(1-\dfrac{\theta\lambda_1/((\theta-1)\lambda_2)}{\|\tilde{\boldsymbol{\eta}}_{ij}^{s+1}\|_2}\right)_+\tilde{\boldsymbol{\eta}}_{ij}^{s+1} & \text{if } \lambda_1+\dfrac{\lambda_1}{\lambda_2}\leq\|\tilde{\boldsymbol{\eta}}_{ij}^{s+1}\|_2<\theta\lambda_1,\\
%\left(1-\dfrac{\lambda_1/\lambda_2}{\|\tilde{\boldsymbol{\eta}}_{ij}^{s+1}\|_2}\right)_+\tilde{\boldsymbol{\eta}}_{ij}^{s+1} & \text{if }\|\tilde{\boldsymbol{\eta}}_{ij}^{s+1}\|_2<\lambda_1+\dfrac{\lambda_1}{\lambda_2}.
%\end{array}\right.
%\end{equation}

If the minimization function of $\boldsymbol{\eta}^{s+1}$ is non-convex, assigning appropriate initial values becomes essential. A proper start will lead to an ideal solution. Inspired by \cite{Zhu&Qu:2018},  the clustering method can be initialized as shown in the following algorithm.

\bigskip

\begin{algorithm}[H]
	\caption{F-clust Algorithm }\label{algorithm}
	\SetAlgoLined
	%\begin{algorithmic}
		\textbf{Initialization:}
		
		$\boldsymbol{\xi}^0=\mathbf{0},~\boldsymbol{\gamma}^0=\left(W'W\right)^{-1}\left(W'\mathbf{y}\right),~\boldsymbol{\eta}^0=\arg\min_{\boldsymbol{\eta}}Q_{\lambda_2}(\boldsymbol{\gamma},\boldsymbol{\eta},\boldsymbol{\xi})$, where $\lambda_2$ and $\theta>1/\lambda_2$ are fixed.
		
		%\hspace{5mm}
		
		\For{$s=0,1,2,\cdots$}{
			
			$\boldsymbol{\gamma}^{s+1}=\left(W'W+\lambda_2D'D\right)^{-1}\left\{W'\mathbf{y}+\lambda_2D'(\boldsymbol{\eta}^s+\boldsymbol{\xi}^s/\lambda_2)\right\}$.
			
			$\boldsymbol{\eta}^{s+1}=\arg\min_{\boldsymbol{\eta}}Q_{\lambda_2}(\boldsymbol{\gamma}^{s+1},\boldsymbol{\eta},\boldsymbol{\xi}^{s})$,
			
			$\boldsymbol{\xi}_{ij}^{s+1}=\boldsymbol{\xi}_{ij}^{s}+\lambda_2(\boldsymbol{\eta}_{ij}^{s+1}-\boldsymbol{\gamma}_i^{s+1}+\boldsymbol{\gamma}_j^{s+1})$, for all $1\leq i<j\leq n$.
			
			\If{the stopping criteria are true}{
				Break
			}}
			
		%\\\hrulefill
		%\State part 2 here
	%\end{algorithmic}
\end{algorithm}
The estimated number of groups, $\widehat{G}$, can be obtained by $\boldsymbol{\eta}$. 
%At the step $s^*$ when the algorithm terminates, $\boldsymbol{\gamma}_i$ and $\boldsymbol{\gamma}_j$ are in the same group if $\|\boldsymbol{\eta}_{ij}^{s^*}\|_2^2=0$. 
If $\widehat{\boldsymbol{\gamma}}_i=\widehat{\boldsymbol{\gamma}}_j$,  $\boldsymbol{\gamma}_i$ and $\boldsymbol{\gamma}_j$ are expected to be in the same cluster. However, as a penalty $\boldsymbol\eta_{ij}$ has been imposed in the clustering algorithm, the equality of two estimated parameters is not achievable. 
As a result, the MCP penalty is utilized on $\widehat{\boldsymbol\eta}_{ij}$. 
Two parameters $\boldsymbol{\gamma}_i$ and $\boldsymbol{\gamma}_j$ are clustered in the same group if $\widehat{\boldsymbol\eta}_{ij}=\boldsymbol{0}$. 
%Only if the tuning parameter $\lambda_1$ is given, $\widehat{G}$ and the estimated coefficients $\widehat{\gamma}$ can be evaluated. Hence,  different values are assigned to $\lambda_1$, and calculate the corresponding BIC's shown in (\ref{BIC}). $\lambda_1$ is selected when BIC reaches the minimum. 

In Algorithm \ref{algorithm},  the stopping criteria are defined as the following. Let $\boldsymbol{\kappa}^{s+1}_{ij}=\boldsymbol{\gamma}^{s+1}_{i}-\boldsymbol{\gamma}^{s+1}_{j}-\boldsymbol{\eta}^{s+1}_{ij}$, $\boldsymbol{\kappa}=(\boldsymbol{\kappa}'_{12},\cdots,\boldsymbol{\kappa}'_{n-1,n})'$ and $\boldsymbol{\tau}^{s+1}_{k}=-\lambda_2\{\sum_{i=k}(\boldsymbol{\eta}^{s+1}_{ij}-\boldsymbol{\eta}^{s}_{ij})-\sum_{j=k}(\boldsymbol{\eta}^{s+1}_{ij}-\boldsymbol{\eta}^{s}_{ij})\}$, $\boldsymbol{\tau}=(\boldsymbol{\tau}_{1},\cdots,\boldsymbol{\tau}_{n})'$. At any step $s^*$, if for some small values $\epsilon^\kappa$ and $\epsilon^\tau$, $\|\boldsymbol{\kappa}^{s^*}\|_2\leq\epsilon^\kappa$ and $\|\boldsymbol{\tau}^{s^*}\|_2\leq\epsilon^\tau$, the algorithm stops. %\todo{This is the stopping criteria.} 
Following \cite{Zhu&Qu:2018}, define$\epsilon^\kappa$ and $\epsilon^\tau$  as
$$\epsilon^\kappa=\sqrt{np}\epsilon^{abs}+\epsilon^{rel}\|D'\boldsymbol{\xi}^{s^*}\|_2,~~ \epsilon^\tau=\sqrt{|\mathcal{I}|p}\epsilon^{abs}+\epsilon^{rel}\max\{\|D\boldsymbol{\eta}^{s^*}\|_2,\|\boldsymbol{\eta}^{s^*}\|_2\},$$
where $\mathcal{I}=\{(i,j):1\leq i<j\leq n\}$, $|\mathcal{I}|$ indicates the cardinality of $\mathcal{I}$. Here, $\epsilon^{abs}$ and $\epsilon^{rel}$ are predetermined small values.

\begin{proposition}\label{thm:admm}
The above clustering algorithm ensures convergence, that is, 
$\|\boldsymbol{\kappa}^{s+1}\|_2^2\rightarrow0~\text{and}~\|\boldsymbol{\tau}^{s+1}\|_2^2\rightarrow0,$
as $s\rightarrow\infty$.
\end{proposition}

\begin{proof}[Proof of Proposition \ref{thm:admm}]
	$\|\boldsymbol{\kappa}^{s+1}\|_2^2\xrightarrow{s\rightarrow\infty}0$ can be shown similarly to the proof of Proposition 1 in \citet{Ma&Huang:2017}. 
The proof of $\|\boldsymbol{\tau}^{s+1}\|_2^2\xrightarrow{s\rightarrow\infty}0$ can be done by ignoring the penalty term in the objective function in the proof of Theorem 3.1 in \cite{Zhu&Qu:2018}. 
\end{proof}

Proposition \ref{thm:admm} demonstrates that the clustering algorithm is convergent as the number of iteration, $s$, approaches infinity. The stopping criteria can be satisfied at some step eventually.

\subsection{Comparable Clustering Methods 1: B\&R-clust}\label{appen:brclust}

Recall that in (\ref{eq:panel1}), the MIDAS regression model without Fourier transformation of each subject is
$$\mathbf{y}_i=Z_i\boldsymbol{\alpha_i}+X_i\boldsymbol{\beta}_i^*+\boldsymbol{\varepsilon}_i,~~~i=1,\cdots,n.$$
For more than one subject,  the penal MIDAS model can be written as
$$\mathbf{y}_i=(Z_i,X_i)\left(\begin{matrix}\boldsymbol{\alpha_i}\\ \boldsymbol{\beta}_i^*\end{matrix}\right)=\widetilde{W}_i\boldsymbol{\gamma}_i^*,~~\text{or}~~\mathbf{y}={\widetilde{W}}\boldsymbol{\gamma}^*+\boldsymbol{\varepsilon},$$
where $\widetilde{W}_i=(Z_i,X_i)$ is the raw observations, $\boldsymbol{\gamma}^*_i=(\boldsymbol{\alpha_i}',\boldsymbol{\beta^*_i}')'$, $\boldsymbol{\gamma}^*=({\gamma^*_1}',\cdots,{\gamma^*_n}')'$.

Refer to the main idea of \cite{nonparametric}, the cubic smoothing spline penalty is considered. %This penalty, which rejects too sharped changes of parameters when estimating the parameter $\boldsymbol{\gamma}^*$ of the raw data. 
The penalized objective function will be given as
\begin{equation*}
    Q(\boldsymbol{\gamma}^*)=\dfrac{1}{2}\|\mathbf{y}-{W}\boldsymbol{\gamma}^*\|_2^2+\dfrac{1}{2}\theta_{\gamma^*}{\boldsymbol{\gamma}^*}'\mathbf{A}\boldsymbol{\gamma}^*,
\end{equation*}
where $\theta_{\gamma^*}$ is the pre-determined smoothing parameter, $\mathbf{A}=I_n\otimes(A'A)$. $A$ is defined as $$A_{(m-2)\times m}=\left(\begin{matrix}
1 & -2 & 1 & 0 & \cdots & 0\\
0 & 1 & -2 & 1 & \cdots & 0\\
\vdots & \vdots & \vdots & \vdots & \vdots & \vdots\\
0 & 0 & \cdots & 1 & -2 & 1
\end{matrix}\right).$$

According to \cite{Zhu&Qu:2018}, our goal is to solve the constrained optimization function
\begin{equation}\label{ZQandBR}
    Q_{\lambda_2}(\boldsymbol{\gamma}^*,\boldsymbol{\eta},\boldsymbol{\xi})=Q(\boldsymbol{\gamma}^*)+\sum_{i<j}\rho(\boldsymbol{\eta}_{ij},\lambda_1)+\frac{\lambda_2}{2}\sum_{i<j}
||\boldsymbol{\gamma}^*_i-\boldsymbol{\gamma}^*_j-\boldsymbol{\eta}_{ij}||_2^2+\sum_{i<j}\boldsymbol{\xi}_{ij}'(\boldsymbol{\gamma}^*_i-\boldsymbol{\gamma}^*_j-\boldsymbol{\eta}_{ij}).
\end{equation}

The clustering algorithm of (\ref{ZQandBR}) is similar to Algorithm \ref{algorithm}.

\bigskip
\begin{algorithm}[H]
	\caption{B\&R-clust Algorithm}
	\label{algorithm2}
	%\begin{algorithmic}
		\textbf{Initialization:}
		
		$\boldsymbol{\xi}^0=\mathbf{0},~\boldsymbol{\gamma}^0=\left(\widetilde{W}'\widetilde{W}+\theta_{\gamma^*}\mathbf{A}\right)^{-1}\left(\widetilde{W}'\mathbf{y}\right),~\boldsymbol{\eta}^0=\arg\min_{\boldsymbol{\eta}}Q_{\lambda_2}(\boldsymbol{\gamma},\boldsymbol{\eta},\boldsymbol{\xi})$, where $\lambda_2$ and $\theta>1/\lambda_2$ are fixed.
		
		%\hspace{5mm}
		
		\For{$s=0,1,2,\cdots$}{
			
			$\boldsymbol{\gamma}^{s+1}=\left(W'W+\lambda_2D'D+\theta_{\gamma^*}\mathbf{A}\right)^{-1}\left\{W'\mathbf{y}+\lambda_2D'(\boldsymbol{\eta}^s+\boldsymbol{\xi}^s/\lambda_2)\right\}$.
			
			$\boldsymbol{\eta}^{s+1}=\arg\min_{\boldsymbol{\eta}}Q_{\lambda_2}(\boldsymbol{\gamma}^{s+1},\boldsymbol{\eta},\boldsymbol{\xi}^{s})$,
			
			$\boldsymbol{\xi}_{ij}^{s+1}=\boldsymbol{\xi}_{ij}^{s}+\lambda_2(\boldsymbol{\eta}_{ij}^{s+1}-\boldsymbol{\gamma}_i^{s+1}+\boldsymbol{\gamma}_j^{s+1})$, for all $1\leq i<j\leq n$.
			
			\If{the stopping criteria are true}{
				Break
			}}
		
		%\\\hrulefill
		%\State part 2 here
	%\end{algorithmic}
\end{algorithm}

\bigskip
Note that Algorithm \ref{algorithm2} follows the same main idea of \cite{Zhu&Qu:2018}.  However, in \cite{Zhu&Qu:2018}, the model introduces B-splines to approximate observations, while Algorithm \ref{algorithm2} simply uses all high-frequency regressors. Moreover, an additional tuning parameter, $\theta_{\gamma^*}$, is required to be predetermined. Refer to \cite{nonparametric,Zhu&Qu:2018}, the selection of $\theta_{\gamma^*}$ is based on the minimum of AIC given by
$$AIC_{\theta_{\gamma^*}}=\sum_{i=1}^{n}\left\{\log\left(\dfrac{\|\mathbf{y}_i-{W}_i\widehat{\boldsymbol{\gamma}}_i\|^2_2}{T}\right)+\dfrac{2\cdot df_i}{T}\right\},$$
where $df_i=tr\{W_i(W_i'W_i+\theta_{\gamma^*}A'A)^{-1}W_i'\}$. The selection of $\lambda_1$ here, is by minimizing
$$BIC_{\lambda_1}=\log\left(\dfrac{\|\mathbf{y}-W\widehat{\boldsymbol{\gamma}}\|_2^2}{n}\right)+\dfrac{\log(n)\left\{\widehat{G}(\frac{1}{n}\sum_{i=1}^{n}df_i)\right\}}{n}.$$
With fixed $\lambda_1$,  $AIC_{\theta_{\gamma^*}}$ can be obtained for different values of $\theta_{\gamma^*}$. Then, fix $\theta_{\gamma^*}$ with minimum BIC,  $BIC_{\lambda_1}$ can be calculated based on the determined $\theta_{\gamma^*}$.

\medskip

\subsection{Comparable Clustering Methods 2: Fourier-SSP}\label{appen:su}

 \cite{Su:2016} introduced C-Lasso for clusters to identify relatively large differences between parameters and group averages rather than the traditional Lasso for each subject to select relevant covariates.
The penalized profile likelihood (PPL)  function mentioned in \cite{Su:2016} is 
\begin{equation*}
    Q(\boldsymbol{\gamma}^*)=\dfrac{1}{nT}\sum_{i=1}^n\sum_{t=1}^T\phi(w_{it};\boldsymbol{\gamma}^*_i,\widehat{\mu}_i(\boldsymbol{\gamma}_i^*)).
\end{equation*}
By introducing the group Lasso penalty, the PPL criterion function becomes
\begin{equation*}\label{su:penalty}
    Q_{G,\lambda_{PPL}} = Q(\boldsymbol{\gamma}^*)+\dfrac{\lambda_{PPL}}{N}\sum_{i=1}^N\prod_{g=1}^{G_0}\|\boldsymbol{\beta}_i-\boldsymbol{\alpha}_g\|_2,
\end{equation*}
where $\lambda_{PPL}$ is a tuning parameter. 
The C-Lasso estimation $\widehat{\boldsymbol{\gamma}}$ and $\widehat{\boldsymbol{\alpha}}$, respectively. 
Without any prior knowledge of the true clusters, PPL C-Lasso estimation requires a predetermination of a reasonable maximum value, $G_0$, of groups. 
An appropriate choice of $(\lambda_{PPL},G_0)$ can be found by minimizing IC based on all possible values of clusters less than $G_0$ as long as predetermined values of $\lambda_{PPL}$. 
To start the algorithm, \cite{Su:2016} suggested a natural initial value as $\widehat{\boldsymbol{\alpha}}_g^{(0)}=0$ for all $g=1,\cdots,G_0$ and ${\widehat{\boldsymbol{\gamma}}}^{*{(0)}}$ as the quasi-maximum likelihood estimation (QMLE) of $\boldsymbol{\gamma}^*_i$ in each subjects. 
More details can be found in \cite{Su:2016}.

\begin{algorithm}[H]
	\caption{SSP -- PPL Algorithm Given $G_0$ and $\lambda_{PPL}$ }
	\label{algorithm3}
	\SetAlgoLined
	%\begin{algorithmic}
		\textbf{Initialization:}
		${\widehat{\boldsymbol{\alpha}}^{(0)}=(\widehat{\boldsymbol{\alpha}}_1^{(0)},\cdots,\widehat{\boldsymbol{\alpha}}_{G_0}^{(0)})}'$, ${\widehat{\boldsymbol{\gamma}}}^{*^{(0)}}={({\widehat{\boldsymbol{\gamma}}}_1^{*^{(0)}},\cdots,{\widehat{\boldsymbol{\gamma}}}_n^{*^{(0)}})}'$ s.t. $\sum_{i=1}^n\|{\widehat{\boldsymbol{\gamma}}}_i^{*^{(0)}}-\widehat{\boldsymbol{\alpha}}_g^{(0)}\|\neq0$ for all $g=2,\cdots,G_0$.
		
		%\hspace{5mm}
		
		\For{$s=1,2,\cdots$}{
		    \For{$g=1,2,\cdots G_0$}{
		    Obtain the estimator $({\widehat{\boldsymbol{\gamma}}}^{*^{(s,G)}},\widehat{\boldsymbol{\alpha}}_g^{(s)})$ of $(\boldsymbol{\gamma}^*,\boldsymbol{\alpha}_g)$ by minimizing the following objective function $Q_{G,\lambda_{PPL}}^{(s,g)}(\boldsymbol{\gamma}^*,\boldsymbol{\alpha}_g)$.
		    
		    \uIf{$g=1$}{
		    $Q_{G,\lambda_{PPL}}^{(s,g)}(\boldsymbol{\gamma}^*,\boldsymbol{\alpha}_g) = Q(\boldsymbol{\gamma}^*)+\dfrac{\lambda_{PPL}}{N}\sum_{i=1}^N\|\boldsymbol{\gamma}_i^*-\boldsymbol{\alpha}_g\|\prod_{k=2}^G\|{\boldsymbol{\gamma}_i}^{*^{(s-1,k)}}-\boldsymbol{\alpha}_k^{(s-1)}\|$ \;   }
		    \uElseIf{$g\neq G$}{
		    $Q_{G,\lambda_{PPL}}^{(s,g)}(\boldsymbol{\gamma}^*,\boldsymbol{\alpha}_g) = Q(\boldsymbol{\gamma}^*)+\dfrac{\lambda_{PPL}}{N}\sum_{i=1}^N\|\boldsymbol{\gamma}_i^*-\boldsymbol{\alpha}_g\|\prod_{j=1}^{g-1}\|{\widehat{\boldsymbol{\gamma}}_i}^{*^{(s,j)}}-\boldsymbol{\alpha}_j^{(s)}\|\prod_{k=g+1}^G\|{\boldsymbol{\gamma}_i}^*{^{(s-1,k)}}-\boldsymbol{\alpha}_k^{(s-1)}\|$\;  }
		    \Else{
		    $Q_{G,\lambda_{PPL}}^{(s,g)}(\boldsymbol{\gamma}^*,\boldsymbol{\alpha}_g) = Q(\boldsymbol{\gamma}^*)+\dfrac{\lambda_{PPL}}{N}\sum_{i=1}^N\|\boldsymbol{\gamma}_i^*-\boldsymbol{\alpha}_g\|\prod_{k=1}^{G-1}\|{\widehat{\boldsymbol{\gamma}}_i}^{*^{(s,k)}}-\boldsymbol{\alpha}_k^{(s)}\|$ \;  }
  
			}
			\If{the stopping criteria are true}{Break}
			}
\end{algorithm}

\cite{Su:2016} provided a stopping criteria for this algorithm:
\begin{equation*}
    \widehat{Q}_{G,\lambda_{PPL}}^{(s-1)}-\widehat{Q}_{G,\lambda_{PPL}}^{(s)}\leq\epsilon_{tl}\text{ and }
    \dfrac{\sum_{g=1}^G\left\|\widehat{\boldsymbol\alpha}_g^{(s)}-\widehat{\boldsymbol\alpha}_g^{(s-1)}\right\|^2}{\sum_{g=1}^G\left\|\widehat{\alpha}_g^{(s-1)}\right\|^2+0.0001}\leq\epsilon_{tl},
\end{equation*}
where $\epsilon_{tl}$ is a predetermined small value indicating the tolerance level.

\subsection{Algorithm for dropping a part of regressors in clustering}
\label{appendix:Cmat}

In the framework shown in Section \ref{sec:panelMIDAS}, the procedure concentrates on clustering weights of $Z_i$ and $\widetilde{X}_i$ at the same time. To cluster part of weights, a selection matrix $C_s$ is introduced\footnote{Although we do not provide a formal proof for this argument, the validity of this algorithm can be proved in a similar manner, following \cite{ma2016estimating}'s argument. To keep the paper concise, we do not present the detail in this paper.}.
The modified penalized objective function:
\begin{equation*}\label{objf2}
Q(\boldsymbol{\gamma})=\frac{1}{2}||\mathbf{y}-W\boldsymbol{\gamma}||_2^2+\sum_{1\leq i<j\leq n}\rho(C_s\boldsymbol{\gamma}_i-C_s\boldsymbol{\gamma}_j,\lambda_1),
\end{equation*}
where $C_s$ is a matrix of 1s and 0s that picks up the coefficient of interest. For example,  if one is interested in clustering  Fourier transformed weights only, the matrix $C_s$ is the same as $\mathbf{D}$ in (\ref{beta}).
The group-specified parameter is $\widetilde{\boldsymbol{\eta}}_{ij}=C_s\boldsymbol{\gamma}_i-C_s\boldsymbol{\gamma}_j$, and the constrained optimization problem is
\begin{equation*}
\begin{aligned}
Q_{\lambda_2}(\boldsymbol{\gamma},\boldsymbol{\eta},\boldsymbol{\xi})=&\frac{1}{2}||\mathbf{y}-W\boldsymbol{\gamma}||_2^2
+\sum_{i<j}\rho(\boldsymbol{\eta}_{ij},\lambda_1)\\
&+\frac{\lambda_2}{2}\sum_{i<j}
||C_s\boldsymbol{\gamma}_i-C_s\boldsymbol{\gamma}_j-\boldsymbol{\eta}_{ij}||_2^2
+\sum_{i<j}\boldsymbol{\xi}_{ij}'(C_s\boldsymbol{\gamma}_i-C_s\boldsymbol{\gamma}_j-\boldsymbol{\eta}_{ij}).
\end{aligned}
\end{equation*}
Equivalently,
\begin{equation*}\label{Obj1}
    Q_{\lambda_2}^{\boldsymbol{\gamma}}(\boldsymbol{\gamma},\boldsymbol{\eta},\boldsymbol{\xi})=\dfrac{1}{2}\|\mathbf{y}-W\boldsymbol{\gamma}\|_2^2+\dfrac{\lambda_2}{2}\|\widetilde{D}\boldsymbol{\gamma}-(\boldsymbol{\eta}+\boldsymbol{\xi}/\lambda_2)\|_2^2,
\end{equation*}
where $D_{ij}=(\boldsymbol{e}_i-\boldsymbol{e}_j)'\otimes I_{p}$ and $\widetilde{D}=(D_{12}'C_s',D_{13}'C_s',\cdots,D_{n-1,n}'C_s')'$.
The corresponding algorithm can be summarized as Algorithm \ref{algorithm5}.

\begin{algorithm}
	\caption{F-clust excluding some coefficients from clustering}
	\label{algorithm5}
	\SetAlgoLined
	%\begin{algorithmic}
		\textbf{Initialization:}
		
		$\boldsymbol{\xi}^0=\mathbf{0},~\boldsymbol{\gamma}^0=\left(W'W\right)^{-1}\left(W'\mathbf{y}\right),~\boldsymbol{\eta}^0=\arg\min_{\boldsymbol{\eta}}Q_{\lambda_2}(\boldsymbol{\gamma},\boldsymbol{\eta},\boldsymbol{\xi})$, where $\lambda_2$ and $\theta>1/\lambda_2$ are fixed.
		
		%\hspace{5mm}
		
		\For{$s=0,1,2,\cdots$}{
			
			$\boldsymbol{\gamma}^{s+1}=\left(W'W+\lambda_2\widetilde{D}'\widetilde{D}\right)^{-1}\left\{W'\mathbf{y}+\lambda_2\widetilde{D}'(\boldsymbol{\eta}^s+\boldsymbol{\xi}^s/\lambda_2)\right\}$.
			
			$\boldsymbol{\eta}^{s+1}=\arg\min_{\boldsymbol{\eta}}Q_{\lambda_2}(\boldsymbol{\gamma}^{s+1},\boldsymbol{\eta},\boldsymbol{\xi}^{s})$,
			
			$\boldsymbol{\xi}_{ij}^{s+1}=\boldsymbol{\xi}_{ij}^{s}+\lambda_2(\boldsymbol{\eta}_{ij}^{s+1}-C\boldsymbol{\gamma}_i^{s+1}+C\boldsymbol{\gamma}_j^{s+1})$, for all $1\leq i<j\leq n$.
			
			\If{the stopping criteria are true}{
				Break
			}}
		
		%\\\hrulefill
		%\State part 2 here
	%\end{algorithmic}
\end{algorithm}

\section{Proofs}\label{appen:proofs}
\subsection{Lemmas}\label{appen:lemmasection}
Assumptions on regressors (in our setting, $W$) made in  \cite{ma2016estimating} and related papers can be somewhat too strong for our panel setting. For example, (C3) in \cite{ma2016estimating} assumes that each column of $W$, taking only the rows that correspond to the $k$-th group, should be nonrandom, and the sum of squares of all its elements is assumed to be equal to the size of $k$-th group, i.e., $|\mathcal{G}_k|$.
This type of assumption could be realistic for data involved with an experimental design, but not suitable for panel data setting, where columns of $W$ generally consists of random variables. In this proof, we circumvent this issue by using the following lemmas.

\begin{lemma}\label{appendix:lem0}
	Suppose a random vector $\boldsymbol{\varepsilon}=(\varepsilon_{1,1},\varepsilon_{1,2},\ldots,\varepsilon_{n,T})'$ of length $nT$ as in (\ref{eq:panel2}) satisfies Assumption \ref{ass:subgauss}. 
	Let $A\in\mathbf{R}^{a\times nT}$ be a nonrandom matrix with a positive integer $a$. Let $\Sigma=A'A$.
	For any $\zeta>0$,
	$$P\left[\|A\boldsymbol{\varepsilon}\|_2^2>2\tilde{c}\{{\rm tr}(\Sigma)+2\sqrt{{\rm tr}(\Sigma^2)}\zeta+2\|\Sigma\|_2\zeta\}\right]\leq e^{-\zeta}.$$
\end{lemma}
\begin{proof}[Proof of Lemma \ref{appendix:lem0}]
		When $a=nT$, this lemma is a special case of Theorem 2.1 in \cite{hsu2012tail}. This can be easily seen by recognizing their $\mu$, $\sigma^2$, and $\alpha$ are $0$, $2\tilde{c}$, and $(\nu_{1,1},\nu_{1,2},\ldots,\nu_{n,T})'$, respectively.
	
	If $a<nT$, a similar argument can still be used.  Consider a singular value decomposition of $A=USV'$, where $U$ and $V$ are $a\times a$ and $nT\times nT$ orthogonal matrices, respectively. Let $\rho=(\rho_1,\ldots,\rho_a)'$ denote 
	the nonzero eigenvalues of $A'A$ and $AA'$. $S$ is an $a\times nT$ matrix, where its diagonal elements are equal to $\sqrt{\rho_i}$ for $i=1,\ldots,a$ and all other entries are zero.
	Let $z$ be a vector of $a$ independent standard Gaussian random variables.
	Since $U$ is orthogonal, $y=U'z$ is also an $a\times 1$ vector of $a$ independent standard Gaussian random variables. Let $y=(y_1,\ldots,y_a)'$. Applying Lemma 2.4 of \cite{hsu2012tail} on 
	$\|A'z\|^2=Z'AA'z=z'USV'VS'U'z=ySS'y'=\sum_{i=1}^a\rho_iy_i^2,$
	then
	\begin{equation}\label{lem:eq1}E\left\{\exp\left(\gamma\|A'z\|^2\right)\right\}\leq \exp\left(\|\rho\|_1\gamma+\frac{\|\rho\|_2^2\gamma^2}{1-2\|\rho\|_\infty\gamma}\right)\end{equation}
	for any $0\leq\gamma<1/(2\|\rho\|_\infty)$.
		For any $\lambda\in\mathbf{R}$ and $\delta\geq0$, using  similar arguments as in (2.3) and (2.4) of \cite{hsu2012tail}, Assumption \ref{ass:subgauss}, and (\ref{lem:eq1}),
	$$P(\|A\boldsymbol{\varepsilon}\|^2>\delta)\leq \exp\left(-\frac{\lambda^2\delta}{2}\right)\exp\left\{\|\rho\|_1{(\lambda^2\tilde{c})}+\frac{\|\rho\|_2^2(\lambda^2\tilde{c})^2}{1-2\|\rho\|_\infty(\lambda^2\tilde{c})}\right\}.$$
Let $\delta=2\tilde{c}(\|\rho\|_1+\tau$), $\lambda^2=\frac{1}{\tilde{c}}\frac{1}{2\|\rho\|_\infty}\left(1-\sqrt{\frac{\|\rho\|_2^2}{\|\rho\|_2^2+2\|\rho\|_\infty\tau}}\right)$, and $\tau=2\sqrt{\|\rho\|_2^2\zeta}+2\|\rho\|_\infty\zeta$. The desired proof is concluded by using similar arguments as \cite{hsu2012tail} and observing 
$\|\rho\|_1=\sum_{i=1}^a\rho_i={\rm tr}(\Sigma)$, $\|\rho\|_2^2=\sum_{i=1}^a\rho_i^2={\rm tr}(\Sigma^2)$, and $\|\rho\|_\infty=\max_i \rho_i=\|\Sigma\|_2$.

A similar proof works for $a>nT$. In this case, without loss of generality, the only nonzero element in $S$ are the first $nT$ diagonal elements of $S$. Let $\sqrt{\rho_i}$, $i=1,\ldots,nT$, be the nonzero diagonal elements of $S$. Then $||A'z||^2=\sum_{i=1}^{nT} \rho_iy_i^2$, where $y_i$ are independent standard Gaussian random variables. The rest of the proof is the same.
	\end{proof}
	
\begin{lemma}\label{appendix:lem1}
Suppose conditions of Lemma \ref{appendix:lem0} hold.
For any $nT\times np$ matrix  $W$ satisfying Assumption \ref{ass:lambda},
	$$\begin{aligned}
	&P\left[\|W'\boldsymbol{\varepsilon}\|_2^2>
	    2\tilde{c} (np+2\sqrt{np\zeta^*}+2\zeta^*)\|{W}'{W}\|_2 \Bigm\vert W\right]\leq e^{-\zeta^*}~~~{\rm and}\\
    &P\left[\|\Gamma'W'\boldsymbol{\varepsilon}\|_2^2>2\tilde{c}(Gp+2\sqrt{Gp\zeta}+2\zeta)\|\Gamma'W'W\Gamma\|_2\Bigm\vert W\right]\leq e^{-\zeta}
    \end{aligned}$$
    hold for any $\zeta^*>0$ and $\zeta>0$.
\end{lemma}
\begin{proof}[Proof of Lemma \ref{appendix:lem1}]
Fix a $nT\times nP$ matrix $W$ that satisfies Assumption \ref{ass:lambda}. Using Lemma \ref{appendix:lem0}, for any $\zeta^*>0$,
$$
\begin{aligned}
&P\left[\|W'\varepsilon\|_2^2>2\tilde{c}({\rm tr}(WW')+2\sqrt{{\rm tr}((WW')^2)\zeta^*}+2\|WW'\|_2\zeta^*)\bigm\vert W\right]\leq e^{-\zeta^*},~~~{\rm and}\\
&P\left[\|\Gamma'W'\varepsilon\|_2^2>2\tilde{c}({\rm tr}(\Gamma WW'\Gamma')+2\sqrt{{\rm tr}((\Gamma WW'\Gamma')^2)\zeta}+2\|\Gamma WW'\Gamma'\|_2\zeta)\bigm\vert W\right]\leq e^{-\zeta}.
\end{aligned}
$$
Since $\|WW'\|_2$ is the maximum eigenvalue of $WW'$, using the fact that $WW'$ is symmetric and positive definite with rank $np$, it can be easily seen that  $\lambda_{max}(WW')=\lambda_{max}(W'W)$, 
$$
    \|{W}{W}'\|_2=\|{W}'{W}\|_2=\|diag(W_1'W_1,\cdots,W_n'W_n)\|_2\leq\max_i\|W_i'W_i\|_2,~~{\rm and}
$$
$${\rm tr}({W}{W}')=tr({W}'{W})\leq{np}\|{W}'{W}\|_2,~~{\rm tr}(({W}{W}')^2)=tr(({W}'{W})^2)\leq{np}\|{W}'{W}\|^2_2.$$
Therefore
$$
    {\rm tr}({W}{W}')+2\sqrt{{\rm tr}[({W}{W}')^2]\zeta^*}+2\|{W}{W}'\|_2\zeta^* \leq (np+2\sqrt{np\zeta^*}+2\zeta^*)\|{W}'{W}\|_2.
$$
Similarly, $\|W\Gamma\Gamma'W'\|_2=\|\Gamma'W'W\Gamma\|_2$,
%	$${\rm tr}(\Sigma)={\rm tr}(\Gamma'W'W\Gamma)\geq (Gp)\lambda_{\min}(\Gamma'W'W\Gamma)\leq (Gp)(cg_{\min}T).$$
%	$${\rm tr}(\Sigma^2)={\rm tr}\{(\Gamma'W'W\Gamma)^2\}\geq (Gp)\{\lambda_{\min}(\Gamma'W'W\Gamma)\}^2\leq (Gp)(cg_{\min}T)^2.$$
\begin{equation*}\label{appendix:eq1-2}
{\rm tr}(W\Gamma\Gamma'W')={\rm tr}(\Gamma'W'W\Gamma)\leq Gp\lambda_{\max}(\Gamma'W'W\Gamma)=Gp\|\Gamma'W'W\Gamma\|_2,~~{\rm and}
\end{equation*}
\begin{equation*}\label{appendix:eq1-3}
{\rm tr}\{(W\Gamma\Gamma'W')^2\}={\rm tr}\{(\Gamma'W'W\Gamma)^2\}\leq Gp\{\lambda_{\max}(\Gamma'W'W\Gamma)\}^2=Gp\|\Gamma'W'W\Gamma\|_2^2.
\end{equation*}
Therefore for any $\zeta>0$,
\begin{equation*}\label{appendix:eq1-4}
{\rm tr}(\Gamma'W'W\Gamma)+2\sqrt{{\rm tr}\{(\Gamma'W'W\Gamma)^2\}}\sqrt{\zeta}+2\|\Gamma'W'W\Gamma\|_2\zeta
\leq(Gp+2\sqrt{Gp\zeta}+2\zeta)\|\Gamma'W'W\Gamma\|_2.
\end{equation*}
As a result, given any matrix $W$, the inequalities in the statement have been validated.
\end{proof}

\begin{lemma}\label{appendix:lem2}
Suppose Assumptions \ref{ass:lambda} and \ref{ass:subgauss} hold for $W$ and $\boldsymbol{\varepsilon}$. Define $$
\begin{aligned}
S_{\zeta}=&2\tilde{c}(Gp+2\sqrt{Gp\zeta}+2\zeta)g_{\max}m\tilde{M}\sqrt{GpT}B_{q,m},\\
S_{\zeta^*}=&2\tilde{c}(np+2\sqrt{np\zeta^*}+2\zeta^*)m\tilde{M}\sqrt{T}B_{q,m}\sqrt{p},
\end{aligned}$$
%where $B_{q,m}=\sqrt{q}+\sqrt{m}(L+1+2K)$, $\tilde{M}=\max(M_1,M_2,M_3,M_4)$. %Here, 
%$M_1,\ldots,M_4$ and $\tilde{c}$ are as given in 
%Assumptions \ref{ass:lambda} and \ref{ass:subgauss}.
%Recall that $p=q+L+1+2K$.
%$P\left[\|W'\boldsymbol{\varepsilon}\|_2^2>S_{\zeta^*}\right]$ and $P\left[\|\Gamma'W'\boldsymbol{\varepsilon}\|_2^2>S_{\zeta}\right]$ will be bounded by small values related to $\zeta^*$ and $\zeta^*$ in Lemma \ref{appendix:lem1}. Furthermore, let $M^*=m\sqrt{pT}B_{q,m}\sqrt{q+L+1+2K}$ where $B_{q,m}=(q^{1/2}+m^{1/2}(L+1+2K))$,  
%there exist $\iota>0$ and $\iota^*>0$ such that
%$$P\left[\|W'\boldsymbol{\varepsilon}\|_2^2>S_{\zeta^*}\right]\leq e^{-\iota^*}~~~{\rm and}~~~ P\left[\|\Gamma'W'\boldsymbol{\varepsilon}\|_2^2>S_{\zeta}\right]\leq e^{-\iota}.$$
%where $\iota=\min(\zeta,-\log(\epsilon))-\log(2)$ and $\iota^*=\min(\zeta^*,-\log(\epsilon))-\log(2)$ for any 
where $B_{q,m}=(q^{1/2}+m^{1/2}(L+1+2K))$, $p=q+L+1+2K$, $\tilde{M}=\max(M_1,M_2,M_3,M_4)$ and $\tilde{c}$ given in Assumption \ref{ass:lambda} and \ref{ass:subgauss},
%$P\left[\|W'\boldsymbol{\varepsilon}\|_2^2>S_{\zeta^*}\right]$ and $P\left[\|\Gamma'W'\boldsymbol{\varepsilon}\|_2^2>S_{\zeta}\right]$ will be bounded by small values related to $\zeta^*$ and $\zeta^*$ in Lemma \ref{appendix:lem1}. Furthermore, let $M^*=m\sqrt{pT}B_{q,m}\sqrt{q+L+1+2K}$ where $B_{q,m}=(q^{1/2}+m^{1/2}(L+1+2K))$,  
then $P\left[\|W'\boldsymbol{\varepsilon}\|_2^2>S_{\zeta^*}\right]\leq e^{-\iota^*}$ and $P\left[\|\Gamma'W'\boldsymbol{\varepsilon}\|_2^2>S_{\zeta}\right]\leq e^{-\iota}$ where $\iota=\min(\zeta,-\log(\epsilon))-\log(2)$ and $\iota^*=\min(\zeta^*,-\log(\epsilon))-\log(2)$ for any $\zeta$ and $\zeta^*$ in Lemma \ref{appendix:lem1}.
\end{lemma}

\begin{proof}[Proof of Lemma \ref{appendix:lem2}]
Using the law of iterated expectations, 
$$\begin{aligned}
E\left[P\left(\|W'\boldsymbol{\varepsilon}\|^2_2>S_{\zeta^*}\bigm\vert W\right)\right]=&P\left[\|W'\boldsymbol{\varepsilon}\|_2>S_{\zeta^*}\right]\\
=&E\left[I_{\{\|W'\boldsymbol{\varepsilon}\|^2_2>S_{\zeta^*}\}}\bigm\vert \|WW'\|_2\leq M^*\right] P(\|WW'\|_2\leq M^*)\\
&+E\left[I_{\{\|W'\boldsymbol{\varepsilon}\|^2_2>S_{\zeta^*}\}}\bigm\vert \|WW'\|_2> M^*\right]P(\|WW'\|_2> M^*)\\
=&P\left[\|W'\boldsymbol{\varepsilon}\|^2_2>S_{\zeta^*}\bigm\vert \|WW'\|_2\leq M^*\right] P(\|WW'\|_2\leq M^*)\\
&+P\left[\|W'\boldsymbol{\varepsilon}\|^2_2>S_{\zeta^*}\bigm\vert \|WW'\|_2> M\right]P(\|WW'\|_2> M^*).\\
\end{aligned}$$
Since $\|{M}\|_\infty\leq m$ and $\|{M}'\|_\infty\leq L+1+2K$ as all elements of ${M}$ in (\ref{M}) smaller than 1 in magnitude, 
\begin{equation*}
\begin{aligned}
&\left\|\sum_{i\in\mathcal{G}_g}Z_i'Z_i\right\|_\infty
=\sum_{i\in\mathcal{G}_g}\left\|Z_i'Z_i\right\|_\infty
\leq M_1|\mathcal{G}_g|\sqrt{qT},\\
&\left\|\sum_{i\in\mathcal{G}_g}Z_i'\tilde{X}_i\right\|_\infty
%=\left\|\sum_{i\in\mathcal{G}_g}Z_i'X_i{M}'\right\|_\infty
\leq\sum_{i\in\mathcal{G}_g}\left\|Z_i'X_i\right\|_\infty\|{M}'\|_\infty
\leq M_3|\mathcal{G}_g|\sqrt{mT}(L+1+2K),\\
&\left\|\sum_{i\in\mathcal{G}_g}\tilde{X}_i'Z_i\right\|_\infty
\leq\|{M}\|_\infty\sum_{i\in\mathcal{G}_g}\left\|Z_i'X_i\right\|_\infty
\leq M_4|\mathcal{G}_g|m\sqrt{qT},~~~{\rm and}\\
&\left\|\sum_{i\in\mathcal{G}_g}\tilde{X}_i'\tilde{X}_i\right\|_\infty
\leq\|{M}\|_\infty\sum_{i\in\mathcal{G}_g}\left\|X_i'X_i\right\|_\infty\|{M}'\|_\infty
\leq M_2|\mathcal{G}_g|m\sqrt{mT}(L+1+2K)
\end{aligned}
\end{equation*}
hold with probability at least $1-\epsilon$ 
for any $\epsilon>0$ defined in Assumption \ref{ass:lambda}.
%With probability at least $1-\epsilon$, 
%$$\begin{aligned}\left\|\sum_{i\in\mathcal{G}_g}Z_i'Z_i\right\|_\infty\leq \tilde{M}|\mathcal{G}_g|\sqrt{qT},~~\left\|\sum_{i\in\mathcal{G}_g}Z_i'\tilde{X}_i\right\|_\infty\leq \tilde{M}|\mathcal{G}_g|\sqrt{mT}(L+1+2K),\\ \left\|\sum_{i\in\mathcal{G}_g}\tilde{X}_i'Z_i\right\|_\infty\leq \tilde{M}|\mathcal{G}_g|m\sqrt{qT},~~\left\|\sum_{i\in\mathcal{G}_g}\tilde{X}_i'\tilde{X}_i\right\|_\infty\leq \tilde{M}|\mathcal{G}_g|m\sqrt{mT}(L+1+2K).\end{aligned}$$
Therefore, with probability at most $1-\epsilon$,
$$\begin{aligned}
    \|{W}{W}'\|_2=\|{W}'{W}\|_2&=\|diag(W_1'W_1,\cdots,W_n'W_n)\|_2\leq\sup_i\|W_i'W_i\|_2\\
    &\leq \sqrt{p}\sup_i\|W_i'W_i\|_\infty=\sqrt{p}\sup_i\left\|\begin{aligned}
        {Z}_i'{Z}_i & & {Z}_i'\tilde{{X}}_i\\
        \tilde{{X}}_i'{Z}_i & & \tilde{{X}}_i'\tilde{{X}}_i
    \end{aligned}\right\|_\infty\\
        &\leq \tilde{M}m\sqrt{T}B_{q,m}\sqrt{p}.
\end{aligned}$$
Since $tr({W}{W}')=tr({W}'{W})\leq{np}\|{W}'{W}\|_2$ and $tr(({W}{W}')^2)=tr(({W}'{W})^2)\leq{np}\|{W}'{W}\|^2_2,$
$$\begin{aligned}
    &tr({W}{W}')+2\sqrt{tr[({W}{W}')^2]\zeta^*}+2\|{W}{W}'\|_2\zeta^*
    \leq (np+2\sqrt{np\zeta^*}+2\zeta^*)\|WW'\|_2.
%    \leq& {(np+2\sqrt{np\zeta^*}+2\zeta^*)}mT^{1/2}p^{1/2}\tilde{B}_{q,m},
\end{aligned}$$
%where $\tilde{B}_{q,m}=B_{q,m}\sqrt{q+L+1+2K}$. 
Since $||WW'||_2$ is bounded in probability, for any $\epsilon>0$, there exists some $M^*=\tilde{M}m\sqrt{T}B_{q,m}\sqrt{p}$ such that $P[\|WW'\|_2>M^*]\leq\epsilon$. Therefore
$$\begin{aligned}
P\left[\|W'\boldsymbol{\varepsilon}\|^2_2>S_{\zeta^*}\bigm\vert W,\|WW'\|_2\leq M^*\right]\leq e^{-\zeta^*},&~~1-\epsilon<P(\|WW'\|_2\leq M^*)\leq1,\\
P\left[\|W'\boldsymbol{\varepsilon}\|^2_2>S_{\zeta^*}\bigm\vert W,\|WW'\|_2> M^*\right]\leq 1&,~~P(\|WW'\|_2> M^*)\leq\epsilon,
\end{aligned}$$
and $P\left[\|W'\boldsymbol{\varepsilon}\|^2_2>S_{\zeta^*}\right]\leq e^{-\zeta^*}+\epsilon$ where $S_{\zeta^*}=2\tilde{c}(np+2\sqrt{np\zeta^*}+2\zeta^*)M^*$.

Without loss of generality, let $\tilde{\zeta}^*=\min\{\zeta^*,-\log(\epsilon)\}$ for a large constant $\zeta^*>1$ and for small positive constant $\epsilon\leq1$, then 
$e^{-\zeta^*}+\epsilon=e^{-\zeta^*}+e^{\log(\epsilon)}=e^{-\tilde{\zeta}^*}(1+e^{-|\zeta^*+\log(\epsilon)|})\leq 2e^{-\tilde{\zeta}^*}=e^{\log(2)-\tilde{\zeta}^*}$. 
Take $\iota^*=\tilde{\zeta}^*-\log(2)$, then $P\left[\|W'\boldsymbol{\varepsilon}\|_2^2>S_{\zeta^*}\right]\leq e^{-\iota^*}$. For large enough $\tilde{\zeta}^*$, $\log(2)$ is negligible. 
Similarly, $S_{\zeta}$ in $P\left[\|\Gamma'W'\boldsymbol{\varepsilon}\|_2^2>S_{\zeta}\right]\leq e^{-\iota}$ can be found as the following.

A straightforward calculation derives that  $$\Gamma'W'W\Gamma={\rm diag}\left(\sum_{i\in\mathcal{G}_1} W_i'W_i,\ldots, \sum_{i\in\mathcal{G}_G} W_i'W_i\right).	$$
	It follows that, with probability $1-\epsilon$,
	$$\begin{aligned}
\|\Gamma'W'W\Gamma\|_\infty
&=\max_{1\leq g\leq G}\left\|\sum_{i\in\mathcal{G}_g}W_i'W_i\right\|_\infty 
\leq \max_{1\leq g\leq G}\sum_{i\in\mathcal{G}_g}\left\|W_i'W_i\right\|_\infty
\leq g_{max}\sup_{1\leq i\leq n}\|W_i'W_i\|_\infty\\
&\leq g_{\max}m\tilde{M}\sqrt{T}B_{q,m},
	\end{aligned}$$
and therefore,
\begin{equation*}\label{appendix:eq1-1}
\|\Gamma'W'W\Gamma\|_2\leq \sqrt{Gp} \|\Gamma'W'W\Gamma\|_\infty\leq  g_{\max}m\tilde{M}\sqrt{GpT}B_{q,m}.\end{equation*}

For any $\epsilon>0$, there exists some $M=g_{\max}m\tilde{M}\sqrt{T}B_{q,m}$, such that $P[\|W\Gamma\Gamma'W'\|_2>M]\leq\epsilon$, then
$$\begin{aligned}
P\left[\|\Gamma'W'\boldsymbol{\varepsilon}\|^2_2>S_{\zeta}\bigm\vert W,\|W\Gamma\Gamma'W'\|^2_2\leq M\right]\leq e^{-\zeta},&~~1-\epsilon<P(\|W\Gamma\Gamma'W'\|_2\leq M)\leq1,\\
P\left[\|W'\boldsymbol{\varepsilon}\|_2>S_{\zeta}\bigm\vert W,\|W\Gamma\Gamma'W'\|_2> M\right]\leq 1&,~~P(\|W\Gamma\Gamma'W'\|_2> M)\leq\epsilon.
\end{aligned}$$

Therefore, $P\left[\|\Gamma'W'\boldsymbol{\varepsilon}\|_2^2>S_{\zeta}\right]\leq e^{-\zeta}+\epsilon$ where $S_{\zeta}=2\tilde{c}(Gp+2\sqrt{Gp\zeta}+2\zeta)M$.
Similarly, take $\iota=\min\{\zeta,-\log(\epsilon)\}-\log(2)$, then $P\left[\|\Gamma'W'\boldsymbol{\varepsilon}\|_2^2>S_{\iota}\right]\leq e^{-\iota}$.
%Without loss of generality, for large $\zeta>1$ as well as $\epsilon\leq1$, Let $\tilde{\zeta}=\min\{\zeta,-\log(\epsilon)\}$. Then 
%$$e^{-\zeta}+\epsilon=e^{-\zeta}+e^{\log(\epsilon)}=e^{-\tilde{\zeta}}(1+e^{-|\zeta+\log(\epsilon)|})\leq 2e^{-\tilde{\zeta}}=e^{\log(2)-\tilde{\zeta}}.$$ Let $\iota=\tilde{\zeta}-\log(2)$, then 
%$$\begin{aligned}
%P\left[\|W'\boldsymbol{\varepsilon}\|_2^2>S_{\iota^*}\right]\leq e^{-\iota^*},~~P\left[\|\Gamma'W'\boldsymbol{\varepsilon}\|_2^2>S_{\iota}\right]\leq e^{-\iota}
%\end{aligned}$$
\end{proof}

%%%%%%%%%%%%%%%%     Theorem 2    %%%%%%%%%%%%%%%%%%%%%%%%%%

\subsection{Convergence of the Oracle Estimator}\label{pf:oracle}
Theorem \ref{thm:oracle} and Corollary \ref{cor1} are proved in this section.
\begin{proof}[Proof of Theorem \ref{thm:oracle}]
The definition of $\Gamma$ and  $\mathbf{y}=W\boldsymbol{\gamma}^{or}+\boldsymbol{\varepsilon}$ lead to
$$ \begin{array}{lll}
\hat{\boldsymbol{\gamma}}^{or}- {\boldsymbol{\gamma}}^{0}&=&\Gamma(\Gamma'W'W\Gamma)^{-1}\Gamma'W'\boldsymbol{\varepsilon}
\\&=&\Gamma\left\{ {\rm diag}\left(\sum_{i\in\mathcal{G}_1} W_i'W_i,\ldots, \sum_{i\in\mathcal{G}_G} W_i'W_i\right)\right\}^{-1}\left(\begin{matrix}\sum_{i\in\mathcal{G}_1} W_i'\boldsymbol{\varepsilon}_i
\\ \vdots
\\ \sum_{i\in\mathcal{G}_G} W_i'\boldsymbol{\varepsilon}_i
\end{matrix}\right),
\end{array}$$
where  for any $g\in\{1,\ldots,G\}$,
$$\sum_{i\in \mathcal{G}_g}W_i'W_i=\left(\begin{matrix}\sum_{i\in\mathcal{G}_g}Z_i'Z_i & (\sum_{i\in\mathcal{G}_g}Z_i'X_i)\mathbf{M}' \\\mathbf{M}(\sum_{i\in\mathcal{G}_g}X_i'Z_i)&\mathbf{M}(\sum_{i\in\mathcal{G}_g}X_i'X_i)\mathbf{M}'\end{matrix}\right)~~{\rm and}~~\sum_{i\in \mathcal{G}_g}W_i'\boldsymbol{\varepsilon}_i=\left(\begin{matrix}\sum_{i\in \mathcal{G}_g}Z_i'\boldsymbol{\varepsilon}_i\\ \mathbf{M}(\sum_{i\in \mathcal{G}_g}X_i'\boldsymbol{\varepsilon}_i)\end{matrix}\right).$$
Assumption \ref{ass:lambda} implies that 
$$\lambda_{\min}(\Gamma'W'W\Gamma)\geq cg_{\min}T,$$ so that 
\begin{equation}\label{appendix:eq1}
\|(\Gamma'W'W\Gamma)^{-1}\|_\infty\leq \sqrt{Gp}\|(\Gamma'W'W\Gamma)^{-1}\|_2\leq \sqrt{Gp}(cg_{\min}T)^{-1}.
\end{equation}
For all $p$-norms, $\|A\otimes B\|=\|A\|\|B\|$ holds (for example, see p. 433 of \citet{Langville:Stewart:2004}),
\begin{equation}\label{appendix:eq2}
\|\Gamma\|_\infty\leq\|\Pi\|_\infty\|I_p\|_\infty=1.
\end{equation}
Lemma \ref{appendix:lem2}, equations (\ref{appendix:eq1}) and (\ref{appendix:eq2}), and the triangle inequality imply that for any $\iota>0$, 
$$ \begin{aligned}
\|\widehat{\boldsymbol{\gamma}}^{or}- {\boldsymbol{\gamma}}^{0}\|_\infty
&\leq \|\Gamma\|_\infty\|(\Gamma'W'W\Gamma)^{-1}\|_\infty\|\Gamma'W'\boldsymbol{\varepsilon}\|_\infty\\
&\leq(Gp)^{1/2}(cg_{\min}T)^{-1}\|\Gamma'W'\boldsymbol{\varepsilon}\|_2
\leq (Gp)^{1/2}(cg_{\min}T)^{-1}S_{\zeta}^{1/2},
\end{aligned}$$
with probability at least $1-e^{\iota}$.
%By Theorem 2.1 of \cite{hsu2012tail} and Assumption \ref{ass:subgauss}, 
%Therefore, 
%$$P\left[\|\Gamma'W'\boldsymbol{\varepsilon}\|_2^2>\tilde{c}^2Cg_{\max}m\sqrt{GpT}\{\sqrt{q}+\sqrt{m}(L+1+2K)\}(Gp+2\sqrt{Gp\zeta}+2\zeta)\Bigm\vert W \right]\leq e^{-\zeta}.$$
%Further, we can get that
%$$P\left[\|\Gamma'W'\boldsymbol{\varepsilon}\|_2^2>\tilde{c}^2Cg_{\max}m\sqrt{GpT}\{\sqrt{q}+\sqrt{m}(L+1+2K)\}(Gp+2\sqrt{Gp\zeta}+2\zeta)\Bigm\vert W \right]\leq e^{-\zeta},$$
%where $C=O_p(1)$.
%Furthermore, the result follows Lemma \ref{appendix:lem2} %by plugging in $\zeta=\zeta_{n,T}$ and observing that
Therefore,
\begin{equation}\label{phi_n1}
\phi_{n,T,G,\zeta}:=\frac{\sqrt{2\tilde{c}}}{c}\frac{(m\tilde{M}g_{\max})^{1/2}(Gp)^{3/4}}{g_{\min}T^{3/4}}B_{q,m}^{1/2}(Gp+2\sqrt{Gp}\sqrt{\zeta}+2\zeta)^{1/2},
\end{equation}
where $B_{q,m}$ is defined in Lemma \ref{appendix:lem2}.
Therefore, with probability at least $1-e^{-\iota}$,
$$\|\widehat{\boldsymbol{\gamma}}^{or}-\boldsymbol{\gamma}^0\|_\infty\leq\phi_{n,T,G,\zeta}.$$

This proves the first part of Theorem \ref{thm:oracle}. The remaining proof is for the asymptotic normality of $\widehat{\boldsymbol{\gamma}}^{or}$.
Let $V_i=W_i(\Pi_{i\cdot}\otimes I_p)$ be a $T\times Gp$ matrix, where $\Pi_{i\cdot}$ is the $i$-th row of the matrix $\Pi$, $V=W\Gamma=(V_1',\cdots,V_n')'$. Then, for any $c_n\in\mathbb{R}^{Gp}$ with $\|c_n\|_2=1$,
\begin{equation*}
\begin{aligned}
    c_n'(\hat{\boldsymbol{\gamma}}^{or}-\boldsymbol{\gamma}^0)=\sum_{i=1}^n c_n'(V'V)^{-1}V_i'\boldsymbol{\varepsilon}_i=\sum_{i=1}^n c_n'(V'V)^{-1}\sum_{t=1}^T\mathbf{v}_{it}'{\varepsilon}_{it}.
\end{aligned}
\end{equation*}
Since $\{\boldsymbol{\varepsilon}_{i}\}$ is assumed to be an i.i.d. subgaussian distributed sequence with mean 0 and variance proxy $2\tilde{c}$, then $E(\boldsymbol{\varepsilon}_{i})=\mathbf{0}$. Hence,
$$E\left[c_n'(\hat{\boldsymbol{\gamma}}^{or}-\boldsymbol{\gamma}^0)\right]=0.$$
Suppose that Assumption \ref{ass:lambda} and \ref{ass:subgauss} hold where $\lambda_{\max}(V'V)=\lambda_{\max}(\Gamma'W'W\Gamma)\leq c^*|\mathcal{G}_g|T\leq c^*g_{\max}T$ and $Var(\varepsilon_{it})=O(2\tilde{c})$, then 
\begin{equation*}\label{var}
\begin{aligned}
    \sigma_\gamma^2:=Var[c_n'(\hat{\boldsymbol{\gamma}}^{or}-\boldsymbol{\gamma}^0)]&\geq\dfrac{Var(\varepsilon_{it})}{{c^*}g_{\max}T}.
\end{aligned}
\end{equation*}
Moreover, for any $\epsilon>0$, applying Cauchy-Schwarz inequality, 
\begin{equation}\label{main}
\begin{aligned}
    &\sum_{i=1}^n E\left((c_n'(V'V)^{-1}{V}_{i}'\boldsymbol\varepsilon_{i})^2\mathbbm{1}\{|c_n'(V'V)^{-1}{V}_{i}\boldsymbol\varepsilon_{i}|>\epsilon \sigma_\gamma\}\right)\\
    \leq& \sum_{i=1}^n\left\{E(c_n'(V'V)^{-1}{V}_{i}'\boldsymbol\varepsilon_{i})^4\right\}^{1/2} \left\{E\left(\mathbbm{1}\{|c_n'(V'V)^{-1}{V}_{i}'\boldsymbol\varepsilon_{i}|>\epsilon \sigma_\gamma\}^2\right)\right\}^{1/2}\\
    =& \sum_{i=1}^n \left\{E(c_n'(V'V)^{-1}{V}_{i}'\boldsymbol\varepsilon_{i})^4\right\}^{1/2} \left\{E\left(\mathbbm{1}\{|c_n'(V'V)^{-1}{V}_{i}'\boldsymbol\varepsilon_{i}|>\epsilon \sigma_\gamma\}\right)\right\}^{1/2}\\
    =& \sum_{i=1}^n \left\{E(c_n'(V'V)^{-1}{V}_{i}'\boldsymbol\varepsilon_{i})^4\right\}^{1/2} \left\{P(|c_n'(V'V)^{-1}{V}_{i}'\boldsymbol\varepsilon_{i}|>\epsilon \sigma_\gamma)\right\}^{1/2}.
\end{aligned}
\end{equation}
The first term can be derived as
\begin{equation*}
\begin{aligned}
\left[E(c_n'(V'V)^{-1}{V}_{i}'\boldsymbol\varepsilon_{i})^4\right]^{1/2}
&=\left[E(c_n'(V'V)^{-1}{V}_{i}'\boldsymbol\varepsilon_{i}\boldsymbol\varepsilon_{i}'
{V}_{i}'(V'V)^{-1}c_n)^2\right]^{1/2}\\
&=\left[\{c_n'(V'V)^{-1}{V}_{i}\}^2E(\boldsymbol\varepsilon_{i}\boldsymbol\varepsilon_{i}')^2\{{V}_{i}'(V'V)^{-1}c_n\right\}^2]^{1/2}\\
&=c_n'(V'V)^{-1}{V}_{i}[E(\boldsymbol\varepsilon_{i}\boldsymbol\varepsilon_{i}')^2]^{1/2}{V}_{i}'(V'V)^{-1}c_n\\
&\leq\|c_n'(V'V)^{-1}{V}_{i}\|_2^2\left\|E(\boldsymbol\varepsilon_{i}\boldsymbol\varepsilon_{i}')^2\right\|_2^{1/2}.
\end{aligned}
\end{equation*}
For any $n\times n$ matrix $A$, $\|A\|_2\leq\sqrt{n}\|A\|_{\infty}$. Since $E(\varepsilon_{it}^k)\leq (2\sigma^2)^{k/2}k\Gamma(k/2)$ for $k\geq1$, then
\begin{equation*}
\begin{aligned}
\left\|E(\boldsymbol\varepsilon_{i}\boldsymbol\varepsilon_{i}')^2\right\|_2
&\leq\sqrt{T}\left\|E(\boldsymbol\varepsilon_{i}\boldsymbol\varepsilon_{i}')^2\right\|_\infty
=\sqrt{T}\max_{\tau = 1,\cdots,T}E\left(\varepsilon_{i\tau}\sum_{t=1}^T\varepsilon_{it}\sum_{t=1}^T\varepsilon_{it}^2\right)  %% explicit the power
\leq\sqrt{T}(16+T)4\tilde{c}^2.
\end{aligned}
\end{equation*}
According to Assumption \ref{ass:lambda}, $\|{V}_{i}\|_\infty$ is bounded and let the upper bound be some constant $c_2$, then $\|{V}_{i}\|_2\leq\sqrt{Gp}c_2$.
Following \ref{ass:lambda}, $\|(V'V)^{-1}\|_2\geq (cg_{min}T)^{-1}$, 
\begin{equation*}
\begin{aligned}
\left\{E(c_n'(V'V)^{-1}V_{i}\varepsilon_{it})^4\right\}^{1/2}&\leq\|c_n'\|_2^2\|(V'V)^{-1}\|_2^2\|V_{i}\|_2^2T^{1/4}(16+T)^{1/2}2\tilde{c}^2\\
&\leq\dfrac{c_2^2Gp(16+T)^{1/2}2\tilde{c}}{c^2g_{min}^2T^{3/4}}.
\end{aligned}
\end{equation*}
Then, by Chebyshev's inequality, the second term of (\ref{main}) can be derived as
\begin{equation}\label{ChebIneq}
\begin{aligned}
P(|c_n'(V'V)^{-1}V_i\boldsymbol\varepsilon_{i}|>\epsilon \sigma_\gamma)&\leq \dfrac{E[c_n'(V'V)^{-1}V_i\boldsymbol\varepsilon_{i}]^2}{\epsilon^2\sigma_\gamma^2},
\end{aligned}
\end{equation}
where 
\begin{equation*}
\begin{aligned}
E(c_n'(V'V)^{-1}V_i\boldsymbol\varepsilon_{i})^2&=E(c_n'(V'V)^{-1}V_i\boldsymbol\varepsilon_{i}\boldsymbol\varepsilon_{i}'V_i'(V'V)^{-1}c_n)\\
&\leq \|c_n\|_2^2\|(V'V)^{-1}\|_2^2\|V_{i}\|_2^2\|E(\boldsymbol\varepsilon_{i}\boldsymbol\varepsilon_{i}')\|_2\leq \dfrac{c_2^2Gp2\tilde{c}}{c^2g_{min}^2T^2},
\end{aligned}
\end{equation*}
then, (\ref{ChebIneq}) becomes
\begin{equation*}
\begin{aligned}
P(|c_n'(V'V)^{-1}V_i\boldsymbol\varepsilon_{i}|>\epsilon \sigma_\gamma)&
\leq\dfrac{c_2^2Gp2\tilde{c}}{c^2g_{min}^2T^2\epsilon^2\sigma_\gamma^2}.
\end{aligned}
\end{equation*}
Therefore, the following inequality can be derived.
\begin{equation}\label{Lindeberg}
\begin{aligned}
&\sigma_\gamma^{-2}\sum_{i=1}^n E\left((c_n'(V'V)^{-1}V_i\boldsymbol\varepsilon_{i})^2\mathbbm{1}\{|c_n'(V'V)^{-1}V_i\boldsymbol\varepsilon_{i}|>\epsilon \sigma_\gamma\}\right)\\
\leq&\sigma_\gamma^{-2}\sum_{i=1}^n\dfrac{c_2^2Gp(16+T)^{1/2}2\tilde{c}}{c^2g_{min}^2T^{3/4}}\dfrac{c_2(Gp)^{1/2}\sqrt{2\tilde{c}}}{cg_{min}T\epsilon\sigma_\varphi}
=\dfrac{c_2^3p^{3/2}(2\tilde{c})^{3/2}G^{3/2}(16+T)^{1/2}n}{c^3\epsilon g_{min}^3T^{7/4}\sigma_\gamma^3}\\
\leq&C\dfrac{(2\tilde{c})^{3/2}(n/g_{min})^{3/2}n(16+T)^{1/2}}{\sigma_\gamma^3g_{min}^3T^{7/4}}
=C\dfrac{\tilde{c}^3n^{5/2}(16+T)^{1/2}}{\sigma_\varphi^3g_{min}^{9/2}T^{7/4}}\\
=&C\dfrac{n^{5/2}(16+T)^{1/2}c^{*^{3/2}}g_{\max}^{3/2}T^{3/2}}{g_{min}^{9/2}T^{7/4}}
=O\left(\dfrac{g_{\max}^{3/2}n^{5/2}T^{1/4}}{g_{min}^{9/2}}\right)
.
\end{aligned}
\end{equation}
Suppose that $\dfrac{g_{\min}^3}{g_{\max}}\gg n^{5/3}T^{1/6}$, then (\ref{Lindeberg}) further implies that
$$\sigma_\gamma^{-2}\sum_{i=1}^n E\left((c_n'(V'V)^{-1}V_i\boldsymbol\varepsilon_{i})^2\mathbbm{1}\{|c_n'(V'V)^{-1}V_i\boldsymbol\varepsilon_{i}|>\epsilon \sigma_\gamma\}\right)=O(1).$$
By the Lindeberg-Feller Central Limit Theorem,
$c_n'(\hat{\boldsymbol{\gamma}}^{or}-\boldsymbol{\gamma}^0)\to N(0,\sigma_{\gamma}^2).$
\end{proof}

\begin{proof}[Proof of Corollary \ref{cor1}]
In the following proof, let $m$ and $q$ be fixed for simplification. It further indicates that $p$ is fixed. Let $C_{q,m}=\frac{\sqrt{2\tilde{c}}}{c}m^{1/2}p^{3/4}B_{q,m}^{1/2}$, (\ref{phi_n1}) can be simplified as
\begin{equation*}\label{phi_n2}
\phi_{n,T,G}=C_{q,m}\dfrac{g_{\max}^{1/2}G^{3/4}}{g_{\min}T^{3/4}}(Gp+2\sqrt{Gp}\sqrt{\zeta}+2\zeta)^{1/2}.
\end{equation*}

The rest of the  proof suggests a large enough $\zeta$ for each situation that allows $\phi_{n,T,G,\zeta}$ and $\iota$ to approach infinity.
 We often use these somewhat trivial inequalities $g_{max}\leq n$ and $G\leq n/g_{min}$ in the following proofs, particularly when $n\to\infty$.
 
%%%%%%%%%%%%%%%%%%%%%%%%%
\begin{enumerate}%[label=(\roman*)]
    \item Consider $T\rightarrow\infty$ with $n$ fixed. Let $\zeta\to\infty$ and $\zeta=o(T^{3/2})$.
      Since $G\leq n\ll\zeta$, then  $(Gp+2\sqrt{Gp}\sqrt{\zeta}+2\zeta)^{1/2}=O(2\zeta^{1/2})$. Therefore,
$$\phi_{n,T,G}=C_1T^{-3/4}O(\zeta^{1/2})\stackrel{{T\to\infty}}{\longrightarrow}0,$$ 
where $C_1=2C_{q,m}\frac{g_{\max}^{1/2}G^{3/4}}{g_{\min}}$, which is free of $T$.% Let   $\zeta=o(T^{3/2})$ as $T\rightarrow\infty$. %Thus, $\zeta=o(T^{3/2})$.
   \item Consider $n\rightarrow\infty$ with $T$ fixed.
    \begin{enumerate}
        \item Consider $G\ll\zeta\to\infty$.
        \begin{enumerate}
            \item When $G$ is fixed, then $(Gp+2\sqrt{Gp}\sqrt{\zeta}+2\zeta)^{1/2}=O(2\zeta^{1/2})$. For some constant $\tilde{\alpha}_0<1/2$, let $g_{min}=O(n^{1/2+\tilde{\alpha}_0})$, $\zeta=o(n^{2\tilde{\alpha}_0})$ and $\zeta\to\infty$, %such as $\zeta=o(\ln{n})$, 
            then
            $$\phi_{n,T,G}\leq{C}_3\dfrac{n^{1/2}}{g_{min}}O(\zeta^{1/2})\stackrel{{n\to\infty}}{\longrightarrow}0,$$
            where $C_3=2C_{q,m}\frac{G^{3/4}}{T^{3/4}}$, which is free of $n$.

            \item When $G\rightarrow\infty$, for some constant $\tilde{\alpha}_2<2/7$, let $g_{min}=O(n^{5/7+\tilde{\alpha}_2})$, $\zeta=o(n^{7\tilde{\alpha}_2/2})$ and $\zeta\to\infty$, %such as $\zeta=o(\ln{n})$, 
            then $(Gp+2\sqrt{Gp\zeta}+2\zeta)^{1/2}=O((p+2\sqrt{p}+2)^{1/2}\zeta^{1/2})$. Since $G\leq n/g_{min}$, then
            $$\phi_{n,T,G}\leq{C}_4\dfrac{n^{1/2}G^{3/4}}{g_{min}}O(\zeta^{1/2})\leq{C}_4\dfrac{n^{5/4}}{g_{min}^{7/4}}O(\zeta^{1/2})\stackrel{{n,G\to\infty}}{\longrightarrow}0,$$
            where $C_4=C_{q,m}\frac{1}{T^{3/4}}(p+2\sqrt{p}+2)^{1/2}$, which is free of $n$ and $G$.
        \end{enumerate}
        \item Consider $G \rightarrow\infty$. Let $g_{min}=O(n^{7/9+\tilde{\alpha}_1})$ for some $\tilde{\alpha}_1<2/9$, $\zeta=O(G)$ and $\zeta\rightarrow\infty$, then $Gp+2\sqrt{Gp}\sqrt{\zeta}+2\zeta=O((p+2\sqrt{p}+2)G)=O(G)$. Therefore,
        $$\phi_{n,T,G}\leq C_2\dfrac{n^{1/2}G^{3/4}}{g_{min}}O(G^{1/2})\stackrel{{n\to\infty}}{\longrightarrow}0,$$ 
        where $C_2=C_{q,m}\frac{1}{T^{3/4}}(p+2\sqrt{p}+2)^{1/2}$, which is free of $n$.
    \end{enumerate}
    
%%%%%%%%%%%%%%%%%%%%%%%%%%%%%%%%%%%%%%%%
    \item Consider $T,n\to\infty$.
    \begin{enumerate}
        \item Consider $G\ll\zeta\rightarrow\infty$,
        \begin{enumerate}
            \item When $G$ is fixed, then $(Gp+2\sqrt{Gp\zeta}+2\zeta)^{1/2}= O(2\zeta^{1/2})$. Let $g_{min}=O(n^{1/2+\tilde{\alpha_0}})$ for some positive constant $\tilde{\alpha}_0<1/2$ and $\zeta=o(n^{2\tilde{\alpha}_0}T^{3/2})$, $\zeta\to\infty$, then 
            $$\phi_{n,T,G}\leq{C}_6\dfrac{n^{1/2}}{g_{min}T^{3/4}}O(\zeta^{1/2})\stackrel{{n,T\to\infty}}{\longrightarrow}0,$$
            where $C_6=2C_{q,m}G^{3/4}$.
            
            \item When $G\rightarrow\infty$, for some positive constant $\tilde{\alpha}_2<2/7$,  let $g_{min}=O(n^{5/7+\tilde{\alpha}_2})$ and $G\leq n/g_{min}$, $\zeta=o(n^{7\tilde{\alpha}_2/2}T^{3/2})$ and $\zeta\to\infty$, %such as $\zeta=o(\ln{n})$, 
            then $(Gp+2\sqrt{Gp\zeta}+2\zeta)^{1/2}=O((p+2\sqrt{p}+2)^{1/2}\zeta^{1/2})$. Since $G\leq n/g_{min}$, then
            $$\phi_{n,T,G}\leq{C}_7\dfrac{n^{1/2}G^{3/4}}{g_{min}T^{3/4}}O(\zeta^{1/2})\leq{C}_7\dfrac{n^{5/4}}{g_{min}^{7/4}T^{3/4}}O(\zeta^{1/2})\stackrel{{n,T,G\to\infty}}{\longrightarrow}0,$$
            where $C_7=C_{q,m}(p+2\sqrt{p}+2)^{1/2}$, which is freen of $n,T$ and $G$.
        \end{enumerate}
        \item Consider $G\to\infty$. Let $g_{min}=O(n^{7/9+\tilde{\alpha}_1})$ for some constant $\tilde{\alpha}_1<2/9$, $\zeta=O(G)$ and $\zeta\rightarrow\infty$, then  $Gp+2\sqrt{Gp\zeta}+2\zeta=O((p+2\sqrt{p}+2)G)=O(G)$. Since $G\leq n/g_{min}$, 
$$\phi_{n,T,G}\leq{C}_5\dfrac{n^{1/2}G^{3/4}}{g_{min}T^{3/4}}O(G^{1/2})\leq{C}_5\dfrac{n^{7/4}}{g_{min}^{9/4}T^{3/4}}O(1)\stackrel{{n,T,G\to\infty}}{\longrightarrow}0,$$
    where $C_5=C_{q,m}(p+2\sqrt{p}+2p)^{1/2}$, which is free from $n,T$ and $G$.
    \end{enumerate}
\end{enumerate}

Combining items 2 and 3 above, we can summarize the choice of $\zeta$ as follows:
\begin{enumerate}
    \item[Case 1.]\label{case1} The number $n$ is fixed. Let  $\zeta=o(T^{3/2})$ as $T\rightarrow\infty$;
    \item[Case 2.]\label{case2} The number $n\rightarrow\infty$. Whether $T$ is fixed or $T\rightarrow\infty$,
    \begin{enumerate}
        \item\label{case2a} when $G$ is fixed, and $g_{min}=O(n^{1/2+\tilde\alpha_4})$ for some constant $\tilde{\alpha}_4<1/2$. Let $\zeta=o(n^{2\tilde{\alpha}_4}T^{3/2})$ approaching infinity;
        \item\label{case2b} when $G\rightarrow\infty$,
        \begin{enumerate}
            \item\label{case2bi} suppose $g_{min}=O(n^{7/9+\tilde{\alpha}_3})$ for some constant $\tilde{\alpha}_3<2/9$. Let $\zeta=O(G)$ approaching infinity;
            \item\label{case2bii} suppose  $g_{min}=O(n^{5/7+\tilde{\alpha}_5})$ for some constant $\tilde{\alpha}_5<2/7$. Let $\zeta=o(n^{7\tilde{\alpha}_5/2}T^{3/2})\gg G$ approaching infinity.
        \end{enumerate}
    \end{enumerate}
\end{enumerate}
\end{proof}

%%%%%%%%%%%%%%%%     Theorem 3    Heterogeneous    %%%%%%%%%%%%%%%%%%%%%%%%%%

\subsection{Convergence of the Calculated Estimator ($G\geq2$)% for  Heterogeneous Model
}\label{sec:oracle}

\begin{proof}[Proof of Theorem \ref{thm:main}]
This can be done similarly to the proof of Theorem 4.2 in \cite{ma2016estimating}. 

Define $\mathcal{M_G}:=\{\boldsymbol{\gamma}\in\mathbb{R}^{np}: \boldsymbol{\gamma}_i=\boldsymbol{\gamma}_j,\forall i,j\in\mathcal{G}_g,g=1,\cdots,G\}$ 
and the scaled penalty function as $\tilde{\rho}_\theta(\|\boldsymbol\gamma_i-\boldsymbol\gamma_j\|) = \lambda_1^{-1}{\rho}_\theta(\|\boldsymbol\gamma_i-\boldsymbol\gamma_j\|, \lambda_1)$. Let the least-squares objective function and the penalty function be
\begin{equation}\label{obj_partition}
\begin{aligned}
    &L(\boldsymbol{\gamma})=\frac{1}{2}\|\mathbf{y}-{W}\boldsymbol{\gamma}\|_2^2,~~P(\boldsymbol{\gamma})=\lambda_1\sum_{i<j}\tilde{\rho}_\theta(\|\boldsymbol{\gamma}_i-\boldsymbol{\gamma}_j\|_2)\\
    &L^\mathcal{G}(\boldsymbol{\varphi})=\frac{1}{2}\|\mathbf{y}-\mathbf{W}\Gamma\boldsymbol{\varphi}\|_2^2,~~P^\mathcal{G}(\boldsymbol{\varphi})=\lambda_1\sum_{g<g'}|\mathcal{G}_g\|\mathcal{G}_{g'}|\tilde{\rho}_\theta(\|\boldsymbol{\varphi}_g-\boldsymbol{\varphi}_{g'}\|_2).
\end{aligned}
\end{equation}
Let $Q_(\boldsymbol\gamma)=L(\boldsymbol{\gamma})+P(\boldsymbol{\gamma}), ~~ Q_(\boldsymbol\gamma)^\mathcal{G}(\boldsymbol{\varphi})=L^\mathcal{G}(\boldsymbol{\varphi})+P^\mathcal{G}(\boldsymbol{\varphi})$ and define
\begin{itemize}
    \item[$\diamond$] $F: \mathcal{M_G}\rightarrow \mathbb{R}^{Gp}$. The $g$-th vector component of $F(\boldsymbol{\gamma})$ equals to the common value of $\boldsymbol{\gamma}_i$ for $i\in\mathcal{G}_g$. 
    \item[$\diamond$] $F^*: \mathbb{R}^{np}\rightarrow \mathbb{R}^{Gp}$. $F^*(\boldsymbol{\gamma})=\{|\mathcal{G}_g|^{-1}\sum_{i\in\mathcal{G}_g}\boldsymbol{\gamma}_i',g=1,\cdots,G\}'$, which implies the average of each cluster vectors.
\end{itemize}
It results in that $F(\boldsymbol{\gamma})=F^*(\boldsymbol{\gamma})$ if $\boldsymbol{\gamma}\in\mathcal{M_G}$.
Hence, for every $\boldsymbol{\gamma}\in\mathcal{M_G}$, $P(\boldsymbol{\gamma})=P^\mathcal{G}(F(\boldsymbol{\gamma}))$, and
for every $\boldsymbol{\varphi}\in\mathbb{R}^{Gp}$, $P(F^{-1}(\boldsymbol{\varphi}))=P^\mathcal{G}(\boldsymbol{\varphi})$.
Hence,
\begin{equation}\label{objective}
    Q(\boldsymbol{\gamma})=Q^\mathcal{G}(F(\boldsymbol{\gamma})),~~Q^\mathcal{G}(\boldsymbol{\varphi})=Q(F^{-1}(\boldsymbol{\varphi})).
\end{equation}
Theorem \ref{thm:oracle} results in that for some $\iota>0$, 
$$P(\sup_i\|\widehat{\boldsymbol{\gamma}}^{or}_i-\boldsymbol{\gamma}_i^0\|_2\leq p\sup_i\|\widehat{\boldsymbol{\gamma}}_i^{or}-\boldsymbol{\gamma}_i^0\|_\infty= p\|\widehat{\boldsymbol{\gamma}}^{or}-\boldsymbol{\gamma}^0\|_\infty\leq p\phi_{n,T,G,\zeta})\geq1-e^{\iota},$$
there exists an event $E_1$ in which $\sup_i\|\widehat{\boldsymbol{\gamma}}^{or}_i-\boldsymbol{\gamma}_i^0\|_2\leq p\phi_{n,T,G}=\tilde{\phi}_{n,T,G}$, such that $P(E_1^C)\leq e^{-\iota}$. 
%$\sup_i\|\widehat{\boldsymbol{\gamma}}^{or}_i-\boldsymbol{\gamma}_i^0\|_2\leq \phi_{n,T,G}$, and $P(E_1^C)\leq e^{-\iota}$.
Consider the neighborhood of the true parameter $\boldsymbol{\gamma}^0$,
$$\Theta:=\{\boldsymbol{\gamma}\in\mathbb{R}^{np}:\sup_i\|\boldsymbol{\gamma}_i-\boldsymbol{\gamma}_i^0\|_2\leq\tilde{\phi}_{n,T,G}\}.$$
It implies that $\widehat{\boldsymbol{\gamma}}^{or}\in\Theta$ on the event $E_1$. For any $\boldsymbol{\gamma}\in\mathbb{R}^{np}$, let $\boldsymbol{\gamma}^*=F^{-1}(F^*(\boldsymbol{\gamma}))$, then $\boldsymbol{\gamma}^*_i=\frac{1}{|\mathcal{G}_g|}\sum_{i\in\mathcal{G}_g}\boldsymbol{\gamma}_i$ which implies that $\boldsymbol{\gamma}^*$ is a vector with duplicated group average of $\boldsymbol{\gamma}_i$.
Through two steps as the following, the statement can be proved that with probability approximating to 1, $\widehat{\boldsymbol{\gamma}}^{or}$ is a strictly local minimizer of $Q(\boldsymbol\gamma)$.
\begin{enumerate}[label=\roman*.]
    \item In $E_1$, $Q(\boldsymbol{\gamma}^*)>Q(\widehat{\boldsymbol{\gamma}}^{or})$ for any $\boldsymbol{\gamma}\in\Theta$ and $\boldsymbol{\gamma}^*\neq \widehat{\boldsymbol{\gamma}}^{or}$. This indicates that the oracle estimator $\widehat{\boldsymbol{\gamma}}^{or}$ is the minimizer over all duplicated group average $\boldsymbol{\gamma}^*$. \label{i}
    \item There exists an event $E_2$ such that for large enough $\iota^*$, $P(E_2^C)\leq e^{-\iota^*}$. In $E_1\cap E_2$, there exists a neighborhood $\Theta_n$ of $\widehat{\boldsymbol{\gamma}}^{or}$ such that $Q(\boldsymbol{\gamma})\geq Q(\boldsymbol{\gamma}^*)$ for all $\boldsymbol{\gamma}^*\in\Theta_n\cap\Theta$ for sufficiently large $n$. It means that for all $\boldsymbol{\gamma}$, the duplicated group average $\boldsymbol{\gamma}^*$ is the minimizer. \label{ii}
\end{enumerate}
Then, it results in $Q(\boldsymbol{\gamma})>Q(\widehat{\boldsymbol{\gamma}}^{or})$ for any $\boldsymbol{\gamma}\in\Theta_n\cap\Theta$ and $\boldsymbol{\gamma}\neq\widehat{\boldsymbol{\gamma}}^{or}$ in $E_1\cap E_2$.
Hence, over $E_1\cap E_2$, for large enough $\iota$ and $\iota^*$, $\widehat{\boldsymbol{\gamma}}^{or}$ is a strictly local minimizer of $Q(\boldsymbol{\gamma})$ with the probability $P(E_1\cap E_2)\geq 1-e^{-\iota}-e^{-\iota^*}$.

%\medskip
%\begin{proof}[Proof of \ref{i}.]
First, show $P^\mathcal{G}(F^*(\boldsymbol{\gamma}))=C$ for any $\boldsymbol{\gamma}\in\Theta$, where $C$ is a constant which does not depend on $\boldsymbol{\gamma}$.
It implies that when $\boldsymbol{\gamma}$ is close enough to the true parameter $\boldsymbol{\gamma}^0$, the penalty term would not affect the objective function with respect to different values of $\boldsymbol{\gamma}$.
Let $F^*(\boldsymbol{\gamma})=\boldsymbol{\varphi}$. 
%It suffices to show that $\|\boldsymbol{\varphi}_g-\boldsymbol{\varphi}_{g'}\|_2>a\lambda$ for all $g\neq g'$ and some constant $a>0$. Then by Assumption 6, $\rho(\|\boldsymbol{\varphi}_g-\boldsymbol{\varphi}_{g'}\|_2)$ is a constant, and as a result $P^\mathcal{G}(F^*(\boldsymbol{\varphi}))$ is a constant. 
Consider the triangle inequality $\|\boldsymbol{\varphi}_g-\boldsymbol{\varphi}_{g'}\|_2\geq\|\boldsymbol{\varphi}_g^{0}-\boldsymbol{\varphi}_{g'}^{0}\|_2-2\sup_g\|\boldsymbol{\varphi}_g-\boldsymbol{\varphi}_g^{0}\|_2$. 
Since $\boldsymbol{\gamma}\in\Theta$, then
\begin{equation}\label{phin}
\begin{aligned}
    &\sup_g\|\boldsymbol{\varphi}_g-\boldsymbol{\varphi}^{0}_g\|_2^2
    =\sup_g\left\||\mathcal{G}_g|^{-1}\sum_{i\in\mathcal{G}_g}\boldsymbol{\gamma}_i-\boldsymbol{\varphi}^{0}_g\right\|_2^2\\
    =&\sup_g\left\||\mathcal{G}_g|^{-1}\sum_{i\in\mathcal{G}_g}(\boldsymbol{\gamma}_i-\boldsymbol{\gamma}^{0}_i)\right\|_2^2
    =\sup_g|\mathcal{G}_g|^{-2}\left\|\sum_{i\in\mathcal{G}_g}(\boldsymbol{\gamma}_i-\boldsymbol{\gamma}^{0}_i)\right\|_2^2\\
    \leq&|\mathcal{G}_g|^{-1}\sup_g\sum_{i\in\mathcal{G}_g}\left\|(\boldsymbol{\gamma}_i-\boldsymbol{\gamma}^{0}_i)\right\|_2^2
    \leq\sup_i\left\|(\boldsymbol{\gamma}_i-\boldsymbol{\gamma}^{0}_i)\right\|_2^2\leq\tilde{\phi}_{n,T,G}^2,
\end{aligned}
\end{equation}
Since $b_{n,T,G}:=\min_{g\neq g'}\|\boldsymbol{\varphi}^0_g-\boldsymbol{\varphi}^0_{g'}\|$, then for all $g\neq g'$ and $b_{n,T,G}>a\lambda+2\tilde{\phi}_{n,T,G}$, 
$$\|\boldsymbol{\varphi}^0_g-\boldsymbol{\varphi}^0_{g'}\|_2\geq\|\boldsymbol{\varphi}^{0}_g-\boldsymbol{\varphi}^{0}_{g'}\|_2-2\sup_g\|\boldsymbol{\varphi}_g-\boldsymbol{\varphi}_g^{0}\|_2\geq b_{n,T,G}-2\tilde{\phi}_{n,T,G}>a\lambda_1,$$
for some $a>0$. Then by Assumption 6, $\rho(\|\boldsymbol{\varphi}_g-\boldsymbol{\varphi}_{g'}\|_2)$ is a constant, and furthermore, $P^\mathcal{G}(F^*(\boldsymbol{\varphi}))$ is a constant. 
Therefore, $P^\mathcal{G}(F^*(\boldsymbol{\gamma}))=C$, and $Q^\mathcal{G}(F^*(\boldsymbol{\gamma}))=L^\mathcal{G}(T^*(\boldsymbol{\gamma}))+C$ for all $\boldsymbol{\gamma}\in\Theta$. 
Since $\widehat{\boldsymbol{\varphi}}^{or}$ is the unique global minimizer of $L_n^\mathcal{G}(\boldsymbol{\varphi})$% according to the definition (\ref{oracle_gamma})
, then $L^\mathcal{G}(T^*(\boldsymbol{\gamma}))>L^\mathcal{G}(\widehat{\boldsymbol{\varphi}}^{or})$ for all $T^*(\boldsymbol{\gamma})\neq\widehat{\boldsymbol{\varphi}}^{or}$ and hence $Q^\mathcal{G}(T^*(\boldsymbol{\gamma}))>Q^{\mathcal{G}}(\widehat{\boldsymbol{\varphi}}^{or})$ for all $T^*(\boldsymbol{\gamma})\neq\widehat{\boldsymbol{\varphi}}^{or}$.
By the property of the clustering algorithm, for the $g$-th group, $\widehat{\boldsymbol{\varphi}}^{or}_g=|\mathcal{G}_g|^{-1}\sum_{i\in\mathcal{G}_g}\widehat{\boldsymbol{\gamma}}^{or}_i$, which  implies that, along with the definition of operation $F$,  $\widehat{\boldsymbol{\varphi}}^{or}_g$ equals to the $g$-th component of $F(\widehat{\boldsymbol{\gamma}}^{or})$ for all $i\leq g\leq G$. Then, by (\ref{objective}),
$$Q^\mathcal{G}(\widehat{\boldsymbol{\varphi}}^{or})=Q^\mathcal{G}(T(\widehat{\boldsymbol{\gamma}}^{or}))=Q(\widehat{\boldsymbol{\gamma}}^{or}).$$
Furthermore,  $Q_n^\mathcal{G}(T^*(\boldsymbol{\gamma}))=Q(T^{-1}(T^*(\boldsymbol{\gamma})))=Q(\boldsymbol{\gamma}^*)$.
Therefore, $Q(\boldsymbol{\gamma}^*)>Q(\widehat{\boldsymbol{\gamma}}^{or})$ for all $\boldsymbol{\gamma}^*\neq\widehat{\boldsymbol{\gamma}}^{or}$.

Second, for a positive sequence $r_n$, let $\Theta_n:=\{\boldsymbol{\gamma}_i:\sup_i\|\boldsymbol{\gamma}_i-\widehat{\boldsymbol{\gamma}}^{or}_i\|_2\leq r_n\}$. For any $\boldsymbol{\gamma}\in\Theta_n\cap\Theta$, by the first order Taylor's expansion,
$$Q(\boldsymbol{\gamma})-Q(\boldsymbol{\gamma}^*)
=\dfrac{d Q(\boldsymbol{\gamma}^m)}{d \boldsymbol{\gamma}'}(\boldsymbol{\gamma}-\boldsymbol{\gamma}^*)
=\dfrac{d L(\boldsymbol{\gamma}^m)}{d \boldsymbol{\gamma}'}(\boldsymbol{\gamma}-\boldsymbol{\gamma}^*)+\sum_{i=1}^n\dfrac{\partial P(\boldsymbol{\gamma}^m)}{\partial \boldsymbol{\gamma}_i'}(\boldsymbol{\gamma}-\boldsymbol{\gamma}^*),$$
and let $S_1=\dfrac{d L(\boldsymbol{\gamma}^m)}{d \boldsymbol{\gamma}'_i}(\boldsymbol{\gamma}-\boldsymbol{\gamma}^*_i)$ and $S_2=\sum_{i=1}^n\dfrac{\partial P(\boldsymbol{\gamma}^m)}{\partial \boldsymbol{\gamma}_i'}(\boldsymbol{\gamma}_i-\boldsymbol{\gamma}^*_i)$.
Since
$$\begin{aligned}
    \dfrac{d L(\boldsymbol{\gamma})}{\boldsymbol{\gamma}_i}
    &=\dfrac{1}{2}(-2\mathbf{y'}W+2{\boldsymbol{\gamma}}' W'W)
    =-(\mathbf{y'}-{\boldsymbol{\gamma}}'W)W~~~~{\rm and}\\
    \dfrac{\partial P(\boldsymbol{\gamma})}{\partial\boldsymbol{\gamma}_i}
    &=\lambda_1\sum_{i=1}^n\tilde{\rho}_\theta'(\|\boldsymbol{\gamma}_i-\boldsymbol{\gamma}_j\|_2)\dfrac{1}{2\|\boldsymbol{\gamma}_i-\boldsymbol{\gamma}_j\|_2}2(\boldsymbol{\gamma}_i-\boldsymbol{\gamma}_j)\\
    &=\lambda_1\sum_{i=1}^n\tilde{\rho}_\theta'(\|\boldsymbol{\gamma}_i-\boldsymbol{\gamma}_j\|_2)\dfrac{\boldsymbol{\gamma}_i-\boldsymbol{\gamma}_j}{\|\boldsymbol{\gamma}_i-\boldsymbol{\gamma}_j\|_2},
\end{aligned}$$
we have
$$\begin{aligned}
    S_1=-(\mathbf{y'}-{\boldsymbol{\gamma}^m}'W)W(\boldsymbol{\gamma-\gamma^*})~~{\rm and}~~
    S_2=\sum_{i=1}^n\frac{\partial P(\boldsymbol{\gamma}^m)}{\partial \boldsymbol{\gamma}_i'}(\boldsymbol{\gamma}_i-\boldsymbol{\gamma}_i^*).
\end{aligned}$$
Let $\boldsymbol{\gamma}^m=\vartheta\boldsymbol{\gamma}+(1-\vartheta)\boldsymbol{\gamma}^*$ for some constant $\vartheta\in(0,1)$. Then,
\begin{equation}\label{gamma2}
    \begin{aligned}
    S_2=&\lambda_1\sum_{i<j}\tilde{\rho}_\theta'(\|\boldsymbol{\gamma}_i^m-\boldsymbol{\gamma}_j^m\|_2)\|\boldsymbol{\gamma}_i^m-\boldsymbol{\gamma}_j^m\|_2^{-1}(\boldsymbol{\gamma}_i^m-\boldsymbol{\gamma}^m_j)'(\boldsymbol{\gamma}_i-\boldsymbol{\gamma}_i^*)\\
    &+\lambda_1\sum_{i>j}\tilde{\rho}_\theta'(\|\boldsymbol{\gamma}_i^m-\boldsymbol{\gamma}_j^m\|_2)\|\boldsymbol{\gamma}_i^m-\boldsymbol{\gamma}_j^m\|_2^{-1}(\boldsymbol{\gamma}_i^m-\boldsymbol{\gamma}^m_j)'(\boldsymbol{\gamma}_i-\boldsymbol{\gamma}_i^*)\\
    =&\lambda_1\sum_{i<j}\tilde{\rho}_\theta'(\|\boldsymbol{\gamma}_i^m-\boldsymbol{\gamma}_j^m\|_2)\|\boldsymbol{\gamma}_i^m-\boldsymbol{\gamma}_j^m\|_2^{-1}(\boldsymbol{\gamma}_i^m-\boldsymbol{\gamma}^m_j)'(\boldsymbol{\gamma}_i-\boldsymbol{\gamma}_i^*)\\
    &+\lambda_1\sum_{i<j}\tilde{\rho}_\theta'(\|\boldsymbol{\gamma}_j^m-\boldsymbol{\gamma}_i^m\|_2)\|\boldsymbol{\gamma}_j^m-\boldsymbol{\gamma}_i^m\|_2^{-1}(\boldsymbol{\gamma}_j^m-\boldsymbol{\gamma}^m_i)'(\boldsymbol{\gamma}_j-\boldsymbol{\gamma}_j^*)\\
    =&\lambda_1\sum_{i<j}\tilde{\rho}_\theta'(\|\boldsymbol{\gamma}_i^m-\boldsymbol{\gamma}_j^m\|_2)\|\boldsymbol{\gamma}_i^m-\boldsymbol{\gamma}_j^m\|_2^{-1}(\boldsymbol{\gamma}_i^m-\boldsymbol{\gamma}^m_j)'[(\boldsymbol{\gamma}_i-\boldsymbol{\gamma}_i^*)-(\boldsymbol{\gamma}_j-\boldsymbol{\gamma}_j^*)].
\end{aligned}
\end{equation}
%Consider to separate $\Gamma_2$ into two parts, $i$ and $j$ in the same group, i.e. $i,j\in\mathcal{G}_g$, and $i$ and $j$ belongs to different groups, i.e. $i\in\mathcal{G}_g$, $j\in\mathcal{G}_{g'}$ for $g\neq g'$.
Consider separating $S_2$ into two parts, $i,j\in\mathcal{G}_g$, and $i\in\mathcal{G}_g$, $j\in\mathcal{G}_{g'}$ for $g\neq g'$. 
When $i,j\in\mathcal{G}_g$, since $\boldsymbol{\gamma}^*=T^{-1}(T^*(\boldsymbol{\gamma}))\in\mathcal{M_G}$, then $\boldsymbol{\gamma}_i^*=\boldsymbol{\gamma}_j^*$. 
Thus, the RHS of (\ref{gamma2}) becomes
\begin{equation}\label{gamma2-2}
\begin{aligned}
S_2=&\lambda_1\sum_{g=1}^G\sum_{i,j\in\mathcal{G},i<j}\tilde{\rho}_\theta'(\|\boldsymbol{\gamma}_i^m-\boldsymbol{\gamma}_j^m\|_2)\|\boldsymbol{\gamma}_i^m-\boldsymbol{\gamma}_j^m\|_2^{-1}(\boldsymbol{\gamma}_i^m-\boldsymbol{\gamma}_j^m)'(\boldsymbol{\gamma}_i-\boldsymbol{\gamma}_j)\\
&+\lambda_1\sum_{g<g'}\sum_{i\in\mathcal{G}_g,j\in\mathcal{G}_{g'}}\tilde{\rho}_\theta'(\|\boldsymbol{\gamma}_i^m-\boldsymbol{\gamma}_j^m\|_2)\|\boldsymbol{\gamma}_i^m-\boldsymbol{\gamma}_j^m\|_2^{-1}(\boldsymbol{\gamma}_i^m-\boldsymbol{\gamma}_j^m)'[(\boldsymbol{\gamma}_i-\boldsymbol{\gamma}_i^*)-(\boldsymbol{\gamma}_j-\boldsymbol{\gamma}_j^*)].
\end{aligned}
\end{equation}
Furthermore, by (\ref{phin}), for any $\boldsymbol{\gamma}\in\Theta_n\cap\Theta$,  $F^*(\boldsymbol{\gamma})=\boldsymbol{\varphi}$, and therefore,  for all $i\in\mathcal{G}_g$, $\boldsymbol{\gamma}_i^*=\boldsymbol{\varphi}_g$. This lead to
\begin{equation}\label{I-1}
    \sup_i\|\boldsymbol{\gamma}_i^*-\boldsymbol{\gamma}_i^0\|_2^2=\sup_g\|\boldsymbol{\varphi}_g-\boldsymbol{\varphi}^{0}_g\|_2^2\leq\tilde{\phi}_{n,T,G}^2,
\end{equation}
where the inequality in (\ref{I-1}) is obtained by (\ref{phin}).
Since $\boldsymbol{\gamma}_i^m=\vartheta\boldsymbol{\gamma}_i+(1-\vartheta)\boldsymbol{\gamma}_i^*$, by the triangle inequality,
\begin{align*}
    \sup_i\|\boldsymbol{\gamma}_i^m-\boldsymbol{\gamma}^0_i\|_2
    &=\sup_i\|\vartheta\boldsymbol{\gamma}_i+(1-\vartheta)\boldsymbol{\gamma}_i^*-\boldsymbol{\gamma}^0_i\|_2 \nonumber\\
    &=\sup_i\|\vartheta\boldsymbol{\gamma}_i+(1-\vartheta)\boldsymbol{\gamma}_i^*-(\vartheta+1-\vartheta)\boldsymbol{\gamma}^0_i\|_2 \nonumber\\
    &\leq\vartheta\sup_i\|\boldsymbol{\gamma}_i-\boldsymbol{\gamma}^0_i\|_2+(1-\vartheta)\sup_i\|\boldsymbol{\gamma}_i^*-\boldsymbol{\gamma}_i^0\|_2 \nonumber\\
    &\leq\vartheta\tilde{\phi}_{n,T,G}+(1-\vartheta)\tilde{\phi}_{n,T,G}=\tilde{\phi}_{n,T,G}. \label{III}
\end{align*}
Hence, for $g\neq g'$, $i\in\mathcal{G}_g$, $j\in\mathcal{G}_{g'}$, 
$$\begin{aligned}
    \|\boldsymbol{\gamma}_i^m-\boldsymbol{\gamma}_j^m\|_2
    &=\|\boldsymbol{\gamma}_i^m-\boldsymbol{\gamma}_i^0-\boldsymbol{\gamma}_j^m+\boldsymbol{\gamma}_j^0\|_2
    \geq\|\boldsymbol{\gamma}_i^0-\boldsymbol{\gamma}_j^0\|_2-2\max_{1\leq k \leq n}\|\boldsymbol{\gamma}_k^m-\boldsymbol{\gamma}^0_k\|_2\\
    &\geq\min_{i\in\mathcal{G}_g,j'\in\mathcal{G}_{g'}}\|\boldsymbol{\gamma}_i^0-\boldsymbol{\gamma}_j^0\|_2-2\max_{1\leq k \leq n}\|\boldsymbol{\gamma}_k^m-\boldsymbol{\gamma}^0_k\|_2\geq b_{n,T,G}-2\tilde{\phi}_{n,T,G}>a\lambda_1.
\end{aligned}$$
Since $\tilde{\rho}_\theta(x)$ is constant for all $x\geq a\lambda_1$, then $\tilde{\rho}_\theta'(\|\boldsymbol{\gamma}_i^m-\boldsymbol{\gamma}_j^m\|_2)=0$. 
Therefore, following $\boldsymbol{\gamma}_i^m-\boldsymbol{\gamma}_j^m=\vartheta(\boldsymbol{\gamma}_i-\boldsymbol{\gamma}_j)$ for $i,j\in\mathcal{G}_g$, (\ref{gamma2-2}) becomes
\begin{align*}
S_2=&\lambda_1\sum_{g=1}^G\sum_{i,j\in\mathcal{G},i<j}\dfrac{\tilde{\rho}_\theta'(\|\boldsymbol{\gamma}_i^m-\boldsymbol{\gamma}_j^m\|_2)}{\|\boldsymbol{\gamma}_i^m-\boldsymbol{\gamma}_j^m\|_2}(\boldsymbol{\gamma}_i^m-\boldsymbol{\gamma}_j^m)'(\boldsymbol{\gamma}_i-\boldsymbol{\gamma}_j) && \nonumber\\
&+\lambda_1\sum_{g<g'}\sum_{i\in\mathcal{G}_g,j\in\mathcal{G}_{g'}}\dfrac{\tilde{\rho}_\theta'(\|\boldsymbol{\gamma}_i^m-\boldsymbol{\gamma}_j^m\|_2)}{\|\boldsymbol{\gamma}_i^m-\boldsymbol{\gamma}_j^m\|_2}(\boldsymbol{\gamma}_i^m-\boldsymbol{\gamma}_j^m)'[(\boldsymbol{\gamma}_i-\boldsymbol{\gamma}_i^*)-(\boldsymbol{\gamma}_j-\boldsymbol{\gamma}_j^*)] && \nonumber\\
=&\lambda_1\sum_{g=1}^G\sum_{i,j\in\mathcal{G}_g,i<j}\dfrac{\tilde{\rho}_\theta'(\|\boldsymbol{\gamma}_i^m-\boldsymbol{\gamma}_j^m\|_2)}{\|\boldsymbol{\gamma}_i^m-\boldsymbol{\gamma}_j^m\|_2}(\boldsymbol{\gamma}_i^m-\boldsymbol{\gamma}^m_j)'(\boldsymbol{\gamma}_i-\boldsymbol{\gamma}_j) %&& \tilde{\rho}_\theta'(\|\boldsymbol{\gamma}_i^m-\boldsymbol{\gamma}_j^m\|_2)=0 
\nonumber\\
=&\lambda_1\sum_{g=1}^G\sum_{i,j\in\mathcal{G}_g,i<j}\dfrac{\tilde{\rho}_\theta'(\|\boldsymbol{\gamma}_i^m-\boldsymbol{\gamma}_j^m\|_2)}{\|\vartheta(\boldsymbol{\gamma}_i-\boldsymbol{\gamma}_j)\|_2}\vartheta(\boldsymbol{\gamma}_i-\boldsymbol{\gamma}_j)'(\boldsymbol{\gamma}_i-\boldsymbol{\gamma}_j) %&& \boldsymbol{\gamma}_i^m-\boldsymbol{\gamma}_j^m=\vartheta(\boldsymbol{\gamma}_i-\boldsymbol{\gamma}_j)
\nonumber\\
=&\lambda_1\sum_{g=1}^G\sum_{i,j\in\mathcal{G}_g,i<j}\tilde{\rho}_\theta'(\|\boldsymbol{\gamma}_i^m-\boldsymbol{\gamma}_j^m\|_2)\|\boldsymbol{\gamma}_i-\boldsymbol{\gamma}_j\|_2. && 
\end{align*}
Furthermore, similarly to (\ref{phin}), for all $i\in\mathcal{G}_g$, $\boldsymbol{\gamma}_i^*=\boldsymbol{\varphi}_g$,
$
    \sup_i\|\boldsymbol{\gamma}_i^*-\widehat{\boldsymbol{\gamma}}^{or}_i\|^2_2=\sup_g\|\boldsymbol{\varphi}_g-\widehat{\boldsymbol{\varphi}}^{or}_g\|_2^2\leq\sup_i\|\boldsymbol{\gamma}_i-\widehat{\boldsymbol{\gamma}}_i^{or}\|^2_2.
$
Then, since $\boldsymbol{\gamma}^*_i=\boldsymbol{\gamma}^*_j$,
\begin{align*}
    \sup_i\|\boldsymbol{\gamma}_i^m-{\boldsymbol{\gamma}}_j^m\|_2
    &=\sup_i\|\boldsymbol{\gamma}_i^m-\boldsymbol{\gamma}_i^*-{\boldsymbol{\gamma}}_j^m+\boldsymbol{\gamma}_j^*\|_2  && \nonumber\\
    &\leq \|\boldsymbol{\gamma}_i^*-{\boldsymbol{\gamma}}_j^*\|_2+2\sup_i\|\boldsymbol{\gamma}_i^m-\boldsymbol{\gamma}_i^*\|_2
    \leq 2\sup_i\|\boldsymbol{\gamma}_i^m-\boldsymbol{\gamma}_i^*\|_2  && \nonumber\\
    &=2\sup_i\|\vartheta\boldsymbol{\gamma}_i+(1-\vartheta)\boldsymbol{\gamma}_i^*-\boldsymbol{\gamma}_i^*\|_2 \nonumber\\
    &=2\vartheta\sup_i\|\boldsymbol{\gamma}_i-\boldsymbol{\gamma}_i^*\|_2
    \leq2\sup_i\|\boldsymbol{\gamma}_i-\boldsymbol{\gamma}_i^*\|_2  && \nonumber\\
    &\leq2(\sup_i\|\boldsymbol{\gamma}_i-\widehat{\boldsymbol{\gamma}}_i^{or}\|_2+\sup_i\|\boldsymbol{\gamma}^*_i-\widehat{\boldsymbol{\gamma}}_i^{or}\|_2) %&& \mbox{triangular inequality}
    \nonumber\\
    &\leq4\sup_i\|\boldsymbol{\gamma}_i-\widehat{\boldsymbol{\gamma}}_i^{or}\|_2\leq4r_n. && %\mbox{since (\ref{II-2})}
\end{align*}
Hence, $\tilde{\rho}_\theta'(\|\boldsymbol{\gamma}_i^m-\boldsymbol{\gamma}_j^m\|_2)\geq\tilde{\rho}_\theta'(4r_n)$, because $\rho(x)$ is nondecreasing and concave as assumed in Assumption \ref{ass:penalty}. Then,
\begin{equation}\label{Gamma22}
    S_2\geq\lambda_1\sum_{g=1}^G\sum_{i,j\in\mathcal{G}_k,i<j}\tilde{\rho}_\theta'(4r_n)\|\boldsymbol{\gamma}_i-\boldsymbol{\gamma}_j\|_2.
\end{equation}
Let $U=(U_1',\cdots,U_n')'=[(\mathbf{y}-W\boldsymbol{\gamma}^m)'W]'$, then
\begin{align}
    S_1=&-U'(\boldsymbol{\gamma}-\boldsymbol{\gamma}^*)
    =-(U_1',\cdots,U_n')'\left(
    \begin{matrix}
    \boldsymbol{\gamma}_1-\boldsymbol{\gamma}_1^*\\
    \boldsymbol{\gamma}_2-\boldsymbol{\gamma}_2^*\\
    \vdots\\
    \boldsymbol{\gamma}_n-\boldsymbol{\gamma}_n^*\\
    \end{matrix}\right) =-\sum_{i=1}^n U_i'(\boldsymbol{\gamma}_i-\boldsymbol{\gamma}_i^*) && %\forall i\in\mathcal{G}_g, \boldsymbol{\gamma}_i^*=\frac{1}{|\mathcal{G}_g|}\sum_{j\in\mathcal{G}_g}\boldsymbol{\gamma}_j 
    \nonumber\\
    =&-\sum_{g=1}^G\sum_{i\in\mathcal{G}_g}\dfrac{1}{|\mathcal{G}_g|}U_i'\left(|\mathcal{G}_g|\boldsymbol{\gamma}_i-\sum_{j\in\mathcal{G}_g}\boldsymbol{\gamma}_j\right)  \nonumber\\
    =&-\sum_{g=1}^G\sum_{i\in\mathcal{G}_g}\dfrac{1}{|\mathcal{G}_g|}U_i'\sum_{j\in\mathcal{G}_g}\left(\boldsymbol{\gamma}_i-\boldsymbol{\gamma}_j\right)
    =-\sum_{g=1}^G\sum_{i,j\in\mathcal{G}_g}\frac{U_i'(\boldsymbol{\gamma}_i-\boldsymbol{\gamma}_j)}{|\mathcal{G}_g|}  \nonumber\\
    =&-\sum_{g=1}^G\sum_{i,j\in\mathcal{G}_g}\frac{U_i'(\boldsymbol{\gamma}_i-\boldsymbol{\gamma}_j)}{2|\mathcal{G}_g|}+\sum_{g=1}^G\sum_{i,j\in\mathcal{G}_g}\frac{U_j'(\boldsymbol{\gamma}_i-\boldsymbol{\gamma}_j)}{2|\mathcal{G}_g|}  \nonumber\\
    =&-\sum_{g=1}^G\sum_{i,j\in\mathcal{G}_g}\frac{(U_j-U_i)'(\boldsymbol{\gamma}_j-\boldsymbol{\gamma}_i)}{2|\mathcal{G}_g|}  \nonumber\\
    =&-\sum_{g=1}^G\sum_{i,j\in\mathcal{G}_g,i<j}\frac{(U_j-U_i)'(\boldsymbol{\gamma}_j-\boldsymbol{\gamma}_i)}{|\mathcal{G}_g|}. &&% \mbox{when } i=j,~\boldsymbol{\gamma}_i=\boldsymbol{\gamma}_j
    \label{Gamma1}
\end{align}

In addition, 
$U_i={W}_i'(\mathbf{y}_i-{W}_i\gamma_i^m)={W}_i'({W}_i\gamma_i^0+\boldsymbol{\varepsilon}_i-{W}_i\gamma_i^m)={W}_i'(\boldsymbol{\varepsilon}_i+{W}_i(\gamma_i^0-\gamma_i^m)),$
and then,
\begin{align*}
    \sup_i\|U_i\|_2%&\leq\sup_i\{\|{W}_i'(\boldsymbol{\varepsilon}_i+{W}_i(\gamma_i^0-\gamma_i^m))\|_2\} && % \mbox{triangular inequality}\nonumber\\
    &\leq\sup_i\{\|{W}_i'\boldsymbol{\varepsilon}_i\|_2+\|{W}_i'{W}_i(\gamma_i^0-\gamma_i^m)\|_2\} &&  \nonumber\\
   % &\leq\sup_i\|{W}_i'\boldsymbol{\varepsilon}_i\|_2+\sup_i\|{W}_i'{W}_i\|_2\|(\gamma_i^0-\gamma_i^m)\|_2 && %\mbox{since (\ref{III})}  \nonumber\\
    &\leq\sup_i\|{W}_i'\boldsymbol{\varepsilon}_i\|_2+\sup_i\sqrt{p}\|{W}_i'{W}_i\|_\infty\tilde{\phi}_{n,T,G} && %\mbox{remove $i$, use $L$-$\infty$} 
    \nonumber\\
    &\leq\sup_i\|{W}_i'\boldsymbol{\varepsilon}_i\|_2+m\sqrt{pT}(q^{1/2}+m^{1/2}(L+1+2K))\tilde{\phi}_{n,T,G}  && %\mbox{Convert to $\infty$ norm to remove $i$} 
    \nonumber\\
    &\leq\sup_i\sqrt{p}\|{W}_i'\boldsymbol{\varepsilon}_i\|_\infty+m\sqrt{pT}(q^{1/2}+m^{1/2}(L+1+2K))\tilde{\phi}_{n,T,G}  && %\mbox{$\max$ element of $i\leq\max$ of all} 
    \nonumber\\
    &\leq\sqrt{p}\|{W}'\boldsymbol{\varepsilon}\|_2+m\sqrt{pT}{(q^{1/2}+m^{1/2}(L+1+2K))}\tilde{\phi}_{n,T,G}  && %\mbox{vector $L$-$\infty$ norm $\leq$ $L$-2 norm} 
    \nonumber\\
    &=\sqrt{p}\|{W}'\boldsymbol{\varepsilon}\|_2+m\sqrt{pT}B_{q,m}\tilde{\phi}_{n,T,G},
\end{align*}
where $B_{q,m}={q^{1/2}+m^{1/2}(L+1+2K)}$.
By Lemma \ref{appendix:lem2},
$P\left[\|W'\boldsymbol{\varepsilon}\|_2^2>2\tilde{c}(np+2\sqrt{np\zeta^*}+2\zeta^*)m\tilde{M}\sqrt{T}B_{q,m}\sqrt{p}\right]\leq e^{-\iota^*}$,
where $B_{q,m}=(q^{1/2}+m^{1/2}(L+1+2K))$, $p=q+L+1+2K$, $\tilde{M}=\max(M_1,M_2,M_3,M_4)$ and $\tilde{c}$ given in Assumption \ref{ass:lambda} and \ref{ass:subgauss}. $\iota^*$ is defined in Lemma \ref{appendix:lem2}.
%Refer to \citet{hsu2012tail},
%$$P(\|{W}'\boldsymbol{\varepsilon}\|^2_2>tr({W}{W}')+2\sqrt{tr(({W}{W}')^2)\zeta^*}+2\|{W}{W}'\|_2\zeta^*)\leq e^{-\zeta^*}.$$
%Since
%$$\begin{aligned}
%    \|{W}{W}'\|_2=\|{W}'{W}\|_2&=\|diag(W_1'W_1,\cdots,W_n'W_n)\|_2\leq\max_i\|W_i'W_i\|_2\\
%    &\leq \sqrt{q+L+1+2K}\max_i\|W_i'W_i\|_\infty=\sqrt{q+L+1+2K}\max_i\left\|\begin{aligned}
%        {Z}_i'{Z}_i & & {Z}_i'\widetilde{{X}}_i\\
%        \widetilde{{X}}_i'{Z}_i & & \widetilde{{X}}_i'\widetilde{{X}}_i
%    \end{aligned}\right\|_\infty\\
%    &\leq m\sqrt{pT}B_{q,m}\sqrt{q+L+1+2K},
%\end{aligned}$$
%$$tr({W}{W}')=tr({W}'{W})\leq{np}\|{W}'{W}\|_2,$$
%$$tr(({W}{W}')^2)=tr(({W}'{W})^2)\leq{np}\|{W}'{W}\|^2_2,$$
%then
%$$\begin{aligned}
%    &tr({W}{W}')+2\sqrt{tr[({W}{W}')^2]\zeta^*}+2\|{W}{W}'\|_2\zeta^*\\
%    \leq& (np+2\sqrt{np\zeta^*}+2\zeta^*)\|{W}'{W}\|_2\\
%    \leq& {(np+2\sqrt{np\zeta^*}+2\zeta^*)}mT^{1/2}p^{1/2}\tilde{B}_{q,m},
%\end{aligned}$$
%where $\tilde{B}_{q,m}=B_{q,m}\sqrt{q+L+1+2K}$.
%Thus, there is an event $E_2$ such that $P(E_2^C)\leq e^{-\zeta^*}$, and over the event $E_2$,
%$$\sup_i\|Q_i\|_2\leq T^{1/4}(mp)^{1/2}(p^{1/4}\tilde{B}_{q,m}^{1/2}(np+2\sqrt{np\zeta^*}+2\zeta^*)^{1/2}+T^{1/4}m^{1/2}B_{q,m}\tilde{\phi}_{n,T,G}).$$
Then, over the event $E_2$,
\begin{align}
    &\left|\frac{(U_j-U_i)'(\boldsymbol{\gamma}_j-\boldsymbol{\gamma}_i)}{|\mathcal{G}_g|}\right|
    \leq g_{min}^{-1}\|U_j-U_i\|_2\|\boldsymbol{\gamma}_j-\boldsymbol{\gamma}_i\|_2 
    \leq g_{min}^{-1}2\sup_i\|U_i\|_2\|\boldsymbol{\gamma}_i-\boldsymbol{\gamma}_j\|_2 \nonumber\\
    \leq&2g_{min}^{-1}T^{1/4}(mp)^{1/2}\|\boldsymbol{\gamma}_i-\boldsymbol{\gamma}_j\|_2 \nonumber\\
    &\left(p^{1/4}\tilde{B}_{q,m}^{1/2}(np+2\sqrt{np\zeta^*}+2\zeta^*)^{1/2}+T^{1/4}m^{1/2}B_{q,m}\tilde{\phi}_{n,T,G}\right).\label{sum}
\end{align}
Therefore, by (\ref{Gamma22}), (\ref{Gamma1}) and (\ref{sum}), %\color{red} assume that $\zeta^*\leq n$ \color{black},
\begin{align*}
    &Q(\boldsymbol{\gamma})-Q(\boldsymbol{\gamma}^*)  \nonumber\\
    \geq&\sum_{g=1}^G\sum_{i,j\in\mathcal{G}_g,i<j}\|\boldsymbol{\gamma}_i-\boldsymbol{\gamma}_j\|_2  \nonumber\\
    &\left\{\lambda_1\tilde{\rho}_\theta'(4r_n)-2g_{min}^{-1}T^{1/4}(mp)^{1/2}(p^{1/4}\tilde{B}_{q,m}^{1/2}(np+2\sqrt{np\zeta^*}+2\zeta^*)^{1/2}\right. \nonumber\\
    &\left.+T^{1/4}m^{1/2}B_{q,m}\tilde{\phi}_{n,T,G})\right\}  \nonumber\\
    \geq&\sum_{g=1}^G\sum_{i,j\in\mathcal{G}_g,i<j}\|\boldsymbol{\gamma}_i-\boldsymbol{\gamma}_j\|_2  \nonumber\\
    &\left\{\lambda_1\tilde{\rho}_\theta'(4r_n)-B_1g_{min}^{-1}T^{1/4}(np+2\sqrt{np\zeta^*}+2\zeta^*)^{1/2}-B_2g_{min}^{-1}T^{1/2}\tilde{\phi}_{n,T,G}\right\}, \label{last}
\end{align*}
where $B_1=2(mp\tilde{B}_{q,m})^{1/2}p^{1/4}$ and $B_2=2mp^{1/2}B_{q,m}$.

Let $r_n=o(1)$, then $\tilde{\rho}_\theta'(4r_n)\rightarrow1$. Suppose that the following condition is true over the event $E_1\cap E_2$,
\begin{equation}\label{Thm3}
    B_1g_{min}^{-1}(np+2\sqrt{np\zeta^*}+2\zeta^*)^{1/2}T^{1/4}\to0,~~B_2pg_{min}^{-1}T^{1/2}{\phi}_{n,T,G}\to0,
\end{equation}
then $P\left(Q(\boldsymbol{\gamma})-Q(\boldsymbol{\gamma}^*)\geq0\right)\geq1-e^\iota-e^{\iota^*}$. %As $\zeta\to\infty$ and $\zeta^*\to\infty$, $Q_n(\boldsymbol{\gamma})-Q_n(\boldsymbol{\gamma}^*)\geq0$, with a slight abuse of notation.
Once (\ref{Thm3}) holds, $Q(\gamma)-Q(\gamma^*)\geq0$ with probability approaching to 1 as $\iota,\iota^*\to\infty$.

Note that $\zeta^*=\zeta^*_{n,T,G}$ can be chosen as any sequence of numbers, as long as $\zeta^*\to\infty$ to ensure $\iota^*\to\infty$.
In the following argument, conditions on $n$, $T$, $G$, and other numbers that satisfies (\ref{Thm3}) are spelled out: %The main idea is to derive the conditions only for Theorem 3. To show that the proposed estimator converges to the oracle estimator, which converges to the true parameter as well, the conditions in both Theorem 2 and Theorem 3 are supposed to be considered.
\begin{enumerate}
    \item As $T\to\infty$ with $n$ fixed, the proposed estimator does not  converge to the oracle estimator.
    
    \item As $n\to\infty$ with $T$ fixed, if conditions in Corollary \ref{cor1} are satisfied, the second part of (\ref{Thm3}) is true. It is enough discuss the conditions  for first part of (\ref{Thm3}).
  Choose $\zeta^*$ such that $\zeta^*\leq n$ and   $\zeta^*\to\infty$ as $n\to\infty$. Let $g_{min}\gg(p+2\sqrt{p}+2)^{1/2}n^{1/2}$. Since $(np+2\sqrt{np\zeta^*}+2\zeta^*)^{1/2}=(p+2\sqrt{p}+2)^{1/2}O(n^{1/2})$,  
        $$B_1g_{min}^{-1}(np+2\sqrt{np\zeta^*}+2\zeta^*)^{1/2}T^{1/4}\leq B_1T^{1/4}g_{min}^{-1}(p+2\sqrt{p}+2)^{1/2}O(n^{1/2}) \to0.$$

    \item[3-1.] Let $T,n\to\infty$. Consider the first part of (\ref{Thm3}). 
Choose $\zeta^*$ such that $\zeta^*\leq n$ and   $\zeta^*\to\infty$ as $n\to\infty$.
        Let  $g_{min}\gg(p+2\sqrt{p}+2)^{1/2}n^{1/2}T^{1/4}$. Then 
        $$B_1g_{min}^{-1}(np+2\sqrt{np\zeta^*}+2\zeta^*)^{1/2}T^{1/4}\leq B_1g_{min}^{-1}(p+2\sqrt{p}+2)^{1/2}n^{1/2}T^{1/4} \to0.$$

    \item[3-2.] Let $T,n\to\infty$. Consider the second part of (\ref{Thm3}).    
    \begin{enumerate}
     \item Suppose $G$ is fixed. Choose $\zeta$ such that $\zeta=o(n^{4\tilde{\alpha}_1}T^{1/2})$ and $\zeta\to\infty$ as $n,T\to\infty$.
     Let $g_{min}=O(n^{1/4+\tilde{\alpha_1}})$ for some positive constant $\tilde{\alpha}_1<3/4$.
     Then, $(Gp+2\sqrt{Gp\zeta}+2\zeta)^{1/2}= O(2\zeta^{1/2})$, and $$B_2pg_{min}^{-1}T^{1/2}\phi_{n,T,G}
            \leq B_2p{C}_6\dfrac{n^{1/2}}{g_{min}^2T^{1/4}}O(\zeta^{1/2})\stackrel{{n,T\to\infty}}{\longrightarrow}0,$$
            where $C_6=2C_{q,m}G^{3/4}$.

        \item Suppose $G\to\infty$.
        Choose $\zeta$ such that 
        $\zeta\leq G$ and $\zeta\rightarrow\infty$ as $n,T,G\to\infty$.
        Let $\frac{n^{7/13}}{T^{1/13}}\ll g_{min}<n/G$.  Then, $G\ll\frac{T^{1/13}}{n^{6/13}}$ and $Gp+2\sqrt{Gp\zeta}+2\zeta\leq(p+2\sqrt{p}+2)G=O(G)$. Further, since $G\leq n/g_{min}$, 
$$\begin{aligned}
    B_2pg_{min}^{-1}T^{1/2}\phi_{n,T,G}&\leq B_2p{C}_5\dfrac{n^{1/2}G^{3/4}T^{1/2}}{g_{min}^2T^{3/4}}O(G^{1/2})\\
    &\leq B_2p{C}_5\dfrac{n^{7/4}}{g_{min}^{13/4}T^{1/4}}O(1)\stackrel{{n,T,G\to\infty}}{\longrightarrow}0,
\end{aligned}$$
    where $C_5=C_{q,m}(p+2\sqrt{p}+2p)^{1/2}$, which is free from $n,T$ and $G$.
    
   \item Suppose $G\to\infty$. Let $g_{min}=O(n^{5/11+\tilde{\alpha}_7})$ for a positive constant $\tilde{\alpha}_7<6/11$. 
   Choose $\zeta$ such that 
    $G\ll \zeta$ and $\zeta=o(n^{11\tilde{\alpha}_7/2}T^{1/2})$.
Then,
$(Gp+2\sqrt{Gp\zeta}+2\zeta)^{1/2}=o((p+2\sqrt{p}+2)^{1/2}\zeta^{1/2})$. Since $G\leq n/g_{min}$, 
         $$B_2pg_{min}^{-1}T^{1/2}\phi_{n,T,G}
            %\leq B_2pg_{min}^{-1}T^{1/2}{C}_7\dfrac{n^{1/2}G^{3/4}}{g_{min}T^{3/4}}O(\zeta^{1/2})
            \leq B_2p{C}_7\dfrac{n^{5/4}}{g_{min}^{11/4}T^{1/4}}O(\zeta^{1/2})\stackrel{{n,T,G\to\infty}}{\longrightarrow}0,$$
            where $C_7=C_{q,m}(p+2\sqrt{p}+2)^{1/2}$, which is free of $n,T$ and $G$.
             
    \end{enumerate}
    
\end{enumerate}

Combining the above calculations and the proof of Corollary \ref{cor1},  the conditions for (\ref{Thm3}) can be summarized as follows:
 
\begin{enumerate}
    \item Suppose $n\rightarrow\infty$ with $T$ fixed. Let  $(p+2\sqrt{p}+2)^{1/2}n^{1/2}\ll g_{min}=O(n^{7/9+\tilde{\alpha}_0})\leq n/2,$ then (\ref{Thm3}) holds;
    \item Suppose $n,T\rightarrow\infty$ and $G$ is fixed. Let $g_{min}=O(n^{1/2+\tilde{\alpha}_4})$ for some constant $\tilde{\alpha}_4<1/2$. Then, (\ref{Thm3}) holds by choosing $\zeta$ and $\zeta^*$ such that  $\zeta=o(\min(n^{1+4\tilde{\alpha}_4}T^{1/2},\allowbreak n^{2\tilde{\alpha}_4}T^{3/2}))$ approaching infinity and $\zeta^*\leq n$ approaching infinity;

        \item Suppose $n,T,G\to\infty$.
        \begin{enumerate}
            \item Let $\max\left\{\frac{n^{7/13}}{T^{1/13}},(p+2\sqrt{p}+2)^{1/2}n^{1/2}\right\}\ll g_{min}=O(n^{7/9+\tilde{\alpha}_3})$ for some constant $\tilde{\alpha}_3<2/9$. Then, (\ref{Thm3}) holds  by choosing  $\zeta=O(G)$ and $\zeta^*\leq n$ approaching infinity;
            \item Let $g_{min}=O(n^{5/7+\tilde{\alpha}_5})$ for some constant $\tilde{\alpha}_5<2/7$. Then, (\ref{Thm3}) holds by choosing $\zeta=o(\min\{n^{10/7+11/2\tilde{\alpha}_5}T^{1/2},\allowbreak n^{7\tilde{\alpha}_5/2}T^{3/2}\})$.
        \end{enumerate}
    \end{enumerate}

\end{proof}

\bibliography{References}